\documentclass[12pt]{article}
\pdfoutput=1
\usepackage{amssymb,amsmath}
\usepackage{epsfig}
\usepackage{epstopdf}
\usepackage{latexsym}
\usepackage{graphicx}
\usepackage{booktabs}
\usepackage{bbm}
\usepackage{enumitem}
\usepackage[numbers,compress]{natbib}

\usepackage[margin=20pt,small]{caption}
\usepackage{subcaption}

\usepackage[toc]{appendix}

\usepackage{color}
\usepackage{datetime}
\usepackage[
      colorlinks=false,
      linkcolor=darkblue,  
      urlcolor=blue,    
      filecolor=blue,     
      citecolor=red,
linktocpage=true,
      pdfstartview=FitV,
      bookmarksopen=true,
	  hidelinks
      ]{hyperref}

\ifpdf
\fi

\newcommand*{\boxedcolor}{red}
\makeatletter
\renewcommand{\boxed}[1]{\textcolor{\boxedcolor}{%
  \fbox{\normalcolor\m@th$\displaystyle#1$}}}
\makeatother

\definecolor{cardinal}{rgb}{0.6,0,0}
\definecolor{darkgreen}{rgb}{0,0.5,0}
\definecolor{golden}{rgb}{0.92, 0.7, 0}
\definecolor{midnight}{rgb}{0, 0, 0.5}
\definecolor{darkblue}{rgb}{0.2, 0, 0.8}

\def\Re{{\rm Re}}

\def\im{{\rm i}}

\def\ds{\displaystyle}

\def\cN{{\cal N}}
\def\cO{{\cal O}}
\def\cP{{\cal P}}

\newcommand{\reef}[1]{(\ref{#1})}
\newcommand{\be}{\begin{equation}}
\newcommand{\ee}{\end{equation}}
\newcommand{\bea}{\begin{eqnarray}}
\newcommand{\eea}{\end{eqnarray}}

\newcommand{\df}{\Delta_{\Phi}}

\begin{document}  
%%%%%%%%%%%%%%%%%%%%%%%%%%%%%%

\begin{titlepage}
 
\medskip
\begin{center} 
{\Large \bf  Bootstrapping SCFTs with Four Supercharges}

\bigskip
\bigskip
\bigskip

{\bf Nikolay Bobev,${}^{(1)}$ Sheer El-Showk,${}^{(2,3,4)}$  Dalimil Maz\'a\v c,${}^{(5)}$  and Miguel F. Paulos${}^{(2)}$ \\ }
\bigskip
\bigskip
${}^{(1)}$
Instituut voor Theoretische Fysica, KU Leuven \\
Celestijnenlaan 200D, B-3001 Leuven, Belgium
\vskip 5mm
${}^{(2)}$ Theory Division, CERN, Geneva, Switzerland\\
\vskip 5mm
${}^{(3)}$ Sorbonne Universit\'{e}s, UPMC Univ Paris 06, UMR 7589\\ LPTHE, F-75005, Paris, France\\
\vskip 5mm
${}^{(4)}$ CNRS, UMR 7589, LPTHE, 75005, Paris, France\\
\vskip 5mm
${}^{(5)}$
Perimeter Institute for Theoretical Physics \\
31 Caroline Street North, ON N2L 2Y5, Canada\\
\bigskip
\texttt{nikolay@itf.fys.kuleuven.be,~sheer.el-showk@cern.ch, dalimil.mazac@gmail.com,~miguel.paulos@gmail.com } \\
\end{center}

\bigskip
\bigskip

\begin{abstract}

\noindent  
\end{abstract}

\noindent 

We study the constraints imposed by superconformal symmetry, crossing symmetry, and unitarity for theories with four supercharges in spacetime dimension $2\leq d\leq 4$. We show how superconformal algebras with four Poincar\'{e} supercharges can be treated in a formalism applicable to any, in principle continuous, value of $d$ and use this to construct the superconformal blocks for any $d\leq 4$. We then use numerical bootstrap techniques to derive upper bounds on the conformal dimension of the first unprotected operator appearing in the OPE of a chiral and an anti-chiral superconformal primary. We obtain an intriguing structure of three distinct kinks. We argue that one of the kinks smoothly interpolates between the $d=2$, $\mathcal N=(2,2)$ minimal model with central charge $c=1$ and the theory of a free chiral multiplet in $d=4$, passing through the critical Wess-Zumino model with cubic superpotential in intermediate dimensions. 
%We argue that two of the kinks are characterized by the chiral ring relation $\Phi^2=0$.

\end{titlepage}

%%%%%%%%%%%%%%%%%%%%%%%%%%%%%%%%%%%%%

\setcounter{tocdepth}{2}
\tableofcontents

\newpage

%%%%%%%%%%%%%%%%%%%%%%%%%%%%%%%%%%%%%
\section{Introduction}
%%%%%%%%%%%%%%%%%%%%%%%%%%%%%%%%%%%%%

Recently, there has been a remarkable revival of interest in the 40-year-old conformal bootstrap idea \cite{Polyakov:1974gs,Ferrara:1974ny,Ferrara:1974pt,Ferrara:1973yt,Ferrara:1971vh,Ferrara:1974nf,Ferrara:1973vz}. The basic method, developed in \cite{Rattazzi:2008pe}, has now matured to the point where it is possible to extract the spectrum of operator dimensions and Wilsonian operator product expansion (OPE) coefficients of particular conformal field theories (CFTs) with great accuracy \cite{ElShowk:2012ht,El-Showk2014a,ElShowk:2012hu,Kos:2014bka}. The  ingredients for the bootstrap program are minimal, namely conformal symmetry, unitarity, and crossing symmetry of the four-point function. Strikingly, this is sufficient to derive highly non-trivial, non-perturbative constraints on the space of generic conformal field theories. Beginning with \cite{Poland:2010wg,Rattazzi:2010yc}, it has been shown that, as it is natural to expect, imposing additional symmetries on the CFT allows one to obtain even stronger constraints. A particularly interesting possibility is to consider supersymmetry. Supersymmetry leads to exact results for specific quantities such as the dimensions of chiral operators. This is a nice complement to the conformal bootstrap approach, which, although very powerful -- it can determine and bound unprotected quantities -- is a somewhat blunt instrument, since it addresses general properties for the space of {\em all} consistent conformal field theories. When supersymmetry is combined with the conformal bootstrap we expect an interesting interplay where exact information is used to restrict bootstrap searches to specific theories or classes of theories whereupon one can obtain accurate information about the unprotected part of the spectrum.  

Bootstrap methods have been previously applied to theories with various amounts of supersymmetry. Theories with maximal supersymmetry are very constrained and thus particularly suited for analysis using bootstrap technology. This has been explored in four \cite{Beem2013b,Alday2014a,Alday:2014qfa,Alday2014} and three \cite{Chester:2014fya} dimensions where various bounds on the spectrum of conformal dimensions and OPE coefficients were found. It turns out that theories with at least sixteen superconformal charges in various dimensions admit a remarkable algebraic structure which leads to the closure of the crossing equations on the space of certain protected operators. This was uncovered in  \cite{Beem2013c} where it was shown that in four-dimensional CFTs with $\mathcal{N}=2$ supersymmetry one can solve analytically for correlation functions of some protected operators by exploiting an underlying chiral algebra. This feature was further explored to great efficacy in six \cite{Beem2014e}, four \cite{Beem2014c, Lemos:2014lua} and three dimensions \cite{Chester:2014mea}. These analytic  results can then be used as input to perform a numerical bootstrap analysis and obtain bounds on the spectrum of unprotected operators in these highly symmetric theories \cite{Beem2014g,Chester:2014mea}.

While theories with eight or more Poincar\'e supercharges are quite rigid and possess deep mathematical properties, their dynamics is highly constrained. Thus it is worthwhile to explore theories with less supersymmetry, which are harder to control, but perhaps of greater phenomenological interest. This motivates our study in this paper of CFTs with four Poincar\'e supercharges (eight superconformal charges). Some of the first papers on the modern incarnation of the bootstrap program studied $\mathcal N=1$ SCFTs in $d=4$ \cite{Poland:2010wg,Poland:2011ey,Berkooz:2014yda}. Very little has been done in two and three dimensions, a notable exception being the work in  \cite{Bashkirov:2013vya} which studied SCFTs with four superconformal charges in $d=3$, i.e. CFTs with $\mathcal{N}=1$ supersymmetry. 

In this paper, we aim to fill this gap by applying bootstrap methods to SCFTs with four Poincar\'e supercharges in any dimension in the range $2\leq d \leq 4$.  This corresponds to $\mathcal{N}=(2,2)$ and $\mathcal{N}=2$ theories in $d=2$ and $d=3$, respectively, and to $\mathcal{N}=1$ SCFTs in $d=4$. Moreover, one of the advantages of bootstrap methods is that they allow for a straightforward analytic continuation into fractional values of the spacetime dimension. This has been explored before in \cite{El-Showk2014}, where the numerical bootstrap results were successfully compared with analytic calculations for the Wilson-Fisher fixed point in $d < 4$. Here we follow a similar approach, tracking the four-supercharge version of the Wilson-Fisher fixed point from four to two dimensions. This CFT is simply the critical Wess-Zumino  (cWZ) model, i.e. the theory of a single chiral superfield with cubic superpotential at its infrared fixed point. As compared to the non-supersymmetric case, we have a lot less room for error here, since the conformal dimension of the ``spin'' field is protected and equal to $(d-1)/3$ in any spacetime dimension $d$. Remarkably, we show that bounds on the dimension of the leading scalar operator in a chiral-antichiral OPE exhibit ``kinks'' (as in {\em e.g.} \cite{ElShowk:2012ht,Rychkov:2009ij}) at precisely this point for all $2\leq d<4$. They range from the $(2,2)$ $c=1$ supersymmetric minimal model in $d=2$, where the numerical bootstrap agrees with various exact results, all the way to the free theory in $d=4$. In $d=3$, we find a kink at conformal dimension $2/3$, and are able to read off the dimension of the leading unprotected scalar, which is approximately 1.9098. Also in $d=3$, our bootstrap prediction for the two-point function of the stress-tensor is in close agreement with the exact localization calculation of \cite{Nishioka:2013gza}. Furthermore, for $d\lesssim4$ our results for the dimension of the leading unprotected scalar agree with those of the one-loop $\epsilon$-expansion \cite{Thomas}. This strongly suggests that in 3d our kink does indeed describe the super-Ising model. This theory is of some interest in condensed matter physics \cite{Lee:2006if,Yu:2010zv,Ponte:2012ru,Grover:2013rc}, and we perform a detailed analysis of its properties in a companion paper \cite{BEMP2}.

Part of the difficulty in bootstrapping supersymmetric theories lies in determining the form of the superconformal blocks. Supersymmetry organizes conformal into superconformal multiplets, and accordingly conformal blocks of primaries with different dimensions and spins also become grouped. The calculation of superconformal blocks for general external operators can be a cumbersome technical problem. In this paper, we find the superconformal blocks in theories with four supercharges for external scalar superconformal primary operators with arbitrary scaling dimensions. A crucial ingredient is that at least two of these operators should be chiral primaries. Our approach is facilitated by the existence of a formal dimensional continuation of the superconformal algebra with four supercharges to arbitrary dimension $d\leq 4$. The commutation relations for the ordinary conformal algebra formally make sense when we let the spacetime vector indices run from $1$ to $d$, since Jacobi identities can be verified without specializing to a fixed integer $d$. Here we show that this picture can be extended to include fermionic generators, namely four Poincar\'{e} supercharges and four conformal supercharges. We are able to write down (anti)commutation relations among all generators without specializing to a fixed $d$ and demonstrate the validity of all super-Jacobi identities in an essentially dimension-independent way. The superconformal blocks can then be found in general $d$ using the fact that they are eigenfunctions of the quadratic Casimir operator of this superconformal algebra. This method is similar to the way in which non-supersymmetric conformal blocks were found in \cite{Dolan:2003hv}. Our general results reduce to previously studied cases, namely $d=2$ and $d=4$ \cite{Poland:2010wg, Khandker:2014mpa, Fitzpatrick2014}. Remarkably, we find that for any dimension, the superconformal blocks take the same functional form as ordinary, non-supersymmetric, blocks where the dimensions of external and propagating operators are shifted. This fact was also observed for $d=2,4$ in \cite{Fitzpatrick2014}.

We begin our exploration in the next section, where we describe a construction of the superconformal algebras in $d\leq 4$ in a unified framework. This is necessary in order to properly define the Casimir operator of the algebra and its action on local operators, which is used in Section \ref{sec:superblocks} to find the superconformal blocks.  In Section \ref{sec:WZ}, we provide a short review of the properties of the critical Wess-Zumino model in general dimension. Section \ref{sec:equations} describes the set of crossing equations that we utilize in the numerical bootstrap procedure, the results of which are presented and discussed in Section \ref{sec:results}. We finish with a discussion in Section \ref{sec:discussion}. Several appendices complement the main text with technical results.

%%%%%%%%%%%%%%%%%%%%%%%%%%%%%%%%%%%%%
\section{Superconformal algebra in continuous dimension}
\label{sec:superalgebras}
%%%%%%%%%%%%%%%%%%%%%%%%%%%%%%%%%%%%%

%%%%%%%%%%%%%%%%%
\subsection{General results}
\label{subsec:interpolating}
%%%%%%%%%%%%%%%%%

We begin by presenting what we would like to call the dimensional continuation of the superconformal algebra with four Poincar\'e supercharges to an arbitrary spacetime dimension $d \leq 4$. The superconformal algebras in the traditional sense exist only for integer values of $d$. We will show however that some insight can be gained by considering a set of (anti)commutation relations which formally make sense for any real $0\leq d\leq 4$, such that we obtain the corresponding superconformal algebras for integer $d$. We believe that this language is useful because it allows us to
\begin{itemize}
\item cast the $d=4$ $\mathcal{N}=1$, $d=3$ $\mathcal{N}=2$, $d=2$ $\mathcal{N}=(2,2)$ and $d=1$ $\mathcal{N}=4$ superconformal algebras in a unified way, where $d$ enters only as a real parameter in the (anti)commutation relations (besides defining the range for the spacetime vector index),
\item derive unitarity bounds on highest-weight representations for the whole $d$-dependent family in a unified manner, and verify they reduce to the correct results for the algebras in integer $d$,
\item find the superconformal blocks as analytic functions of $d$.
\end{itemize}
It is a well-known fact that the $d=3$, $\mathcal{N}=2$, and $d=2$, $\mathcal{N}=(2,2)$ Poincar\'e supersymmetry algebras are dimensional reductions of the $\mathcal{N}=1$ algebra in $d=4$, with the extra $U(1)$ R-symmetry in $d=2$ coming from rotations in the two ``transverse" dimensions. Here, we generalize this dimensional reduction to the full superconformal algebra. Imposing the Jacobi identities in a superconformal algebra leads to non-trivial polynomial relations for the generators of the Clifford algebra, this being the essential reason for the scarcity of superconformal algebras \cite{Nahm:1977tg}. Consistency of our approach requires a continuous version of these identities valid in any $d\leq4$. The identities can be checked for any $d=0,1,\ldots,4$ and at the level of traces even for continuous $d$. The superconformal algebra thus exists in continuous dimension in the same sense as the ordinary conformal algebra, where Jacobi identities can be checked formally without fixing spacetime dimension. Moreover, we show in Section \ref{sec:superblocks} that superconformal blocks can be derived as analytic functions of $d$ exactly as in the non-supersymmetric case \cite{Dolan:2000ut,Dolan:2003hv}.

We work in Euclidean signature, with reality conditions equivalent to those imposed by unitarity in Lorentzian signature.\footnote{The four-dimensional Euclidean superconformal algebra with four Poincar\'e supercharges does not admit unitary representations. This is not important for us since we insist on unitarity in Lorentzian signature. We thank Toine Van Proeyen for emphasizing this point.} The unhatted Latin indices will run over the unreduced spacetime directions $i=1,\ldots,d$, while the hatted indices over the reduced ones $\hat i= d+1,\ldots,4$. The bosonic generators include the usual momenta $P_i$, special conformal $K_i$, dilation $D$, and rotation $M_{ij}$ generators, with $i,j=1,\ldots, d$, the $U(1)$ R-symmetry $R$, and finally the rotations in the reduced dimensions $M_{\hat i \hat j}$. Because of our formal approach, it is important to keep the reduced rotations $M_{\hat i \hat j}$ for any $d$, although there are no physical generators for $d>2$. In our conventions, the bosonic commutation relations are
\begin{equation}
\begin{aligned} 
 \lbrack M_{ij},M_{kl}] &= -\im (\delta_{il}M_{jk} + \delta_{jk}M_{il} - \delta_{ik}M_{jl} - \delta_{jl}M_{ik})\,,\\
 \lbrack M_{\hat i \hat j},M_{\hat k \hat l}] &= -\im (\delta_{\hat i \hat l}M_{\hat j \hat k} + \delta_{\hat j \hat k}M_{\hat i \hat l} - \delta_{\hat i \hat k}M_{\hat j \hat l} - \delta_{\hat j \hat l}M_{\hat i \hat k})\,,\\
 [M_{ij},P_k] &= -\im(\delta_{jk}P_i - \delta_{ik}P_j)\,,\\
 [M_{ij},K_k] &= -\im(\delta_{jk}K_i - \delta_{ik}K_j)\,,\\ 
 [D,P_i] &= -\im P_i\,,\\
 [D,K_i] &= \im K_i\,,\\
 [P_i,K_j] &= -2 \im(\delta_{ij}D + M_{ij})\,,
\end{aligned}
\label{bosonicCA}
\end{equation}
with all other commutators vanishing. The Hermitian conjugation rules are
\begin{equation}
D^{\dagger} = -D\;, \qquad R^{\dagger} = R\;, \qquad M_{ij}^{\dagger} = M_{ij}\;, \qquad M_{\hat i \hat j}^{\dagger} = M_{\hat i\hat j}\;, \qquad P_{i}^{\dagger} = K_i\;.
\label{eq:conjbos}
\end{equation}
We note that in our conventions, the action of the dilation generator $D$ on an operator $\mathcal{O}$ is $[D, \mathcal{O}] = -\im\Delta \mathcal{O}$, where $\Delta$ is the conformal dimension of $\mathcal{O}$.

The fermionic generators include four Poincar\'{e} supercharges, $Q$, and four conformal supercharges, $S$. The former will be denoted $Q^{+}_{\alpha}$, $Q^{-}_{\dot\alpha}$, with $\alpha,\dot\alpha=1,2$, where the upper index denotes the R-charge eigenvalue, $\pm 1$, and the lower index transforms under the $SO(d)\times SO(4-d)$ rotations. As indicated by the dot, the supercharges with different R-charge are allowed to transform non-equivalently under rotations, as is the case for $\mathcal{N}=1$ in $d=4$. The conformal supercharges are Hermitian conjugates of the Poincar\'{e} supercharges
\begin{equation}
S^{\alpha -} = (Q_{\alpha}^+)^\dagger\,,\qquad\qquad S^{\dot\alpha +} = (Q_{\dot\alpha}^-)^\dagger\,.
\label{eq:conjSQ}
\end{equation}
With this convention for placement of indices, contraction of an upper and a lower index of the same kind is an invariant operation since any representation of the compact group $SO(d)\times SO(4-d)$ is unitary. Let us now sketch how the structure of the superconformal algebra, with the  generators above, follows from the Jacobi identities. With the exception of the anticommutator of Poincar\'{e} and conformal supercharges, we simply reproduce the $d=4$, $\mathcal{N}=1$ superconformal algebra, but we think it is worthwhile to show how the structure emerges in a $d$-independent language. 

Jacobi identities involving $D$ or $R$ imply that both sides of an (anti)commutation relation must have the same scaling dimension and R-charge, and we always impose these constraints in what follows. Furthermore repeated indices always imply a summation. The basic building block is the anticommutator of Poincar\'{e} supercharges, which takes the form
\begin{equation}
\{Q_{\alpha}^+,Q_{\dot \alpha}^{-}\} = \Sigma_{\alpha\dot\alpha}^i P_{i}\,,
\label{eq:qq}
\end{equation}
where $\Sigma_{\alpha\dot\alpha}^i$ is an, as yet, unspecified tensor. Using the conjugation rules \eqref{eq:conjbos}, \eqref{eq:conjSQ} one finds
\begin{equation}
\{S^{\dot\alpha +},S^{\alpha -}\} = \bar{\Sigma}^{\dot\alpha \alpha}_i K_{i}\,,
\end{equation}
with $\bar{\Sigma}^{\dot\alpha \alpha}_i = (\Sigma_{\alpha\dot\alpha}^i)^*$. The only generators that can appear in the anticommutator of a Poincar\'{e} and a conformal supercharge are $D$, $R$, $M_{ij}$ and $M_{\hat i \hat j}$. Rotation invariance dictates that $D$ comes multiplied with one of the invariant tensors $\delta^{\alpha}_{\phantom{a}\beta}$, $\delta^{\dot\alpha}_{\phantom{a}\dot\beta}$. Let us normalize our supercharges so that
\begin{equation}
\begin{aligned}
\{S^{\alpha -},Q^+_{\beta}\} &= \im\delta^{\alpha}_{\phantom{a}\beta}D + \ldots\,, \\
\{S^{\dot\alpha +},Q^-_{\dot\beta}\} &= \im\delta^{\dot\alpha}_{\phantom{a}\dot\beta}D + \ldots\,,
\end{aligned}
\end{equation}
where the dots stand for the contribution of other generators. The Jacobi identities coming from the triplets $[Q_{\alpha}^+,Q_{\dot \alpha}^{-},K_i]$ and $[S^{\dot\alpha +},S^{\alpha -},P_i]$ then determine the following commutators
\begin{equation}
\begin{aligned}
\lbrack K_i,Q_{\alpha}^+] &= \Sigma_{\alpha\dot\alpha}^i S^{\dot\alpha +}\,, \\
[K_i,Q_{\dot\alpha}^-] &= \Sigma_{\alpha\dot\alpha}^i S^{\alpha -}\,, \\
[P_i,S^{\dot\alpha +}] &=- \bar{\Sigma}^{\dot\alpha \alpha}_i Q_{\alpha}^+\,, \\
[P_i,S^{\alpha -}] &=- \bar{\Sigma}^{\dot\alpha \alpha}_i Q_{\dot\alpha}^-\,.
\end{aligned}
\label{eq:qksp}
\end{equation}
We denote the representation matrices of rotations on the supercharges as $m_{ij}$, $\tilde{m}_{ij}$, i.e.
\begin{equation}
\begin{aligned}
\lbrack M_{ij}, Q_{\alpha}^+ ] &= (m_{ij})^{\phantom{a}\beta}_{\alpha}Q_{\beta}^{+}\,,\\
\lbrack M_{ij}, Q_{\dot\alpha}^- ] &= (\tilde{m}_{ij})^{\dot\beta}_{\phantom{a}\dot\alpha}Q_{\dot\beta}^{-}\,,\\
\lbrack M_{ij}, S^{\dot\alpha+} ] &= -(\tilde{m}_{ij})^{\dot\alpha}_{\phantom{a}\dot\beta}S^{\dot\beta+}\,,\\
\lbrack M_{ij}, S^{\alpha-} ] &= -(m_{ij})^{\phantom{a}\alpha}_{\beta}S^{\beta -}\,,
\end{aligned}
\label{eq:rotations}
\end{equation}
where the latter two follow from the former two using the conjugation rules in \eqref{eq:conjbos} and \eqref{eq:conjSQ}. Note that the matrices $m_{ij}$, $\tilde{m}_{ij}$ are necessarily antisymmetric in the space-time indices. The Jacobi identities for the triplets $[P_i,K_j,Q_{\alpha}^+]$ and $[P_i,K_j,Q_{\dot \alpha}^{-}]$ imply
\begin{equation}
\begin{aligned}
\Sigma_j\bar{\Sigma}_i &= \delta_{ij} + 2 \im m_{ij}\,,\\
\bar{\Sigma}_i\Sigma_j &= \delta_{ij} + 2 \im \tilde{m}_{ij}\,.
\end{aligned}
\end{equation}
Taking the symmetric parts implies that the $\Sigma_i$ tensors satisfy the Clifford algebra
\begin{equation}
\begin{aligned}
\Sigma_i\bar{\Sigma}_j + \Sigma_j\bar{\Sigma}_i &= 2\delta_{ij}\,,\\
\bar{\Sigma}_i\Sigma_j + \bar{\Sigma}_j\Sigma_i&= 2\delta_{ij}\,,
\end{aligned}
\label{eq:clifford1}
\end{equation}
while taking the antisymmetric parts leads to explicit formulas for the rotation generators in terms of $\Sigma_i$
\begin{equation}
\begin{aligned}
m_{ij} &= -\frac{\im}{4}(\Sigma_j\bar{\Sigma}_i - \Sigma_i\bar{\Sigma}_j)\,,\\
\tilde{m}_{ij} &= -\frac{\im}{4}(\bar{\Sigma}_i\Sigma_j - \bar{\Sigma}_j\Sigma_i)\,.
\end{aligned}
\label{eq:mgenerators}
\end{equation}
Since we would like our algebras to be related by the dimensional reduction, we will take \eqref{eq:rotations}, \eqref{eq:clifford1}, and \eqref{eq:mgenerators} to hold also for the hatted indices $\hat{i},\hat{j}=d+1,\ldots,4$, thus defining the action of the extra R-symmetry. It remains to determine the anticommutators between Poincar\'{e} and conformal supercharges, i.e. $\{S^{\alpha -},Q^+_{\beta}\}$ and $\{S^{\dot\alpha +},Q^-_{\dot\beta}\}$. It follows from the $[S^{\alpha-},Q_{\beta}^+,P_i]$, $[S^{\dot\alpha+},Q_{\dot\beta}^-,P_i]$ Jacobi identities that $M_{ij}$ appears contracted with the corresponding tensor $m_{ij}$ or $\tilde{m}_{ij}$ with unit coefficient. The most general form of the anticommutators which can be checked, using only \eqref{eq:clifford1}, to be consistent with all Jacobi identities except for those involving three fermionic generators, is
\begin{equation}
\begin{aligned}
\{S^{\alpha -},Q^+_{\beta}\} &= \delta^{\alpha}_{\phantom{a}\beta}(\im D-aR) + (m_{ij})^{\phantom{a}\alpha}_{\beta} M_{ij} + b (m_{\hat{i}\hat{j}})^{\phantom{a}\alpha}_{\beta} M_{\hat{i}\hat{j}}\,, \\
\{S^{\dot\alpha +},Q^-_{\dot\beta}\} &= \delta^{\dot\alpha}_{\phantom{a}\dot\beta}(\im D+aR) + (\tilde{m}_{ij})^{\dot\alpha}_{\phantom{a}\dot\beta} M_{ij} + b (\tilde{m}_{\hat{i}\hat{j}})^{\dot\alpha}_{\phantom{a}\dot\beta} M_{\hat{i}\hat{j}}\,,
\end{aligned}
\end{equation}
for some real constants $a,b$. It remains to check whether the Jacobi identities involving three fermionic generators are satisfied. The Jacobi identity coming from the triplet $[Q_{\alpha}^+,Q_{\dot\alpha}^-,S^{\beta-}]$ leads to
\begin{equation}
\bar{\Sigma}_i^{\dot\alpha\alpha}\Sigma^i_{\beta\dot\beta} = \frac{2a+1}{2}\delta^{\alpha}_{\phantom{a}\beta}\delta^{\dot\alpha}_{\phantom{a}\dot\beta}+
(m_{ij})^{\phantom{a}\alpha}_{\beta}(\tilde{m}_{ij})^{\dot\alpha}_{\phantom{a}\dot\beta} +b
(m_{\hat i \hat j})^{\phantom{a}\alpha}_{\beta}(\tilde{m}_{\hat i \hat j})^{\dot\alpha}_{\phantom{a}\dot\beta}\, .
\label{eq:id1}
\end{equation}
The rotation generators are traceless in the spinor indices, so taking the trace of this equation with respect to both pairs of indices and noting that \eqref{eq:clifford1} implies
\begin{equation}
\mathrm{tr}(\Sigma^i\bar{\Sigma}_i) =2 d\,, 
\end{equation}
we find
\begin{equation}
a = \frac{d-1}{2}\,.
\end{equation}
To fix $b$, we consider the Jacobi identity of the triplet $[Q_{\alpha}^+,Q_{\beta}^+,S^{\gamma-}]$, which leads to
\begin{equation}
\frac{d-2}{2}\delta^{\alpha}_{\phantom{a}\beta}\delta^{\gamma}_{\phantom{a}\delta} + (\alpha\leftrightarrow\gamma) = (m_{ij})^{\phantom{a}\alpha}_{\beta}(m_{ij})^{\phantom{a}\gamma}_{\delta} + b (m_{\hat i \hat j})^{\phantom{a}\alpha}_{\beta}(m_{\hat i \hat j})^{\phantom{a}\gamma}_{\delta} + (\alpha\leftrightarrow\gamma)\,.
\label{eq:id2}
\end{equation}
There is an analogous identity with dotted indices. Contracting all spinor indices and using that \eqref{eq:clifford1} and \eqref{eq:mgenerators} imply 
\begin{equation}
(m_{i j})^{\phantom{a}\alpha}_{\beta}(m_{ i j})^{\phantom{a}\beta}_{\alpha} = \frac{d(d-1)}{2}\,,\quad
(m_{\hat i \hat j})^{\phantom{a}\alpha}_{\beta}(m_{\hat i \hat j})^{\phantom{a}\beta}_{\alpha} = \frac{(4-d)(3-d)}{2}\,,
\end{equation}
so we find
\begin{equation}
3(d-2) = \frac{d(d-1)}{2} + b\frac{(4-d)(3-d)}{2}\,,
\end{equation}
which holds in continuous $d$ for $b = -1$. The final form of the sought anticommutators is thus
\begin{equation}
\begin{aligned}
\{S^{\alpha -},Q^+_{\beta}\} &= \delta^{\alpha}_{\phantom{a}\beta}\left(\im D-\frac{d-1}{2}R\right) + (m_{ij})^{\phantom{a}\alpha}_{\beta} M_{ij} - (m_{\hat{i}\hat{j}})^{\phantom{a}\alpha}_{\beta} M_{\hat{i}\hat{j}}\,,\\
\{S^{\dot\alpha +},Q^-_{\dot\beta}\} &= \delta^{\dot\alpha}_{\phantom{a}\dot\beta}\left(\im D+\frac{d-1}{2}R\right) + (\tilde{m}_{ij})^{\dot\alpha}_{\phantom{a}\dot\beta} M_{ij} - (\tilde{m}_{\hat{i}\hat{j}})^{\dot\alpha}_{\phantom{a}\dot\beta} M_{\hat{i}\hat{j}}\,.
\end{aligned}
\label{eq:sq}
\end{equation}
We have demonstrated that identities \eqref{eq:id1}, \eqref{eq:id2} are satisfied in any $d$ after contracting the spinor indices. We do not know of a $d$-independent way to argue for their validity in their uncontracted form. However, one can make an explicit choice of the 4d $\Sigma$ matrices satisfying \eqref{eq:clifford1}, and check the identities for all dimensions of interest. Indeed, they are satisfied for any consistent choice of $\Sigma$ matrices for any $d = 0, 1,\ldots,4$. This exhausts all the constraints imposed by Jacobi identities, thus showing that our algebra is consistent in any $d\leq 4$.

%%%%%%%%%%%%%%%%%%%%%%%%%%%%%
\subsection{Realizations in integer $d\leq 4$}
\label{subsec:SCArealization}
%%%%%%%%%%%%%%%%%%%%%%%%%%%%%

In this section, we illustrate that our interpolation reduces to the expected algebras in integer number of dimensions. This is of course necessary since they are the unique superconformal algebras with four Poincar\'{e} supercharges and a $U(1)$ R-symmetry in the respective number of spacetime dimensions. 

For $d=4$, our algebra is manifestly the $\mathcal{N}=1$ superconformal algebra with complexification $sl(4|1)$, with two pairs of Poincar\'{e} supercharges, with opposite R-charge, transforming in the two inequivalent Weyl representations. The value $a=(d-1)/2=3/2$ leads to the well-known chirality condition on scalar superconformal primaries, $\Delta = 3q/2$, with $\Delta$ the dimension and $q$ the R-charge.

For $d=3$, we reproduce the $\mathcal{N}=2$ superconformal algebra, whose complexification is $osp(2|4)$. Choosing $\Sigma^i_{\alpha\dot\alpha} = (\sigma_i)^{\dot\alpha}_{\phantom{a}\alpha}$ for $i=1,2,3$, where $\sigma_i$ are the usual Pauli matrices, we find that $Q_{\dot\alpha}^{-}$ transforms as a complex conjugate of $Q_{\alpha}^{+}$, so that a lower (upper) dotted index is equivalent to an upper (lower) undotted index, and all indices can be raised and lowered using $\epsilon^{\alpha\beta}$, $\epsilon_{\alpha\beta}$. The complete algebra is presented in Appendix \ref{App:opeder}.

The relevant superconformal algebra in two dimensions is the global part of the $\mathcal{N}=(2,2)$ superconformal algebra in the NS-NS sector. The complexified Lie superalgebra is $sl(2|1)_l\oplus sl(2|1)_r$. Working on the holomorphic (left-moving) side, we have the usual bosonic generators $L_n$, $n=-1,0,1$, R-symmetry $\Omega$, and fermionic generators $G_{\pm1/2}^{\pm}$. They satisfy the following (anti)commutation relations
\begin{equation}\label{N=2algebra2d}
\begin{aligned}
\, [L_m, L_n ] &=  (m-n) \, L_{m+n} \,,\\
[L_m, G_r^\pm] &= \left( \frac m2 - r \right) G^\pm_{m+r}  \,,\\
[\Omega, G_r^\pm ] &= \pm G_{r}^\pm \,,\\
\{ G_r^+, G_s^- \} &= 2 L_{r+s} + (r-s) \Omega\,,
\end{aligned}
\end{equation}
with all other (anti)commutators vanishing. The anti-holomorphic generators, which we denote with a bar, satisfy exactly the same algebra. Making the traditional choice $\Sigma^i_{\alpha\dot\alpha} = (\sigma_i)^{\alpha}_{\phantom{a}\dot\alpha}$ for $i=1,2,3$ and $\Sigma^4_{\alpha\dot\alpha} = \im \delta^{\alpha}_{\phantom{a}\dot\alpha}$, our interpolating algebra from Section \ref{subsec:interpolating} reproduces \eqref{N=2algebra2d} after the following identification for the bosonic generators
\begin{equation}
\begin{aligned}
P_1 &= -\im (L_{-1}+\bar{L}_{-1}) \,,\\
K_1 &= \im (L_{1}+\bar{L}_{1}) \,,\\
D &= -\im(L_0 + \bar{L}_0) \,,\\
R &= \Omega + \bar{\Omega} \,,
\end{aligned}\qquad
\begin{aligned}
P_2 &= L_{-1}-\bar{L}_{-1}\,,\\
K_2 &= L_{1}-\bar{L}_{1}\,,\\
M_{12} &= -L_0 + \bar{L}_0\,,\\
M_{34} &= \frac{\Omega - \bar{\Omega}}{2}\,,
\end{aligned}
\label{2dbosonic}
\end{equation}
and the following for the fermionic generators
\begin{equation}
\begin{aligned}
Q_{1}^+ &= G_{-1/2}^{+}\,,\\
Q_{1}^-  &= \bar{G}_{-1/2}^{-}\,,\\
S^{1+}   &= \bar{G}_{1/2}^{-}\,,\\
S^{1-}    &= G_{1/2}^{-}\,,
\end{aligned}\qquad
\begin{aligned}
Q_{2}^+ &= \bar{G}_{-1/2}^{+}\,,\\
Q_{2}^-  &= G_{-1/2}^{-}\,,\\
S^{2+}   &= G_{1/2}^{+}\,,\\
S^{2-}    &= \bar{G}_{1/2}^{-}\,.
\end{aligned}
\end{equation}
Finally consider the case $d=1$. In this dimension, the generator $R$ drops out from equations (\ref{eq:sq}), and indeed from the expression for the conformal Casimir as shown in Section \ref{ssec:casimir}. Hence, it may be safely dropped from the algebra. Since $4=1+3$, the situation is quite similar to the $d=3$ case detailed above and in appendix \ref{App:opeder}. In particular, the rotation generators in the directions $2,3,4$ describe an $su(2)$ algebra acting as an R-symmetry on the supercharges. Analogously to $d=3$, we make the choice $\Sigma^{\hat{i}}_{\alpha\dot\beta} = (\sigma_{\hat{i}-1})^{\dot\beta}_{\phantom{a}\alpha}$ for $\hat{i}=2,3,4$, where $\sigma_a$ are the usual Pauli matrices and $\Sigma^{1}_{\alpha\dot\beta} = \im \delta^{\dot\beta}_{\phantom{a}\alpha}$. A lower dotted index has the same transformation under the R-symmetry as an upper undotted index and vice versa, so that we can write
\begin{equation}
Q^{\alpha - }\equiv Q_{\dot\alpha}^{-}\,,\quad S_{\alpha}^{+}\equiv S^{\dot\alpha +}\,.
\end{equation}
Spinor indices can be raised and lowered using $\epsilon^{\alpha\beta}$, $\epsilon_{\alpha\beta}$. We also define
\bea
H\equiv P_1\;, \qquad K\equiv K_1\,, \qquad R_{\hat{i}-1}\equiv \frac 12\varepsilon_{\hat{i}-1,\hat{j}-1,\hat{k}-1} M_{\hat{j}\hat{k}}\;,
\eea
and find that the algebra becomes $psu(1,1|2)$, described by the non-zero commutators:
\begin{align}
[H,K] &=-2\im D\;, &[D,H]&= - \im H\;,  &[D,K] &= \im K\;, \nonumber \\
\{ Q_\alpha^+,Q^{\beta-}\}&= \im \delta^{\phantom{a}\beta}_{\alpha}\,H\;, &[D,Q^{+}_\alpha]&=-\frac{\im}2 Q^+_\alpha\;, &[D,Q^{\alpha-}]&=-\frac{\im}2 Q^{\alpha-}\;,  \nonumber \\
\{ S_{\alpha}^{+},S^{\beta-}\}&= - \im\delta^{\phantom{a}\beta}_{\alpha}\,K \;, &[D,S^{+}_{\alpha}]&=\frac{\im}2 S^{+}_{\alpha} \;, &[D,S^{\alpha-}]&=\frac{\im}2 S^{\alpha-} \;, \nonumber \\
\{ S^{\alpha-}, Q^{+}_{\beta} \} &= \im D\, \delta^{\alpha}_{\phantom{a}\beta} - (\sigma_{i})^{\alpha}_{\phantom{a}\beta} R_i, &[K,Q^{+}_\alpha]&= \im S^{+}_{\alpha}\;, &[K,Q^{\alpha -}]&= \im S^{\alpha -}\;,\nonumber \\
\{ S_{\alpha}^{+}, Q^{\beta-} \} &= \im D\, \delta^{\phantom{a}\beta}_{\alpha} - (\sigma_{i})^{\phantom{a}\beta}_{\alpha} R_i, &[H,S^{+}_\alpha]&= \im Q^{+}_{\alpha}\;, &[H,S^{\alpha -}]&= \im Q^{\alpha -}\;,\nonumber \\
[R_i,R_j]&=\im \epsilon_{ijk} R_k\;,  &[R_i, X^{+}_\alpha]&=\frac 12 (\sigma_{i})_{\phantom{a}\alpha}^{\beta} X^+_\beta, &[R_i, X^{\alpha-}]&=\frac 12(\sigma_{i})^{\phantom{a}\alpha}_{\beta} X^{\beta-}\,,
\end{align}
where in the last line $X$ stands for either $Q$ or $S$.

%%%%%%%%%%%%%%%%%%%%%%%%%%%%%%%%%%%%
\subsection{Unitarity bounds in general $d$}
\label{ssec:unitarity}
%%%%%%%%%%%%%%%%%%%%%%%%%%%%%%%%%%%%

It will be useful for Section \ref{ssec:chiralOPE} to work out the unitarity bounds, in general dimension, for the symmetric traceless representation and the representation in the tensor product of the symmetric traceless and the spinors. The unitary representations in integer $d$ were found in \cite{Dobrev:1985qv,Minwalla:1997ka} but our goal here is to offer a derivation that formally makes sense in general $d$. It was noted in \cite{Hogervorst:2014rta} that the free scalar CFT in non-integer $d$ contains negative-norm states. This casts doubt on whether one can have unitary CFTs in fractional dimensions. Here, we will be modest and will derive necessary conditions for unitarity in general $d$, by focusing on the lowest levels, assuming that superconformal primaries have positive norm.

Suppose $|\mathcal{O}_A\rangle$ is a superconformal primary of dimension $\Delta$, R-charge $q$, transforming in a representation $R$ of $SO(d)\times SO(4-d)$. Using a standard argument, \cite{Dobrev:1985vh,Dobrev:1985qz,Petkova,Dobrev:1985qv}, it follows from \eqref{eq:sq} that the $Q_\alpha^+$ descendants of $|\mathcal{O}_A\rangle$ have non-negative norm when
\begin{equation}
\Delta \geq \frac{d-1}{2}q + a_R + a_r - \min_{R'\subset R\otimes r}\left(a_{R'}\right)\,,
\label{eq:unitarity1}
\end{equation}
where $a_{R}$ denotes the eigenvalue of representation $R$ under the following Casimir
\begin{equation}
\frac{1}{2}M_{ij}M_{ij} - \frac{1}{2}M_{\hat i\hat j}M_{\hat i\hat j}\,,
\label{rotcasimir}
\end{equation}
and $r$ denotes the spinor representation in which $Q_{\alpha}^+$ transforms, with $a_r$ its eigenvalue under \eqref{rotcasimir}. Similarly, the $Q_{\dot\alpha}^-$ descendants of $|\mathcal{O}_A\rangle$ have non-negative norm whenever
\begin{equation}
\Delta \geq - \frac{d-1}{2}q + a_R + a_{\bar r} - \min_{R'\subset R\otimes \bar r}\left(a_{R'}\right)\,,
\label{eq:unitarity2}
\end{equation}
where $\bar r$ is the representation in which $Q_{\dot \alpha}^-$ transforms. In Euclidean signature, the bar operation on $SO(d)\times SO(4-d)$ representations corresponds to parity, rather than complex conjugation, but we will refer to it as conjugation for simplicity. It remains to evaluate $a_R$ on the representations of interest. Suppose $R=S_s$ is the symmetric traceless tensor of spin $s$. Since it has no indices in the reduced dimensions, we find just the $SO(d)$ eigenvalue
\begin{equation}
a_{S_s} = s(s+d-2)\,.
\end{equation}
Consider now the spinor representations $r$, $\bar{r}$. It follows from \eqref{eq:clifford1} that their eigenvalue under $\frac{1}{2}M_{ij}M_{ij}$ is $\frac{d(d-1)}{8}$, which matches the expected values $\frac{1}{4},\frac{3}{4},\frac{3}{2}$ in $d=2,3,4$, respectively. The spinor indices $\alpha$, $\dot\alpha$ necessarily transform also under the reduced rotations $M_{\hat i\hat j}$, and indeed, the eigenvalue under $\frac{1}{2}M_{\hat i\hat j}M_{\hat i\hat j}$ is related to the eigenvalue under $\frac{1}{2}M_{ij}M_{ij}$ by replacing $d\mapsto 4 - d$. Hence the eigenvalue under \eqref{rotcasimir} is
\begin{equation}
a_{r} = a_{\bar r} = \frac{d(d-1)}{8} - \frac{(4-d)(3-d)}{8} = \frac{3(d - 2)}{4}\,. 
\end{equation}
Let us move on to the representation $|\mathcal{O}_{\alpha i_1\ldots i_{s}}\rangle$, symmetric and traceless in the $s$ vector indices, and satisfying the irreducibility criterion $\bar{\Sigma}_{i_1}^{\dot \alpha\alpha}|\mathcal{O}_{\alpha i_1\ldots i_{s}}\rangle = 0$. We denote this representation as $P_s$. The action of the $SO(d)$ Casimir can be evaluated in general $d$ using the dimensional continuation of the superconformal algebra from the previous sections with the result
\begin{equation}
\frac{1}{2}M_{jk}M_{jk} |\mathcal{O}_{\alpha i_1\ldots i_{s}}\rangle = \left[\frac{d(d-1)}{8} + s(s+d-1)\right] |\mathcal{O}_{\alpha i_1\ldots i_{s}}\rangle\,.
\end{equation}
This expression reduces to the expected $\left(s + \frac{1}{2}\right)^2$ in $d=2$; $j(j+1)$, with $j=s + \frac{1}{2}$, in $d=3$; and $2[j_1(j_1+1) + j_2(j_2+1)]$, with $j_1 = \frac{s+1}{2}$, $j_2 = \frac{s}{2}$, in $d=4$. We have already seen that
\begin{equation}
\frac{1}{2}M_{\hat j\hat k}M_{\hat j\hat k} |\mathcal{O}_{\alpha i_1\ldots i_{s}}\rangle = \frac{(4-d)(3-d)}{8}|\mathcal{O}_{\alpha i_1\ldots i_{s}}\rangle\,,
\end{equation}
leading to
\begin{equation}
a_{P_s} = s(s + d -1) + \frac{3(d-2)}{4}\,.
\end{equation}
The same formula is valid for the conjugate representation $|\mathcal{O}_{\dot\alpha i_1\ldots i_{s}}\rangle$.

Finally, we will need the Casimir for the representation $|\mathcal{O}_{\alpha\beta i_1\ldots i_{s}}\rangle$, symmetric in $\alpha\beta$, symmetric and traceless in the vector indices, and satisfying $\bar{\Sigma}_{i_1}^{\dot \alpha\alpha}|\mathcal{O}_{\alpha \beta i_1\ldots i_{s}}\rangle = 0$. We denote this representation by $Q_s$. The superconformal algebra is not sufficient to find the individual eigenvalues of the $SO(d)$ and $SO(4-d)$ Casimir operators because of the cross-term occurring when the two rotation generators each act on one spinor index. Fortunately, the identity in \eqref{eq:id2} is precisely what is needed to evaluate this cross-term in the difference of the two Casimirs. The final result is
\begin{equation}
a_{Q_s} = s(s+d) + 2(d-2)\,. \label{eq:qscasimir}
\end{equation}
This result is in harmony with the results in $d=3,4$, where the contribution of the $SO(4-d)$ Casimir must vanish. Indeed, in $d=3$, $a_{Q_s} = (s+1)(s+2)$, corresponding to the $j=s+1$ representation, and in $d=4$, $a_{Q_s} = (s+2)^2$, corresponding to the $j_1 = \frac{s}{2} + 1$, $j_2 = \frac{s}{2}$ representation. Formula \eqref{eq:qscasimir} applies also to the conjugate representation $|\mathcal{O}_{\dot\alpha\dot\beta i_1\ldots i_{s}}\rangle$.

We are now in a position to derive the unitarity bounds at level one for the $S_s$ and $P_s$ representations. Due to the covariance of the $\bar\Sigma$ tensor, we have the decompositions $S_s \otimes r = \bar{P}_{s-1}\oplus P_{s+1}$, $S_s \otimes \bar{r} = P_{s-1}\oplus \bar{P}_{s+1}$,  valid for $s>0$. The first direct summand has a smaller value of $a$, leading to the unitarity bound for the symmetric traceless representation
\begin{equation}
\Delta_{S_s} \geq \frac{d-1}{2}|q| + s + d - 2\,,\qquad s>0 \,.
\label{eq:unitaritys}
\end{equation}
This formula reduces to the well-known results in integer $d$. For the special case $s=0$, unitarity at the first level implies only $\Delta\geq \frac{d-1}{2}|q|$. However, this condition is not sufficient for unitarity, since the level-two state $|\mathcal{P}\rangle = \epsilon^{\alpha \beta}Q_{\alpha}^+Q_{\beta}^+|\mathcal{O}_A\rangle$ has norm
\begin{equation}
\langle\mathcal{P}|\mathcal{P}\rangle = 4\left(\Delta - \frac{d-1}{2}q\right)\left(\Delta - \frac{d-1}{2}q - d + 2\right)\,.
\end{equation}
An analogous result holds for $|\tilde{\mathcal{P}}\rangle = \epsilon^{\dot\alpha \dot\beta}Q_{\dot\alpha}^-Q_{\dot\beta}^-|\mathcal{O}_A\rangle$, leading to the unitary representations
\begin{equation}
\begin{aligned}
\Delta_{S_0} &= \frac{d-1}{2}|q|\,,\\
\Delta_{S_0} &\geq \frac{d-1}{2}|q| + d - 2\,.
\end{aligned}
\label{scalarunitarity}
\end{equation}
Consider now the representation $P_s$. We have $P_s\otimes r =S_s \oplus Q_s$, $P_s\otimes \bar{r} = Q_{s-1}\oplus S_{s+1}$, with the first direct summand having the smaller value of $a_{R'}$. Equations \eqref{eq:unitarity1}, \eqref{eq:unitarity2} thus lead to the bound
\begin{equation}
\Delta_{P_s} \geq \left|\frac{d-1}{2} (q+1)-1 \right |+ s + d - \frac{3}{2}\,.
\label{eq:unitarityps}
\end{equation}
In $d=3$, this reduces to the expected $\Delta \geq |q| + s + \frac{3}{2} = |q| + j + 1$, with $j=s + \frac{1}{2}$. In $d=4$, it becomes $\Delta  \geq \left|\frac{3}{2} q + \frac{1}{2} \right |+ s + \frac{5}{2} = \left|\frac{3}{2} q + j_1 - j_2 \right |+ j_1 + j_2 + 2 $, with $j_1 = \frac{s+1}{2}$, $j_2 = \frac{s}{2}$, in perfect agreement with \cite{Dobrev:1985qv,Minwalla:1997ka}. The unitarity bound for the conjugate representation $\bar{P}_s$ is obtained simply by flipping the sign of $q$ in \eqref{eq:unitarityps}
\begin{equation}
\Delta_{\bar{P}_s} \geq \left|\frac{d-1}{2} (q-1)+1 \right |+ s + d - \frac{3}{2}\,.
\label{eq:unitaritybps}
\end{equation}

The notions of short and semi-short representations are useful because operators in these representations often have dimensions protected from quantum corrections. A superconformal chiral scalar operator belongs to a short multiplet and obeys the first equation in \eqref{scalarunitarity} with $q>0$, an anti-chiral operator obeys the same equation with $q<0$. For the symmetric traceless representation of spin $s$, the semi-short multiplets are those for which the superconformal primary saturates \eqref{eq:unitaritys}, which for $s=0$ is the same as saturating the second equation in \eqref{scalarunitarity}. 

Necessary conditions for unitary representations of the nonsupersymmetric conformal algebra in general dimension are presented in Section 6 of \cite{Minwalla:1997ka} (see also \cite{Mack:1975je}) and they are generally weaker than the ones for the superconformal algebra discussed here. It is also useful to recall that a free scalar field in $d$ dimensions necessarily has
\begin{equation}
\Delta_{\rm free} = \frac{d-2}{2}\,,
\label{eq:freescalar}
\end{equation}
and a conserved current of spin $s$ in any CFT obeys
\begin{equation}
\Delta_{\rm cc} = s+d-2\,.
\label{eq:conscurrent}
\end{equation}
%

%%%%%%%%%%%%%%%%%%%%%%%%%%%%%%%%%%%%%
\section{Superconformal blocks}
\label{sec:superblocks}
%%%%%%%%%%%%%%%%%%%%%%%%%%%%%%%%%%%%%

A standard method for finding conformal and superconformal blocks is by utilizing the Casimir equation \cite{Dolan:2003hv,Fitzpatrick2014}. Having formulated a dimensional continuation of the superconformal algebra with four Poincar\'{e} supercharges, we are in a position to find the Casimir equation and its solution for a large class of superconformal blocks in general $d$. This is done in Sections \ref{ssec:casimir}, \ref{ssec:sblocks}. Remarkably, as was already noticed in \cite{Fitzpatrick2014}, superconformal blocks are closely related to non-supersymmetric conformal blocks with shifted scaling dimensions and we provide an interpretation of this fact in Section \ref{ssec:susynosusy}. In section \ref{ssec:chiralOPE} we deal with an important case not captured by the Casimir approach, namely that of superconformal blocks in the chiral channel.

%%%%%%%%%%%%%%%%%%%%%%%%%%%%%%
\subsection{Superconformal Casimir}
\label{ssec:casimir}
%%%%%%%%%%%%%%%%%%%%%%%%%%%%%%

In this section, we find the quadratic Casimir of the relevant superconformal algebra in general dimension. The quadratic Casimir must be a linear combination of the quadratic Casimir $C_b$ of the bosonic conformal subalgebra $so(d+1,1)$, the Casimir of the reduced rotations, $so(4-d)$, $R^2$, and terms quadratic in the fermionic generators. Invariance under $D$, $R$, $M_{ij}$, and $M_{\hat i \hat j}$ reduces the possibilities to
\begin{equation}
 C = C_b + c_1 S^{\alpha+}Q_{\alpha}^{-} + c_2 Q_{\alpha}^{-}S^{\alpha+} + c_3 S^{\alpha-}Q_{\alpha}^{+} + c_4 Q_{\alpha}^{+}S^{\alpha-} + c_5 R^2 +c_6 M_{\hat i \hat j}M_{\hat i \hat j}\,,
\end{equation}
with $c_i$ so far undetermined constants. Note that
\begin{equation}
 C_b = -D^2 - \frac{1}{2}(P_iK_i + K_iP_i) + \frac{1}{2}M_{ij}M_{ij}\,.
\end{equation}
The coefficients $c_i$ can be determined by requiring $[C,Q_{\alpha}^+] =[C,Q_{\dot\alpha}^-]= 0$. Using \eqref{eq:sq} and looking at the coefficients of $DQ_{\alpha}^+$, $DQ_{\dot\alpha}^{-}$, $Q_{\alpha}^+ D$ and $Q_{\dot\alpha}^{-} D$ leads to $c_1 = c_3 = -c_2 = -c_4 = 1/2$. Similarly, the coefficient of $RQ_{\alpha}^{+}$ determines $c_5 = -(d-1)/4$. Finally, the coefficient of $M_{\hat i \hat j} Q_{\alpha}^+$ fixes $c_6 = -1/2$, leading to the final result
\begin{equation}
 C = -D^2 - \frac{1}{2}(P_iK_i + K_iP_i) + \frac{1}{2}M_{ij}M_{ij} - \frac{1}{2}M_{\hat i \hat j}M_{\hat i \hat j} - \frac{d-1}{4}R^2 + \frac{1}{2}\left(\left[S^{\dot\alpha+},Q_{\dot\alpha}^{-}\right] + \left[S^{\alpha-},Q_{\alpha}^{+}\right] \right)\,.
 \label{eq:casimir}
\end{equation}
It is instructive to study the contribution of the R-symmetries in $d=2$, where, after using \eqref{2dbosonic}, one finds
\begin{equation}
\frac{1}{2}M_{\hat i \hat j}M_{\hat i \hat j} + \frac{d-1}{4}R^2 = \frac{1}{4}\left[(\Omega - \bar\Omega)^2 + (\Omega + \bar\Omega)^2\right] = \frac{1}{2}\left(\Omega^2 + \bar\Omega^2\right)\,.
\end{equation}
The contribution is a sum of a holomorphic and an antiholomorphic part as expected. The eigenvalue of $C$ when acting on a superconformal family, where the superconformal primary has dimension $\Delta$, R-charge $q$, and transforms as a symmetric traceless tensor of spin $s$ under $M_{ij}$ and as a singlet under $M_{\hat i \hat j}$, is
\begin{equation}
 \lambda_C = \Delta(\Delta - d + 2) + s (s + d - 2) - \frac{d-1}{4}q^2\,.
\end{equation} 
To find this, we have used the eigenvalue of the usual conformal Casimir operator $C_b$ when acting on a conformal primary \cite{Dolan:2003hv}
\begin{equation}
 \lambda_{C_b} = \Delta(\Delta - d) + s (s + d - 2)\,.
\end{equation} 
%

%%%%%%%%%%%%%%%%%%%%%%%%%%%%%%
\subsection{Casimir equation and its solution}
\label{ssec:sblocks}
%%%%%%%%%%%%%%%%%%%%%%%%%%%%%%

Here, we derive a formula for the superconformal blocks in theories invariant under the superconformal algebra in Section \ref{sec:superalgebras}, for the four-point function $\langle\phi_1\phi_2\phi_3\phi_4\rangle$, where $\phi_i$ are scalar superconformal primaries with dimensions $\Delta_i$ and R-charges $q_i$. In addition, we assume that $\phi_1$ and $\phi_3$ are  chiral, i.e. $Q^+_\alpha \phi_{1,3} = 0$, or equivalently
\begin{equation}
 \Delta_{1,3} = \frac{d-1}{2}q_{1,3}\,.
\end{equation} 
A superconformal block corresponds to the contribution of a single superconformal family produced in the OPE of $\phi_1$ and $\phi_2$. It is therefore an eigenfunction of the superconformal Casimir \eqref{eq:casimir} applied to the first two operators. Due to the appearance of supercharges, the resulting equation will relate the superconformal block of $\langle\phi_1\phi_2\phi_3\phi_4\rangle$ to the one of $\langle\psi^{\dot\alpha}_1\psi^{\alpha}_2\phi_3\phi_4\rangle$, where $\psi_{1,2}$ is a supersymmetric descendant of $\phi_{1,2}$. In the limit $|x_4|\rightarrow\infty$, we can use a supersymmetric Ward identity to reduce the latter correlator to a differential operator acting on $\langle\phi_1\phi_2\phi_3\phi_4\rangle$ and thus derive a differential equation for the original superconformal block. Consider the action of the fermionic part of the superconformal Casimir on the product $\phi_1(x_1)\phi_2(x_2)$. Using the chirality of $\phi_1$ and the superconformal algebra, one can show that
\begin{equation}
\begin{aligned}
\frac{1}{2}&\left(\left[S^{\dot\alpha+},Q_{\dot\alpha}^{-}\right] + \left[S^{\alpha-},Q_{\alpha}^{+}\right] \right)\left(\phi_1(x_1)\phi_2(x_2)\right)|0\rangle = \\
 &= \left[(S^{\alpha-}\phi_1(x_1))(Q_\alpha^{+}\phi_2(x_2)) - (Q^{-}_{\dot\alpha}\phi_1(x_1))(S^{\dot\alpha +}\phi_2(x_2)) + 2 (\Delta_1 + \Delta_2)\phi_1(x_1)\phi_2(x_2)\right]\,|0\rangle\,,
\end{aligned}
\end{equation}
where the action of conserved charges on local operators is the usual one via the commutator. From
\begin{equation}
 \phi(x) = e^{\im x\cdot P}\phi(0)e^{-\im x\cdot P}\,,
\end{equation}
and using \eqref{eq:qksp}, it follows that
\begin{equation}
(S^{\alpha-}\phi_1(x_1)) = \im x_1^i\bar{\Sigma}^{\dot\alpha\alpha}_i(Q^{-}_{\dot\alpha}\phi_1(x_1))\,,\qquad
(S^{\dot\alpha+}\phi_2(x_2)) =  \im x_2^ i\bar{\Sigma}^{\dot\alpha\alpha}_i(Q^{+}_\alpha\phi_2(x_2))\,.
\label{eq:saction}
\end{equation}
It remains to relate the correlator
\begin{equation}
 \left\langle (Q^{-}_{\dot\alpha}\phi_1(x_1))(Q^{+}_{\alpha}\phi_2(x_2))\phi_3(x_3)\phi_4(x_4) \right\rangle\,,
\end{equation}
to $\langle\phi_1(x_1)\phi_2(x_2)\phi_3(x_3)\phi_4(x_4)\rangle$. Starting from the Ward identity
\begin{equation}
 \left\langle \left[Q^{+}_{\alpha},(Q^{-}_{\dot\alpha}\phi_1(x_1))\phi_2(x_2)\phi_3(x_3)\phi_4(x_4)\right] \right\rangle = 0\,,
\end{equation}
and using the anticommutator of Poincar\'{e} supercharges, \eqref{eq:qq}, and the chirality of $\phi_3$, we find
\begin{equation}\label{QQWardidentity}
\begin{aligned}
 &\left\langle (Q^{-}_{\dot\alpha}\phi_1(x_1))(Q^{+}_{\alpha}\phi_2(x_2))\phi_3(x_3)\phi_4(x_4) \right\rangle + \left\langle (Q^{-}_{\dot\alpha}\phi_1(x_1))\phi_2(x_2)\phi_3(x_3)(Q^{+}_{\alpha}\phi_4(x_4)) \right\rangle = \\
 &= -\im\Sigma^i_{\alpha\dot\alpha}\partial^{x_1}_i\langle\phi_1(x_1)\phi_2(x_2)\phi_3(x_3)\phi_4(x_4)\rangle\,.
\end{aligned}
\end{equation}
Conformal invariance ensures that no information is lost if we take the limit $x_4\rightarrow\infty$. The leading behavior of a correlation function containing a primary $\mathcal{O}(x)$ of dimension $\Delta$ and arbitrary Lorentz quantum numbers as $|x|\rightarrow\infty$ is $|x|^{-2\Delta}$. Thus the second term on the left-hand side of \eqref{QQWardidentity} is subleading in the limit $|x_4|\rightarrow \infty$, since $Q^{+}_{\alpha}$ increases the dimension of $\phi_4$ by $1/2$. In the derivation of the differential equation, we can then replace
\begin{equation}
 \left\langle (Q^{-}_{\dot\alpha}\phi_1(x_1))(Q^{+}_{\alpha}\phi_2(x_2))\phi_3(x_3)\phi_4(x_4) \right\rangle\;,
\end{equation}
with
\begin{equation}
-\im\Sigma^i_{\alpha\dot\alpha}\partial^{x_1}_i\langle\phi_1(x_1)\phi_2(x_2)\phi_3(x_3)\phi_4(x_4)\rangle\,,
\end{equation}
while remembering that we must send $|x_4|$ to infinity at the end of the calculation. Combining this with \eqref{eq:saction}, and using 
\begin{equation}
\mathrm{tr}\left(\Sigma^i\bar{\Sigma}_j\right) = 2 \delta^{i}_{\phantom{i}j}\,,
\end{equation}
which follows from the Clifford algebra, we find the action of the fermionic part of the superconformal Casimir on the four-point function to be
\begin{equation}
\begin{aligned}
&\left\langle \frac{1}{2}\left(\left[S^{\dot\alpha+},Q_{\dot\alpha}^{-}\right] + \left[S^{\alpha-},Q_{\alpha}^{+}\right] \right) \left(\phi_1(x_1)\phi_2(x_2)\right) \phi_3(x_3) \phi_4(x_4) \right\rangle \sim\\
&\sim 2\left(x_{12}\cdot\partial_{x_1} + \Delta_1 + \Delta_2\right)\left\langle\phi_1(x_1)\phi_2(x_2)\phi_3(x_3)\phi_4(x_4)\right\rangle\,,
\label{eq:foperator}
\end{aligned}
\end{equation}
where $x_{ij}\equiv x_i-x_j$, and the $\sim$ symbol means equality up to terms subleading as $|x_4|\rightarrow\infty$.

The contribution of a single superconformal family of the superconformal primary $\mathcal{O}$ to the four-point function takes the form
\begin{equation}
 \left\langle\phi_1(x_1)\phi_2(x_2)\phi_3(x_3)\phi_4(x_4)\right\rangle\!|_{\mathcal{O}} = \frac{c_{\phi_1\phi_2}^{\mathcal{O}}c_{\phi_3\phi_4\mathcal{O}}}{|x_{12}|^{\Delta_1 + \Delta_2}|x_{34}|^{\Delta_3+\Delta_4}}\frac{|x_{24}|^{\Delta_{12}}|x_{14}|^{\Delta_{34}}}{|x_{14}|^{\Delta_{12}}|x_{13}|^{\Delta_{34}}}\;\mathcal{G}^{\Delta_{12},\Delta_{34}}_{\Delta_{\mathcal{O}},s_{\mathcal{O}}}(z,\bar{z})\,,
 \label{eq:sfcontribution}
\end{equation}
where $\mathcal{G}^{\Delta_{12},\Delta_{34}}_{\Delta,s}(z,\bar{z})$ is the superconformal block and $\Delta_{ij} \equiv \Delta_{i}-\Delta_{j}$. Here $z$ and $\bar z$ are related to the usual conformally invariant cross-ratios $u,v$ as
\bea
u\equiv\frac{x_{12}^2\,x_{34}^2}{x_{13}^2\,x_{24}^2}=z\bar z\,, \qquad
v\equiv\frac{x_{14}^2\,x_{23}^2}{x_{13}^2\,x_{24}^2}=(1-z)(1-\bar z)\,.\label{eq:crossratios}
\eea
The operator \eqref{eq:foperator} translates into the following action on the superconformal block
\begin{equation}
 2\left[z(1-z)\partial + \bar{z}(1-\bar{z})\bar{\partial}\right]\mathcal{G}^{\Delta_{12},\Delta_{34}}_{\Delta,s}(z,\bar{z}) - \Delta_{34}(z + \bar{z})\mathcal{G}^{\Delta_{12},\Delta_{34}}_{\Delta,s}(z,\bar{z})\,,
\end{equation}
where $\partial\equiv \partial_{z}$ and $\bar{\partial}\equiv \partial_{\bar{z}}$. The action of the R-symmetry cancels on the two sides of the superconformal Casimir equation, and using the result for the conformal Casimir \cite{Dolan:2003hv}, the differential equation for the superconformal block becomes
\begin{equation}
 \mathcal{D}\mathcal{G}^{\Delta_{12},\Delta_{34}}_{\Delta,s}(z,\bar{z}) = \left[\Delta(\Delta - d + 2) + s(s + d - 2)\right]\mathcal{G}^{\Delta_{12},\Delta_{34}}_{\Delta,s}(z,\bar{z})\,,
\end{equation}
where the differential operator $\mathcal{D}$ is given by
\begin{equation}
\begin{aligned}
\mathcal{D} \equiv & \; 2z^2(1-z)\partial^2 + 2\bar{z}^2(1-\bar{z})\bar{\partial} + (\Delta_{12} - \Delta_{34} - 4)(z^2\partial + \bar{z}^2\bar{\partial})+ 2\left(z\partial + \bar{z}\bar{\partial}\right) + \\
&+\frac{1}{2}(\Delta_{12} - 2)\Delta_{34}(z+\bar{z}) + 2(d-2)\frac{z\bar{z}}{z-\bar{z}}\left[(1-z)\partial - (1-\bar{z})\bar{\partial}\right]\,.
\end{aligned}
\end{equation}
It turns out that this equation has a simple solution in terms of the ordinary non-supersymmetric conformal blocks. This has also been pointed out for $d=4$ and $\phi_1=\phi_3=\bar{\phi}_2=\bar{\phi}_4$ in \cite{Fitzpatrick2014}. Indeed, the solution with the correct $z,\bar{z}\rightarrow0$ behavior is
\begin{equation}
 \mathcal{G}^{\Delta_{12},\Delta_{34}}_{\Delta,s}(u,v) = u^{-1/2}G^{\Delta_{12}-1,\Delta_{34}-1}_{\Delta+1,s}(u,v)\,,
 \label{eq:superblock}
\end{equation}
where $G^{\Delta_{12},\Delta_{34}}_{\Delta,s}(u,v)$ is the non-supersymmetric conformal block and we switched to the usual cross-ratios $u,v$. We comment on the relationship between conformal and superconformal blocks in the next section. It is also possible to decompose the superconformal block into conformal blocks using a relation found in \cite{Dolan:2011dv}. Using the convention where the $u\rightarrow 0$, $v\rightarrow 1$ behavior of the conformal blocks is
\begin{equation}
 G^{\Delta_{12},\Delta_{34}}_{\Delta,s}(u,v) \sim\frac{(-1)^s}{2^s}u^{\frac{\Delta - s}{2}}(1-v)^s\,,
 \label{eq:cbasymptotics}
\end{equation}
the decomposition reads
\begin{equation}
\mathcal{G}^{\Delta_{12},\Delta_{34}}_{\Delta,s} = G^{\Delta_{12},\Delta_{34}}_{\Delta,s} +
a_1G^{\Delta_{12},\Delta_{34}}_{\Delta+1,s+1}+
a_2G^{\Delta_{12},\Delta_{34}}_{\Delta+1,s-1}+
a_3G^{\Delta_{12},\Delta_{34}}_{\Delta+2,s}\,,
\label{eq:sbdecomposition}
\end{equation}
where
\begin{eqnarray}
\label{eq:sbcoefficients}
a_1 &\equiv& -\ds\frac{(\Delta+\Delta_{12}+s)(\Delta+\Delta_{34}+s)}{2(\Delta+s)(\Delta+s+1)}\,,\notag\\
a_2 &\equiv& -\ds\frac{s(s+d-3)(\Delta+\Delta_{12}-s-d+2)(\Delta+\Delta_{34}-s-d+2)}{2(2s+d-4)(2s+d-2)(\Delta-s-d+2)(\Delta-s-d+3)}\,,\\
a_3 &\equiv& \ds\frac{\Delta(\Delta-d+3)(\Delta+\Delta_{12}+s)(\Delta+\Delta_{34}+s)(\Delta+\Delta_{12}-s-d+2)(\Delta+\Delta_{34}-s-d+2)}{4(2\Delta-d+4)(2\Delta-d+2)(\Delta+s)(\Delta+s+1)(\Delta-s-d+2)(\Delta-s-d+3)}\,.\notag
\end{eqnarray}
It follows that whenever a superconformal family contributes to a given four-point function, as in \eqref{eq:sfcontribution}, it is through the superconformal primary, $\mathcal{O}$, and three other conformal primaries (and all their conformal descendants). The three conformal primaries are supersymmetric descendants of $\mathcal{O}$, with dimensions and spins that can be read off from \eqref{eq:sbdecomposition}. 

A few comments are in order. Notice that for $\Delta_{12}=\Delta_{34}=0$ the coefficients do not have poles for dimension and spin consistent with the  unitarity bounds, and furthermore their sign is consistent with unitarity. For $\Delta_{12}$ and $\Delta_{34}$ different from zero, there can be poles, for $\Delta,s$ saturating the unitarity bound, but this is expected since the leading block itself diverges. This is related to the fact that conserved currents can only couple to scalars with identical dimensions.

It is useful to pause for a moment and compare our solution for $ \mathcal{G}^{\Delta_{12},\Delta_{34}}_{\Delta,s}(u,v)$  to previous results in integer dimensions. For the special case of the $d=4$, $\mathcal{N}=1$ superconformal blocks studied in \cite{Poland:2010wg,Fitzpatrick2014}, our solution is in agreement with their result since conformal blocks are invariant under $\Delta_{12}\leftrightarrow-\Delta_{34}$. For $d=2$, the explicit form of the solution is (up to an overall constant)
\begin{equation}
\mathcal{G}^{\Delta_{12},\Delta_{34}}_{\Delta,s}(z,\bar{z}) = j_{\frac{\Delta+s}{2}}(z)j_{\frac{\Delta-s}{2}}(\bar{z}) + z\leftrightarrow\bar{z}\,,
\end{equation}
where
\begin{equation}
 j_{h}(z) \equiv z^{h} {}_2F_1\left(h-\frac{\Delta_{12}}{2}+1,h+\frac{\Delta_{34}}{2};2h+1;z\right)\,.
\end{equation}
This also agrees with the result found in \cite{Fitzpatrick2014}, up to the transformation $z\leftrightarrow z/(z-1)$, or equivalently $x_1\leftrightarrow x_2$, and after taking into account the following identity
\begin{equation}
 z^{h}{}_2F_1(h+1,h,2h+1;z)=\left(\frac{z}{1-z}\right)^h{}_2F_1\left(h,h,2h+1;\frac{z}{z-1}\right)\,.
\end{equation}
As a cross-check on the Casimir approach, appendix \ref{App:opeder} contains a derivation of the coefficients in \eqref{eq:sbcoefficients} for $d=3$, using the constraints of superconformal symmetry and chirality of $\phi_{1,3}$ on the OPE. It is conceivable that this type of OPE derivation of the superconformal blocks can be carried out in general $d$ using the superconformal algebra of Section \ref{sec:superalgebras}.

Finally, we would like to point out some curious relations between the coefficients in \eqref{eq:sbcoefficients}. For $d=2$ and $d=4$ one has $a_3=a_1a_2$. This identity is not true in general dimension. However, if one considers $a_i$ as a formal function $a_i(\Delta,s,d,\Delta_{12},\Delta_{34})$ one finds
\begin{equation}
 a_3(\Delta,s,d,\Delta_{12},\Delta_{34}) = a_1(\Delta,s,d,\Delta_{12},\Delta_{34}) a_2(-s,-\Delta,d,\Delta_{12},\Delta_{34})\,.
\end{equation}
%

%%%%%%%%%%%%%%%%%%%%%%%%%%%%%%%%
\subsection{The relationship between conformal and superconformal blocks}
\label{ssec:susynosusy}
%%%%%%%%%%%%%%%%%%%%%%%%%%%%%%%%

The relation in \eqref{eq:superblock} between superconformal and ordinary conformal blocks can be given a simple interpretation. Consider the contribution of the superconformal family of the superconformal primary $\mathcal{O}$ to the correlator $\langle\phi_1\phi_2\phi_3\phi_4\rangle$ as in \eqref{eq:sfcontribution}. It can be rewritten, via \eqref{eq:superblock}, as
\begin{equation}
\begin{aligned}
&\left\langle\phi_1(x_1)\phi_2(x_2)\phi_3(x_3)\phi_4(x_4)\right\rangle\!|_{\mathcal{O}} = \\
&= |x_{24}|^2 \frac{c_{\phi_1\phi_2}^{\mathcal{O}}c_{\phi_3\phi_4\mathcal{O}}}{|x_{12}|^{\Delta_1 + \Delta_2+1}|x_{34}|^{\Delta_3+\Delta_4+1}}\left(\frac{|x_{24}|}{|x_{14}|}\right)^{\Delta_{12}-1}\left(\frac{|x_{14}|}{|x_{13}|}\right)^{\Delta_{34}-1}G^{\Delta_{12}-1,\Delta_{34}-1}_{\Delta_{\mathcal{O}}+1,s_{\mathcal{O}}}(u,v)\,.
\end{aligned}
\end{equation}
Up to the $|x_{24}|^2$ prefactor, this has the form of the contribution of the \textit{conformal} family of a \textit{conformal} primary $\tilde{\cO}$ to the four-point function of some new fields $\tilde{\phi}_i$
\begin{equation}
\begin{aligned}
\left\langle\phi_1(x_1)\phi_2(x_2)\phi_3(x_3)\phi_4(x_4)\right\rangle\!|_{\mathcal{O}} = |x_{24}|^{2}\langle\tilde{\phi}_1(x_1)\tilde{\phi}_2(x_2)\tilde{\phi}_3(x_3)\tilde{\phi}_4(x_4)\rangle\!|_{\tilde{\mathcal{O}}}\,,
\end{aligned}
\label{eq:susynosusy}
\end{equation}
where the quantum numbers of operators with a tilde are related to the original ones as
\begin{equation}
\begin{aligned}
\Delta_{\tilde{\phi}_{1}} &= \Delta_{\phi_1} \,,\\
\Delta_{\tilde{\phi}_{2}} &= \Delta_{\phi_2}+1\,,\\
\Delta_{\tilde{\phi}_{3}} &= \Delta_{\phi_3}\,,\\
\Delta_{\tilde{\phi}_{4}} &= \Delta_{\phi_4}+1\,,\\
\Delta_{\tilde{\mathcal{O}}} &= \Delta_{\mathcal{O}}+1\,,\quad s_{\tilde{\mathcal{O}}} = s_{\mathcal{O}}\,.
\end{aligned}
\end{equation}
Hence, the terms in the \emph{superconformal} block expansion of $\left\langle\phi_1(x_1)\phi_2(x_2)\phi_3(x_3)\phi_4(x_4)\right\rangle$ are in one-to-one correspondence with the terms of the \emph{conformal} block expansion of
\begin{equation}
\langle\tilde{\phi}_1(x_1)\tilde{\phi}_2(x_2)\tilde{\phi}_3(x_3)\tilde{\phi}_4(x_4)\rangle = \frac{1}{|x_{24}|^{2}}\left\langle\phi_1(x_1)\phi_2(x_2)\phi_3(x_3)\phi_4(x_4)\right\rangle\,.
\end{equation}
Moreover, since the only difference between the four-point functions $\langle\phi_1\phi_2\phi_3\phi_4\rangle$ and $\langle\tilde{\phi}_1\tilde{\phi}_2\tilde{\phi}_3\tilde{\phi}_4\rangle$ is the factor $|x_{24}|^2$, we can mimic their relationship by writing
\begin{equation}
\begin{aligned}
\tilde{\phi}_{1,3} &= \phi_{1,3}\,,\\
\tilde{\phi}_{2,4} &=\sigma \phi_{2,4}\,,
\end{aligned}
\label{eq:duality}
\end{equation}
where $\sigma$ is a real scalar conformal primary field of scaling dimension $\Delta_{\sigma} = 1$ not interacting with any of the $\phi_i$. Therefore, there is no regularization needed in defining the composite operators $\sigma\phi_{2,4}$, and the correlation function factorizes as
\begin{equation}
\langle\phi_1(x_1)(\sigma\phi_2)(x_2)\phi_3(x_3)(\sigma \phi_4)(x_4)\rangle =\langle\sigma(x_2)\sigma(x_4)\rangle\langle\phi_1(x_1)\phi_2(x_2)\phi_3(x_3)\phi_4(x_4)\rangle\,.
\end{equation}
It may sound surprising that the conformal block expansion of $\langle\phi_1(\sigma\phi_2)\phi_3(\sigma \phi_4)\rangle$ is the same as the superconformal block expansion of $\langle\phi_1\phi_2\phi_3\phi_4\rangle$. Each superconformal primary $\mathcal{O}^{(0)}$ in the $\phi_1\times\phi_2$ OPE gives rise to four conformal primaries $\mathcal{O}^{(j)}$, $j=0,\ldots, 3$, and each of these gives rise to infinitely many conformal primaries in the $\phi_1\times \sigma\phi_2$ OPE, of the schematic form $\sigma \partial^{n} \mathcal{O}^{(j)}$, $n=0,1,\ldots$. For the proposed relationship between the two expansions to hold, there must occur numerous cancellations among the various conformal primaries, leaving only the contribution of the lowest one $\sigma\mathcal{O}^{(0)}$. Indeed, denoting the conformal descendant of $\mathcal{O}^{(0)}$ with dimension $\Delta_{\mathcal{O}}+1$ and spin $s_{\mathcal{O}}+1$ as $\mathcal{O}^{(1)}$, the contribution of the conformal primary $\sigma\mathcal{O}^{(1)}$ is cancelled by the contribution of $\sigma\overset{\leftrightarrow}{\partial}\mathcal{O}^{(0)}$, which has the same dimension and spin. Remarkably, this cancellation continues to hold for all the higher-lying conformal primaries, leaving only $\sigma\mathcal{O}^{(0)}$. 

It will be curious to study whether the relation in \eqref{eq:susynosusy} between correlation functions in a superconformal field theory and those in a non-supersymmetric conformal field theory can ever be realized for some theories of physical interest.

%%%%%%%%%%%%%%%%%%%%%%%%%%%%%%%%
\subsection{Spectrum in a chiral OPE}
\label{ssec:chiralOPE}
%%%%%%%%%%%%%%%%%%%%%%%%%%%%%%%%

When considering conformal bootstrap for the correlator $\langle\phi_1\phi_2\phi_3\phi_4\rangle$ with $\phi_{1,3}$ chiral primaries and $\phi_{2,4}$ superconformal primaries, there is another possibility for an OPE expansion, namely fusing $\phi_1$ and $\phi_3$. Chirality implies that all conformal primaries appearing in this OPE must be annihilated by $Q_{\alpha}^+$ and it is the goal of this section to derive which components of superconformal multiplets have this property.

Suppose that $\mathcal{P}$ is a conformal primary of dimension $\Delta_{\cP}$ and R-charge $q_{\cP}$ in the symmetric traceless representation of spin $s$, satisfying $[Q_{\alpha}^+,\mathcal{P}] = 0$. Further, assume that $\mathcal{P}$ is a supersymmetric descendant of the superconformal primary $\mathcal{O}$, where $\mathcal{O}$ has dimension $\Delta$ and R-charge $q$. The $SO(d)\times SO(4-d)$ representation $R$ in which $\mathcal{O}$ transforms depends on the precise way $\mathcal{P}$ is obtained from $\mathcal{O}$ through the action of supercharges. The relationship between $\mathcal{O}$ and $\mathcal{P}$ is constrained by observing that the superconformal Casimir \eqref{eq:casimir} must have the same eigenvalue on $\mathcal{O}$ and $\mathcal{P}$. Since $\mathcal{P}$ is annihilated by both $K_i$ and $Q_{\alpha}^+$, it is also annihilated by their anticommutator $S^{\dot\alpha+}$. One can then use the superconformal algebra to evaluate the action of $C$ on $\mathcal{P}$ purely in terms of its quantum numbers, with the resulting eigenvalue
\begin{equation}
\lambda_1 = \Delta_{\mathcal{P}}(\Delta_{\mathcal{P}} - d) + s(s+ d -2) - \frac{d-1}{4}q_{\mathcal{P}}^2 + (d-1)q_{\mathcal{P}}\,,
\end{equation}
where the last term arises from the fermionic generators. Similarly, one can evaluate the eigenvalue of $C$ on $\mathcal{O}$ using the fact that it is a superconformal primary, the result being
\begin{equation}
\lambda_2 = \Delta(\Delta-d + 2) + a_{R} - \frac{d-1}{4}q^2\,,
\end{equation}
where $a_R$ is the $SO(d)\times SO(4-d)$ Casimir familiar from Section \ref{ssec:unitarity}. Moreover, for each conformal primary in the superconformal multiplet, there are relations of the form $\Delta_{\mathcal{P}} = \Delta + \frac{m}{2}$, $m=0,\ldots,4$, $q_{\mathcal{P}} = q + n$, $n=-2,\ldots,2$, and we can proceed case by case and determine whether $\lambda_1 = \lambda_2$ is consistent. We label each case by $(q_{\mathcal{P}},R)$ and use the notation of Section \ref{ssec:unitarity} for the $SO(d)\times SO(4-d)$ representations.

\begin{itemize}
\item At level zero, we have the single case $(q_{\mathcal{P}} = q,R = S_s)$, and $\lambda_1 = \lambda_2$ implies
\begin{equation}
\Delta = \frac{d-1}{2}q\,,
\label{eq:levelzero}
\end{equation}
which corresponds to a unitary representation only if $s=0$, i.e. $\mathcal{P}$ must be a chiral primary. In other words, the chiral superconformal primaries must have $s=0$.
\item There are four cases to consider at level one: $(q+1,\bar{P}_{s-1})$, $(q+1,P_{s})$, $(q-1,P_{s-1})$, $(q-1,\bar{P}_{s})$, corresponding to $\mathcal{P}_{i_1\ldots i_s} = \bar{\Sigma}^{\dot\alpha\alpha}_{i_1}Q_{\alpha}^+\mathcal{O}_{\dot\alpha i_2\ldots i_s}$, $\mathcal{P}_{i_1\ldots i_s} = \epsilon^{\alpha\beta}Q_{\alpha}^+\mathcal{O}_{\beta i_1\ldots i_s}$, $\mathcal{P}_{i_1\ldots i_s} = \bar{\Sigma}^{\dot\alpha\alpha}_{i_1}Q_{\dot\alpha}^-\mathcal{O}_{\alpha i_2\ldots i_s}$, $\mathcal{P}_{i_1\ldots i_s} = \epsilon^{\dot\alpha\dot\beta}Q_{\dot\alpha}^-\mathcal{O}_{\dot\beta i_1\ldots i_s}$ respectively, where in the first and third case we also need to symmetrize with respect to the extra vector index. For the first case, $\lambda_1 = \lambda_2$ implies
\begin{equation}
\Delta = \frac{d-1}{2}q + s - 1 + \frac{d}{2}\,,
\label{eq:levelone}
\end{equation}
which is precisely the unitarity bound \eqref{eq:unitaritybps} for $\bar{P}_{s-1}$. The first case is therefore allowed only if the superconformal multiplet of $\mathcal{O}$ contains the non-trivial null states $Q_{\alpha}^+\mathcal{P}$. We call a null state trivial if it can be seen to vanish without resorting to the computation of its norm. For example, $Q_1^+Q_1^+\mathcal{O}$ is a trivial null state. The shortening condition \eqref{eq:levelone} translates into
\begin{equation}
\Delta_{\mathcal{P}} = \frac{d-1}{2}q_{\mathcal{P}} + s \,,\qquad s>0\,,
\end{equation}
and thus can be thought of as a natural extension of \eqref{eq:levelzero} to $s>0$. The remaining three cases all lead to non-trivial linear relations between $\Delta$, $q$ and $s$, but none of these relations corresponds to the appearance of a non-trivial null-state. Therefore, they all lead to a contradiction since we know that $Q_{\alpha}^+\mathcal{P}$ must be non-trivial null states, since if they were trivial null-states, the condition $\lambda_1 = \lambda_2$ would itself be trivial.
\item We simply state the results for level two. The only case not leading to the type of contradiction we saw for the three disallowed cases at level one is $(q+2,S_s)$, i.e. $\mathcal{P}_{i_1\ldots i_s} = \epsilon^{\alpha \beta}Q_{\alpha}^+Q_{\beta}^+\mathcal{O}_{i_1\ldots i_s}$, which can be easily seen to always satisfy $Q_{\alpha}^+\mathcal{P} = 0$ without the need for a shortening condition on $\mathcal{O}$. There are then two allowed types of unitary representations. Either $\mathcal{O}$ is antichiral, i.e. $s = 0$ and $\Delta = - \frac{d-1}{2}q$, leading to
\begin{equation}
\Delta_{\mathcal{P}} = - \frac{d-1}{2}q_{\cP} + d\,,
\label{eq:antichiral}
\end{equation}
or $\mathcal{O}$ is generic, satisfying \eqref{eq:unitaritys} (including $s=0$), which leads to
\begin{equation}
\Delta_{\mathcal{P}}\geq \left|\frac{d-1}{2}q_{\mathcal{P}}-d + 1\right| + s + d -1\,.
\end{equation}
We must also remember that for $s=0$, the superconformal primary must satisfy the unitarity bound $\Delta\geq\frac{d-2}{2}$.
\item Of the four cases at level three, the condition $\lambda_1 = \lambda_2$ does not lead to an immediate contradiction only for $(q+1,\bar{P}_{s-1})$. Similarly to what happenes at level one, $\lambda_1=\lambda_2$ in this case implies a consistent shortening condition. The novelty here is that this shortening also kills the state $\mathcal{P}$, and thus there are no consistent possibilities at level three.
\item There is only one conformal primary at level four, but $\lambda_1 = \lambda_2$ does not lead to a consistent shortening condition on $\mathcal{O}$, so the primary at level four can not appear in the chiral OPE.
\end{itemize}

Essentially identical results were derived in \cite {Poland:2011ey,Vichi-thesis} for $d=4$, $\mathcal{N}=1$ in the context of the OPE of a chiral operator $\Phi$ with itself. The new feature of the generalization to $d<4$ is the appearance of the level-two descendant of an antichiral primary \eqref{eq:antichiral}. From R-charge conservation we have $\frac{d-1}{2}q_{\mathcal{P}} = 2\Delta_{\Phi}$ which, combined with the unitarity bound $\Delta\geq \frac{d-2}{2}$ for the operator $\mathcal{O}$, implies that the antichiral case can only be included if
\begin{equation}
\Delta_{\Phi}\leq \frac{d}{4}\,.
\label{dover4bound}
\end{equation}
Thus in $d=4$, the antichiral case can only appear when $\Phi$ is the scalar component of a free chiral superfield, where we know it does not appear since there is no coupling between $\Phi$ and any other fields. However, in $d<4$, there is a finite window for $\Delta_{\Phi}$ where the level-two descendant of an antichiral primary can make a contribution, and we will see it plays a crucial role in the Wess-Zumino model, since its appearance corresponds to the Yukawa coupling.

It follows from the above discussion that only one kind of allowed conformal primary $\mathcal{P}$ from the same superconformal multiplet can appear in the $\phi_1 \times \phi_3 $ OPE, and therefore the superconformal blocks coincide with the usual conformal blocks. Supersymmetry plays a role in this channel only through constraints on the spectrum of conformal primaries that can appear in the OPE.

%%%%%%%%%%%%%%%%%%%%%%%%%%%
\section{Intermezzo: review of the Wess-Zumino model}
\label{sec:WZ}
%%%%%%%%%%%%%%%%%%%%%%%%%%%

In this section, we remind the reader of some basic facts about the massless Wess-Zumino model in $d\leq 4$ \cite{wess1974}. A nice review on the subject can be found in \cite{Strassler:2003qg}. The model consists of the theory of a single chiral superfield $\Upsilon$ with cubic superpotential $W(\Upsilon) = \frac{1}{3}\lambda \Upsilon^3$. Equivalently, this is a theory of a complex boson and fermion with the Lagrangian
\begin{equation}\label{WZaction}
\mathcal{L}_{\text{WZ}} = \partial_{\mu}\bar{\phi} \partial^{\mu}\phi +\im \bar{\psi}\gamma^{\mu}\partial_{\mu}\psi + |\lambda|^2|\phi|^4 +  (\lambda\,\phi\psi_{\alpha}\epsilon^{\alpha\beta}\psi_{\beta} + c.c.)\,.
\end{equation}
The classical dimension of the coupling $\lambda$ is $\frac{\epsilon}{2}$, with $\epsilon \equiv 4-d$, and it is convenient to define the dimensionless coupling $\tilde{\lambda} = \mu^{-\epsilon/2} \lambda$, where $\mu$ is the renormalization scale. Supersymmetry implies that the superpotential is not renormalized. Therefore the $\beta$-function of $\tilde{\lambda}$ is determined by the anomalous dimension of the chiral field $\Phi$, which is the lowest component of the superfield $\Upsilon$
\begin{equation}
\beta_{\tilde{\lambda}} = \tilde{\lambda}\left[-\frac{\epsilon}{2} + 3\gamma_{\Phi}(\tilde{\lambda})\right]\,,
\label{eq:betagamma}
\end{equation}
where $\gamma_{\Phi} = - \frac{1}{2}\frac{d \log Z}{d\log\mu}$ and the factor of $3$ comes from the fact that $W(\Upsilon)$ is cubic. Since we know from perturbation theory and unitarity that $\gamma_{\Phi}(\tilde{\lambda}) > 0$ for $\tilde{\lambda}\ll 1$, we expect that for sufficiently small $\epsilon$, the theory has an interacting IR fixed point with unbroken supersymmetry at a coupling $\tilde{\lambda}^*>0$. This CFT is what we refer to as the critical WZ model (cWZ). The exact relation \eqref{eq:betagamma} implies that at the fixed point the anomalous dimension is
\begin{equation}
\gamma_{\Phi}(\tilde{\lambda}^*) = \frac{4-d}{6}\,,
\end{equation}
and hence
\begin{equation}\label{DeltaWZ}
\Delta_{\Phi} = \frac{d-2}{2} + \gamma_{\Phi}(\tilde{\lambda}^*) = \frac{d-1}{3}\,.
\end{equation}
This formula can also be deduced from the exact superconformal relationship between scaling dimension and R-charge of a chiral field $\Delta = \frac{d-1}{2}q$, since the R-charge of the superpotential is $q_{W} = 2$ and thus $q_{\Phi}=2/3$.

An equivalent way to state the result in \eqref{DeltaWZ} is that the $\epsilon$-expansion of the critical exponent $\eta \equiv 2\Delta_{\Phi}- (d-2)$ is exact at the leading order
\begin{equation}
\eta = 2\gamma_{\Phi}(\tilde{y}^*) = \frac{\epsilon}{3}\,.
\end{equation}
The critical exponent $\nu$, characterizing the divergence of the correlation length as the temperature approaches the critical temperature, is related to the scaling dimension of the lowest uncharged scalar, $[\bar\Phi\Phi]$, as follows
\begin{equation}
\nu^{-1} = d - \Delta_{[\bar\Phi\Phi]}\,.
\end{equation}
It is not protected by supersymmetry and has been computed at one loop in the $\epsilon$-expansion \cite{Thomas,Lee:2006if}
\begin{equation}
\nu = \frac{1}{2} + \frac{\epsilon}{4} + \cO(\epsilon^2)\,,
\end{equation}
leading to
\begin{equation}
\Delta_{[\bar\Phi\Phi]} = 2 + \cO(\epsilon^2)\,.
\end{equation}
The critical exponent $\omega$, characterizing the approach to scaling, is related to the scaling dimension of the lowest irrelevant scalar operator, $\mathcal{O}$, as
\begin{equation}
\omega = \Delta_{\mathcal{O}} - d\,.
\end{equation}
It is reasonable to expect that $\mathcal{O}$ is the supersymmetric descendant of $[\bar\Phi\Phi]$ obtained by acting on $[\bar\Phi\Phi]$ with four $Q$ supercharges. This leads to $\Delta_{\mathcal{O}} = \Delta_{[\bar\Phi\Phi]} + 2$, implying the exact relation
\begin{equation}
\omega = 2 - \nu^{-1}\,.
\end{equation}
Finally, let us note that the equation of motion for $\Upsilon$ can be written in superspace language as
\begin{equation}
D_{\alpha}D^{\alpha}\bar{\Upsilon} = \partial_{\Upsilon}W(\Upsilon)\,,
\end{equation}
where $D_{\alpha}$ is the superspace derivative corresponding to the action of the supercharge $Q_{\alpha}^+$ that annihilates the chiral superfield $\Upsilon$. This implies that the chiral ring of the fixed-point theory has the relation
\begin{equation}
\Phi^2 = 0\,.
\end{equation}
In the language of the CFT data, this means that the OPE $\Phi\times\Phi$ does not contain a chiral primary. From the results of Section \ref{ssec:chiralOPE} we can conclude that all operators that appear in the OPE are then exact under $Q_{\alpha}^+$.

There is another piece of data available about the cWZ model in $d=3$ that we will seek to match with the bootstrap results, namely the coefficient of the two-point function of the stress tensor, denoted by $C_T$. In SCFTs with four supercharges, the two-point function of the stress tensor is proportional to the two-point function of the R-current $\tau_{RR}$. In \cite{Closset:2012ru}, it was shown how $\tau_{RR}$ can be computed for $d=3$, $\mathcal{N}=2$ SCFTs from the squashed-sphere partition function $F(b)$
\begin{equation}
\tau_{RR} = \frac{2}{\pi^2}\Re \left.\frac{\partial^2 F(b)}{\partial b^2}\right|_{b=1}\,,
\end{equation}
where $b$ is the squashing parameter, $b=1$ corresponding to the round sphere. A formula for the squashed-sphere partition function of $d=3$, $\mathcal{N}=2$ theories was found using localization in \cite{Imamura:2011wg}. Denoting by $\tau_{RR}^{\textrm{(free)}}$ the two-point function of the R-current in the theory of a single free chiral multiplet, it was found in \cite{Nishioka:2013gza} that\footnote{We are grateful to Simone Giombi, Igor Klebanov, and Silviu Pufu for bringing this result to our attention.}
\begin{equation}
\frac{C_T}{C_T^{\textrm{(free)}}} = \frac{\tau_{RR}}{\tau_{RR}^{\textrm{(free)}}} \simeq 0.7268\,.
\end{equation}
We will comment further on this ratio in Section \ref{sec:results}.

%%%%%%%%%%%%%%%%%%%%%%%%%%%%%%%%%%%%%
\section{Bootstrap setup}
\label{sec:equations}
%%%%%%%%%%%%%%%%%%%%%%%%%%%%%%%%%%%%%

In this section, we review the derivation of a set of crossing symmetry equations which we later solve numerically. The results of the previous sections suggest that the structure of these ``bootstrap equations" should be very similar to those that were studied in the case of $d=4$, $\mathcal{N}=1$ SCFTs in \cite{Poland:2011ey,Vichi-thesis,Rattazzi:2010yc,Vichi2012}, and indeed this is what we find.

We are interested in the crossing symmetry constraints for the four-point function $\langle \Phi\bar \Phi\Phi\bar \Phi\rangle$, where $\Phi$ is a chiral operator with dimension $\df$ and $\bar\Phi$ is its charge conjugate. The chirality condition imposes that the R-charge is given by $q_\Phi=\frac{2}{d-1}\, \df = \frac{2}{d-1} \Delta_{\bar\Phi} = -q_{\bar\Phi}$. Conformal symmetry fixes the four point function to take the form
\bea
\langle \Phi(x_1)\bar \Phi(x_2)\Phi(x_3)\bar \Phi(x_4)\rangle\equiv \frac{g(u,v)}{|x_{12}|^{2\df} |x_{34}|^{2\df}}\,,
\eea
where the cross-ratios $u,v$ are defined in \reef{eq:crossratios}. Let us ignore supersymmetry for the moment but still insist on the presence of a $U(1)$ global symmestry under which $\Phi$ and $\bar{\Phi}$ have opposite charges. The OPE leads to a decomposition of $g(u,v)$ in terms of conformal blocks $G_{\Delta,s}(u,v)$. For instance, in the (12) channel we take $x_1\to x_2$, and get
\bea
g(u,v)=\sum_{\mathcal O} (-1)^s|c_{\Phi \bar \Phi}^{\mathcal O}|^2 G_{\Delta,s}(u,v)\,.
\eea
Recall that we are using the normalization \eqref{eq:cbasymptotics}. Equality of the OPEs in the three channels leads to the constraints
\begin{align}
v^{\df}\sum_{\mathcal O}  (-1)^s |c_{\Phi \bar \Phi}^{\mathcal O}|^2 G_{\Delta,s}(u,v)&= u^{\df}\, \sum_{\mathcal O}   (-1)^s|c_{\Phi \bar \Phi}^{\mathcal O}|^2 G_{\Delta,s}(v,u)\,,\qquad &(12)=(14)\,,\label{1214channel}\\
v^{\df}\sum_{\mathcal O}  |c_{\Phi \bar \Phi}^{\mathcal O}|^2\,G_{\Delta,s}(u,v)&=u^{\df}\sum_{\mathcal P}  |c_{\Phi \Phi}^{\mathcal P}|^2 G_{\Delta,s}(v,u)\,,\qquad &(12)=(13)\,,\label{1213channel}
\end{align}
where $\mathcal{O}$, $\mathcal{P}$ are conformal primaries appearing in the $\Phi\times\bar\Phi$, and $\Phi\times\Phi$ OPE, respectively. Symmetrizing and antisymmetrizing equation \eqref{1213channel} with respect to $u\leftrightarrow v$ allows us to write the equations in \eqref{1214channel}, \eqref{1213channel} as the system
\bea
\sum_{\mathcal  O^+}|c_{\Phi \bar \Phi}^{\mathcal O^+}|^2
\left(\begin{tabular}{c}
$F^{\df}_{\Delta,s}$\\
$F^{\df}_{\Delta,s}$\\
$H^{\df}_{\Delta,s}$
\end{tabular}
\right)
+
\sum_{\mathcal O^-}|c_{\Phi \bar \Phi}^{\mathcal O^-}|^2
\left(\begin{tabular}{c}
$F^{\df}_{\Delta,s}$\\
$-F^{\df}_{\Delta,s}$\\
$-H^{\df}_{\Delta,s}$
\end{tabular}
\right)
+
\sum_{\mathcal P}|c_{\Phi \Phi}^{\mathcal P}|^2
\left(\begin{tabular}{c}
$0$\\
$F^{\df}_{\Delta,s}$\\
$-H^{\df}_{\Delta,s}$
\end{tabular}
\right)=0\,.
\label{eq:bootstrap}
\eea
The first/second sum in \eqref{eq:bootstrap} runs over uncharged conformal primaries with even/odd spin respectively. The third term in \eqref{eq:bootstrap} is a sum over conformal primaries of charge $2q_{\Phi}$ and contains even spins only. The functions $F,H$ in \eqref{eq:bootstrap} are defined as
\bea
F^{\df}_{\Delta,s}&\equiv(-1)^s\left[v^{\df} G_{\Delta,s}(u,v)- u^{\df} G_{\Delta,s}(v,u)\right]\,,\nonumber\\
H^{\df}_{\Delta,s}&\equiv(-1)^s\left[v^{\df} G_{\Delta,s}(u,v)+ u^{\df} G_{\Delta,s}(v,u)\right].
\eea
Including the effects of supersymmetry simply means replacing conformal blocks by the superconformal blocks appropriate for each channel, and taking into account superconformal unitarity bounds. As we showed in Section \ref{sec:superblocks}, the superconformal blocks in the $\Phi\bar\Phi$ channel are linear combinations of four non-supersymmetric conformal blocks, while in the $\Phi \Phi$ channel, at most one conformal primary from a superconformal multiplet can appear, meaning that superconformal blocks are equal to non-supersymmetric conformal blocks. Equations \eqref{eq:sbdecomposition}, \eqref{eq:sbcoefficients} with $\Delta_{12}=\Delta_{34} = 0$, lead us to define
\begin{equation}
\begin{aligned}
\mathcal {F}_{\Delta,s}^{\df} &\equiv F^{\df}_{\Delta,s} + c_1 F^{\df}_{\Delta+1,s+1} + c_2 F^{\df}_{\Delta+1,s-1}+ c_3 F^{\df}_{\Delta+2,s}\,,\\
\tilde{\mathcal {F}}_{\Delta,s}^{\df} &\equiv (-1)^s\left(F^{\df}_{\Delta,s} - c_1 F^{\df}_{\Delta+1,s+1} - c_2 F^{\df}_{\Delta+1,s-1}+ c_3 F^{\df}_{\Delta+2,s}\right)\,,\\
\tilde{\mathcal {H}}_{\Delta,s}^{\df} &\equiv (-1)^s\left(H^{\df}_{\Delta,s} - c_1 H^{\df}_{\Delta+1,s+1} - c_2 H^{\df}_{\Delta+1,s-1}+ c_3 H^{\df}_{\Delta+2,s}\right)\,,
\end{aligned}
\end{equation}
where
\begin{equation}
c_1\equiv-a_{1}|_{\Delta_{12}=\Delta_{34}=0}\,, \qquad c_2\equiv-a_{2}|_{\Delta_{12}=\Delta_{34}=0}\,, \qquad c_3\equiv a_{3}|_{\Delta_{12}=\Delta_{34}=0}\,,
\label{eq:ccoeffs}
\end{equation}
and the $a_i$ were defined in \eqref{eq:sbcoefficients}. The supersymmetric version of equation \eqref{eq:bootstrap} then reads
\bea
\sum_{\mathcal  O^+}|c_{\Phi \bar \Phi}^{\mathcal O^+}|^2
\left(\begin{tabular}{c}
$\mathcal{F}^{\df}_{\Delta,s}$\\
$\tilde{\mathcal {F}}^{\df}_{\Delta,s}$\\
$\tilde{\mathcal {H}}^{\df}_{\Delta,s}$
\end{tabular}
\right)
+
\sum_{\mathcal O^-}|c_{\Phi \bar \Phi}^{\mathcal O^-}|^2
\left(\begin{tabular}{c}
$\mathcal{F}^{\df}_{\Delta,s}$\\
$\tilde{\mathcal {F}}^{\df}_{\Delta,s}$\\
$\tilde{\mathcal {H}}^{\df}_{\Delta,s}$
\end{tabular}
\right)
+
\sum_{\mathcal P}|c_{\Phi \Phi}^{\mathcal P}|^2
\left(\begin{tabular}{c}
$0$\\
$F^{\df}_{\Delta,s}$\\
$-H^{\df}_{\Delta,s}$
\end{tabular}
\right)=0\,,
\label{eq:susybootstrap}
\eea
The first two sums run over superconformal primaries of vanishing R-charge and even/odd spin respectively, while the third sum runs over conformal primaries of R-charge $q_{\mathcal{P}}=2q_{\Phi} = \frac{4}{d-1}\df$. All terms in the sums are constrained by superconformal unitarity bounds, and the third sum also by $[Q_{\alpha}^{+},\mathcal{P}] = 0$, as analyzed in Section \ref{ssec:chiralOPE}. We can summarize the constraints on the spectrum as follows
\begin{subequations}\label{eq:opcontent}
\begin{align}
\mathcal O^+:&\qquad \Delta=0\,,\,\Delta\geq s+d-2\,, &s=0,2,4,\ldots\,,\\
\mathcal O^-: & \qquad \Delta\geq s+d-2\,, &s=1,3,5,\ldots\,,\\
\mathcal P :   &\ \left\{
\begin{tabular}{l}
$\Delta=2\df+s\,,$\\ 
$\Delta=d-2\df\,,$\\
$\Delta\geq |2\df-(d-1)|+s+(d-1)\,,$
\end{tabular}
\right.
&
\begin{tabular}{l}
$s=0,2,\ldots$\,,\\
$s=0, ~~~ \df\leq d/4$\,, \\
$s=0,2,\ldots\,. $
\end{tabular}
\end{align}
\end{subequations}

Equations \reef{eq:susybootstrap}, together with the spectrum specifications \reef{eq:opcontent}, constitute a {\em linear program} for the various OPE coefficients squared. Solving this kind of problem is the basis of the numerical (conformal) bootstrap program, and the procedure has by now been described extensively in the literature. Here, we shall provide a very brief description of how such a problem can be solved, and refer the reader to \cite{El-Showk2014a,Paulos2014a} for further details. 

The first step is to reduce the continuously infinite functional equations to some finite set of constraints. The usual bootstrap procedure is to Taylor expand to some given order in the two cross-ratios $u$ and $v$ (or an alternative coordinate system). The number of derivative components is most conveniently labeled by a parameter $n_{\mbox{\tiny max}}$, in terms of which the total number of constraints is $\frac 12 (n_{\mbox{\tiny max}}+1)(n_{\mbox{\tiny max}}+2)$. Standard algorithms, such as Dantzig's simplex method, can then be used to try to obtain a set of OPE coefficients which solve the equations. This may or may not be possible, depending on the set of operators that we allow in the crossing equations. In particular, to derive bounds, one imposes constraints on the sets of operators allowed in the sum rule \reef{eq:susybootstrap} until a solution can no longer be found. Typically, this constraint is a gap in the set of uncharged scalar operators, so that if a solution cannot be found for a given $n_{\mbox{\tiny max}}$, then it is ruled out definitively. Increasing the parameter  $n_{\mbox{\tiny max}}$ can then only lead to tighter bounds. 
In this work, our calculations were done using a modification of a Python-based arbitrary precision\footnote{In the implementation used in this paper arbitrary-precision arithmetic  was used only for matrix inversion as lower precision generation of conformal blocks proved sufficient at the values of $n_{\textrm{max}}$ presented here.} simplex method solver for semi-infinite linear programs \cite{El-Showk2014a}. The package \cite{Paulos2014a} was also used as a cross-check on some results.

%%%%%%%%%%%%%%%%%%%%%%%%%%%%%%%%%%%%%
\section{Bootstrap results}
\label{sec:results}
%%%%%%%%%%%%%%%%%%%%%%%%%%%%%%%%%%%%%

Having developed the technology to analyze crossing symmetry for SCFTs with
four Poincar\'{e} supercharges in various dimensions, we now apply it to study and constrain
the space of allowed theories. Theories with only four Poincar\'{e} supercharges do
not exist in $d > 4$ and, while the status of SCFTs (and CFTs) in $d < 2$ is
certainly an interesting question, for this study, we choose to restrict
ourselves to $2 \leq d \leq 4$.

Since we made no use of parity invariance in our derivation of superconformal blocks and crossing relations, our bounds also apply to unitary theories which do not preserve parity, such as $\mathcal{N}=2$ superconformal Chern-Simons-matter theories in $d=3$.

Unless otherwise specified, all the plots shown in this section were made using $n_{max}=6$ which gives 84 constraints ($28$ terms in the Taylor expansion of the three-vector identity in \eqref{eq:susybootstrap}).

%%%%%%%%%%%%%%%%%%%%%%%%%
\subsection{Scalar operator bounds}
%%%%%%%%%%%%%%%%%%%%%%%%%

%
\begin{center}
\begin{figure}[h!]
\includegraphics[scale=0.9]{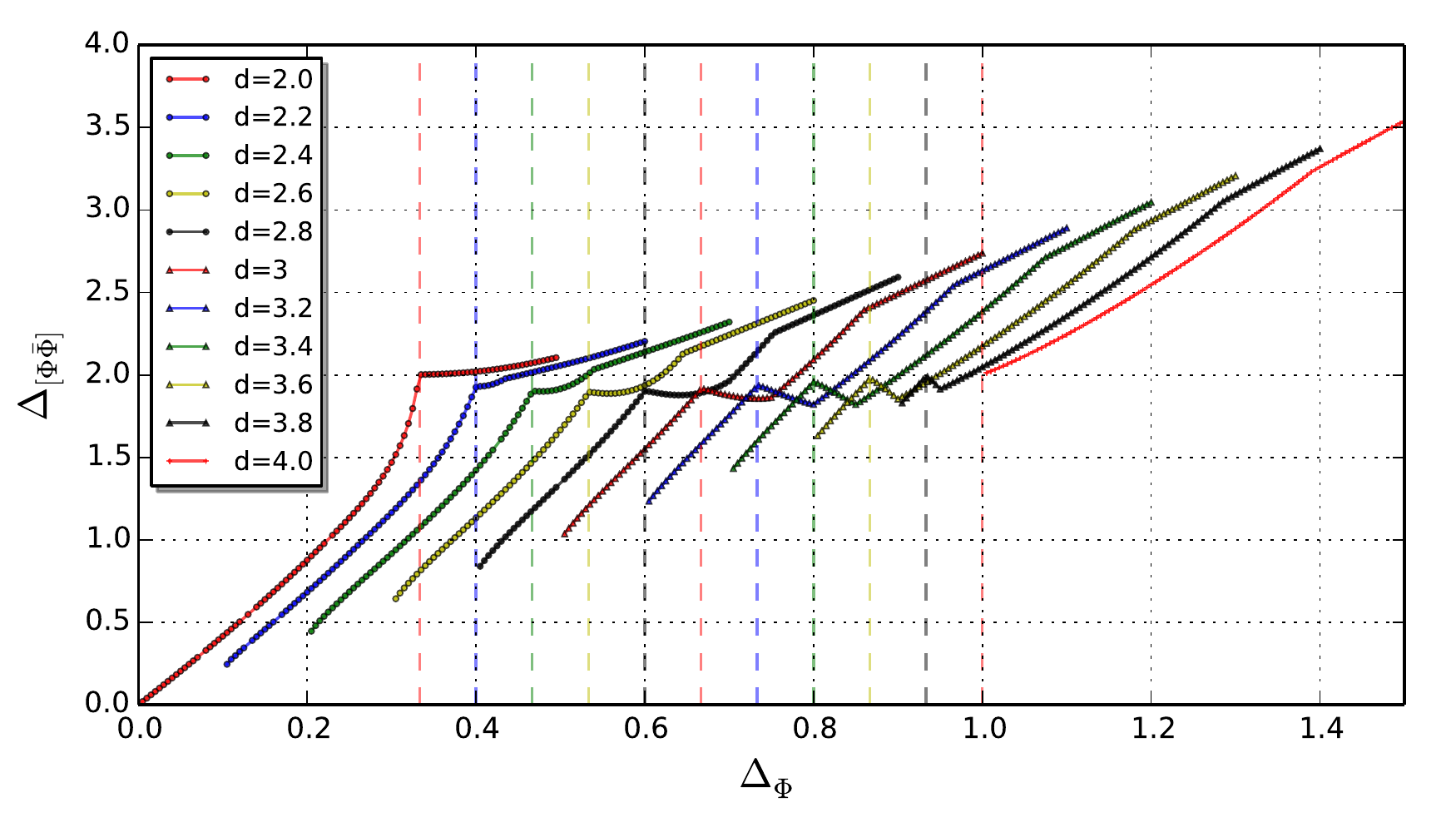}
\centering
\caption{Upper bound on the lowest-dimension neutral scalar operator, $[\Phi \bar{\Phi}]$, appearing in the $\Phi \times \bar{\Phi}$ OPE.  The dashed vertical lines correspond to $\Delta_\Phi=\frac{d-1}{3}$, the protected dimension of $\Phi$ in the cWZ model in dimension $d$. The value of $d$ associated to a line is indicated by its color, which matches the corresponding bound plot.}
\label{fig:epsbound}
\end{figure} 
\end{center}
\begin{center}
\begin{figure}[h!]
\begin{center}
%\centering
\includegraphics[scale=0.8]{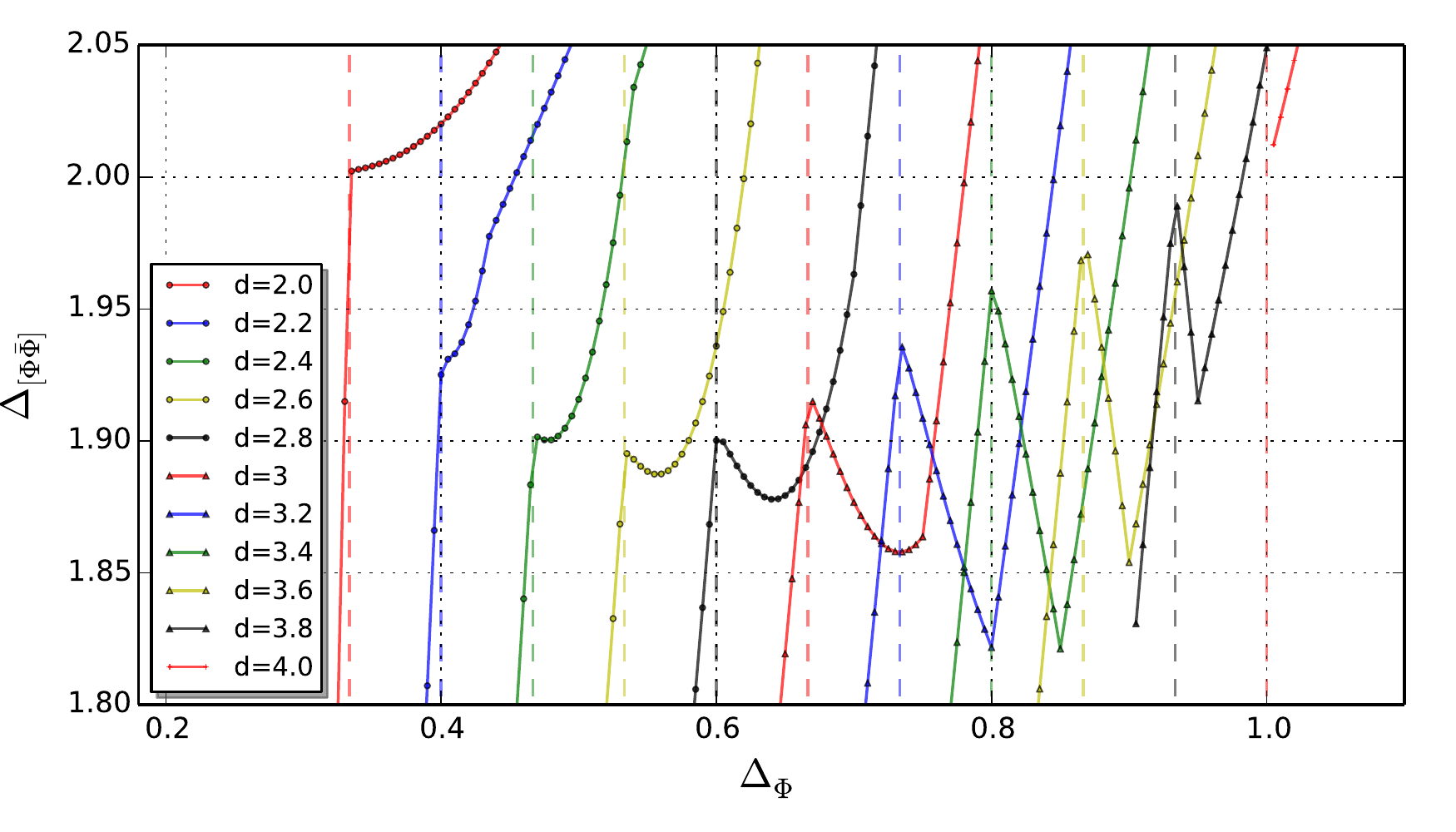}
\caption{A close-up of the bounds in Figure \ref{fig:epsbound}.  Note that the first kink in every dimension corresponds to $\Delta_\Phi=\frac{d-1}{3}$ (the locations of the vertical lines).}
\label{fig:epsbound_zoom}
\end{center}
\end{figure} 
\end{center}

We begin our numerical exploration by determining bounds on the scaling
dimension of the first scalar operator in
the $\Phi \times \bar{\Phi}$ OPE as a function of $\Delta_\Phi$.  This
corresponds to the lowest dimension scalar in the ${\cal O}^+$ channel in
equation \eqref{eq:susybootstrap}.   Throughout this section, we will refer to this
operator schematically as $[\Phi\bar{\Phi}]$, following the weak-coupling
intuition of it being the composite operator of $\Phi$ and $\bar{\Phi}$. Bounds in various dimensions, $d=2,\dots,4$, are shown in Figures \ref{fig:epsbound} and
\ref{fig:epsbound_zoom} for a range of conformal dimensions $\Delta_0 \leq \Delta_\Phi \leq \Delta_0 +
\frac{1}{2}$ (with $\Delta_0 = \frac{d-2}{2}$ the conformal dimension of a free scalar in dimension $d$).

Figures \ref{fig:epsbound} and \ref{fig:epsbound_zoom} exhibit a variety of
interesting features.

\begin{enumerate}
	\item A clear kink at $\Delta_\Phi = \frac{d-1}{3}$ where we conjecture that the bound is saturated by the $d$-dimensional critical Wess-Zumino model with a cubic superpotential.
	\item A second kink located at $\Delta_\Phi = \frac{d}{4}$ that is very
		sharp for $3 \leq d \leq 4$, but seems to soften, and may no
		longer exist, for $d < 3$.
	\item A third kink at some value of $\Delta_\Phi > \frac{d}{4}$.
		In $d=3$ the value is $\Delta_\Phi \approx 0.86$.  In $d=4$ this
		feature appears at $\Delta_\Phi \approx 1.38$ and is likely the same
		feature first observed in \cite{Poland:2011ey}.
\end{enumerate}
The location of the second feature described above, $\Delta_\Phi = \frac{d}{4}$, coincides with a kinematically special point. This is the value of $\Delta_\Phi$  where the scalar operator $\mathcal{P}$ in the $\Phi\times\Phi$ OPE with dimension $d-2\Delta_\Phi$ is a superdescendant of a superconformal primary which   hits the unitarity bound (see \eqref{eq:opcontent} and the discussion around \eqref{dover4bound}).  The third kink, however, does not seem
to correspond to any kinematically special point.  We will discuss these two features in more detail in Section \ref{sec:kinks}.

\subsection{OPE and central charge}\label{subsec:centralcharge}

In addition to placing bounds on operator dimensions, the numerical bootstrap
allows us to extract the spectrum and OPE coefficients associated with the
``extremal'' solution that saturates these bounds \cite{ElShowk:2012hu}.  In
particular, we can use this procedure to deduce $|c_{\Phi\bar\Phi T}|^2$, the
squared OPE coefficient of the stress-tensor in the $\Phi \times \bar{\Phi}$
OPE, from which we can compute $C_T$ (the canonical normalization of the stress-tensor two-point function) associated with the
solutions lying along the bounding curves in Figure \ref{fig:epsbound}.  In two
dimensions, $C_T$ reduces to $2\, c$, where $c$ is the central charge of the left/right Virasoro algebra. In general dimension, $C_T$ is not always related to a conformal anomaly, but we still refer to it as the central charge. In terms of the OPE coefficient in our normalization, equation (\ref{eq:cbasymptotics}), the central charge\footnote{We follow the normalization of \cite{Dolan:2000ut}, in particular equation (4.2) in that reference.}
is
\begin{equation}
	C_T = \frac{\Delta_\Phi^2}{ |c_{\Phi\bar\Phi T}|^2 } \left(\frac{d}{d-1}\right)^2 \,.
\end{equation}
In theories with four Poincar\'{e} supercharges, the stress-tensor is not a superconformal
primary, but rather lies in the supermultiplet of the R-current, so what we
actually read off with our approach is $|c_{\Phi\bar\Phi J}|^2$ with $J$ a conserved
spin-one superconformal primary (of dimension $\Delta_J=d-1$).  From this, we extract the OPE coefficient of the spin-two descendant using \eqref{eq:ccoeffs}.
Note also that unlike in \cite{El-Showk2014}, here we are not maximizing the
stress-tensor (or R-current) OPE coefficient, but rather simply extracting it from a particular solution, characterized by having a maximal allowed dimension of $[\Phi\bar{\Phi}]$.  
\begin{center}
\begin{figure}[h!]
\centering
\includegraphics[scale=0.8]{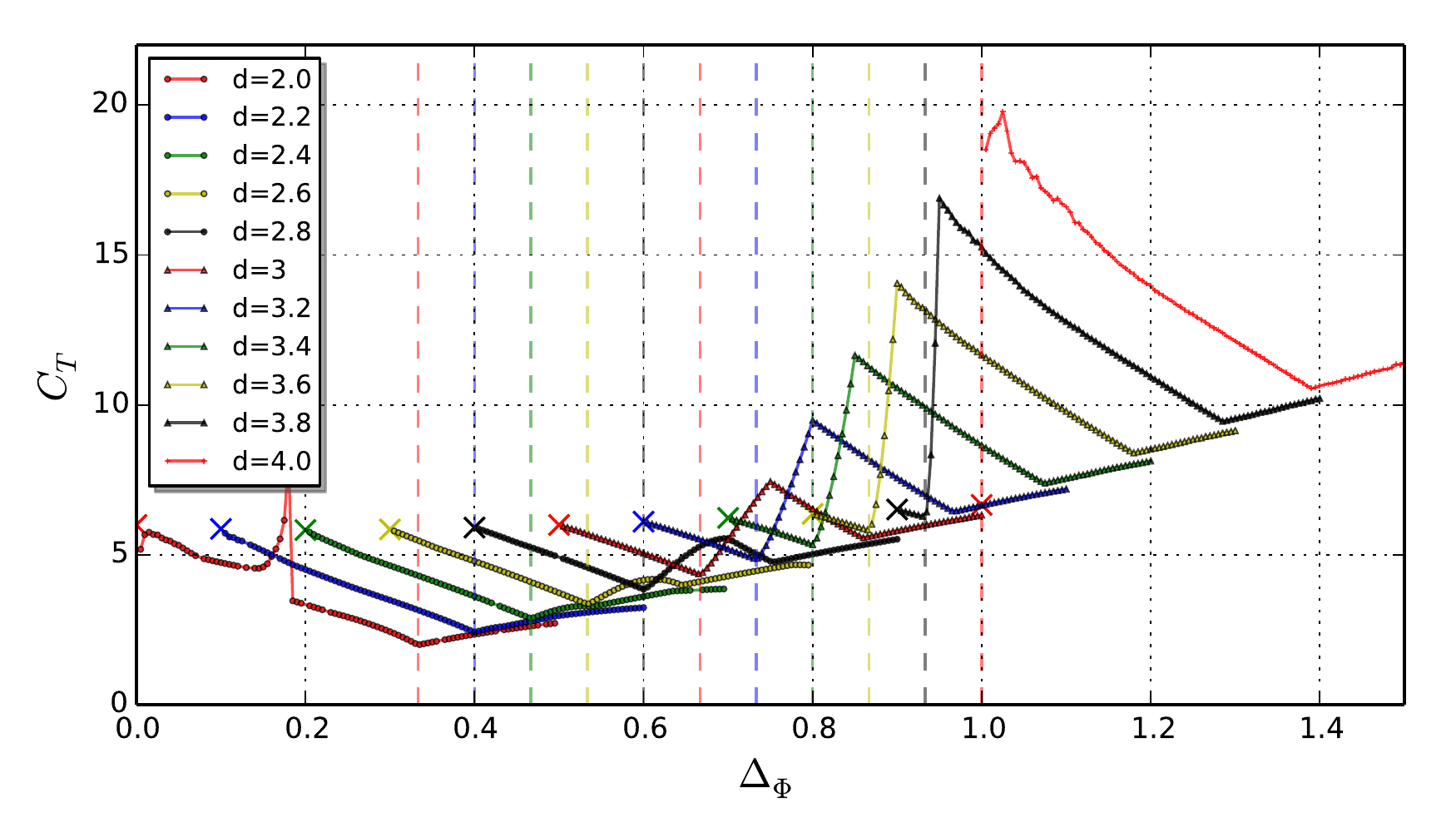}
\caption{The central charge, $C_T$, of the boundary solution, i.e. when $\Delta_{[\Phi \bar{\Phi}]}$ saturates the bounds given in Figure \ref{fig:epsbound}.  The crosses denote the value of $C_T$ for a free chiral multiplet in dimension $d$.  The dashed vertical lines lie at $\Delta_\Phi=\frac{d-1}{3}$, corresponding to the chiral primary field of the cWZ model in dimension $d$.}
\label{fig:CT}
\end{figure} 
\begin{figure}[h!]
\centering
\includegraphics[scale=0.8]{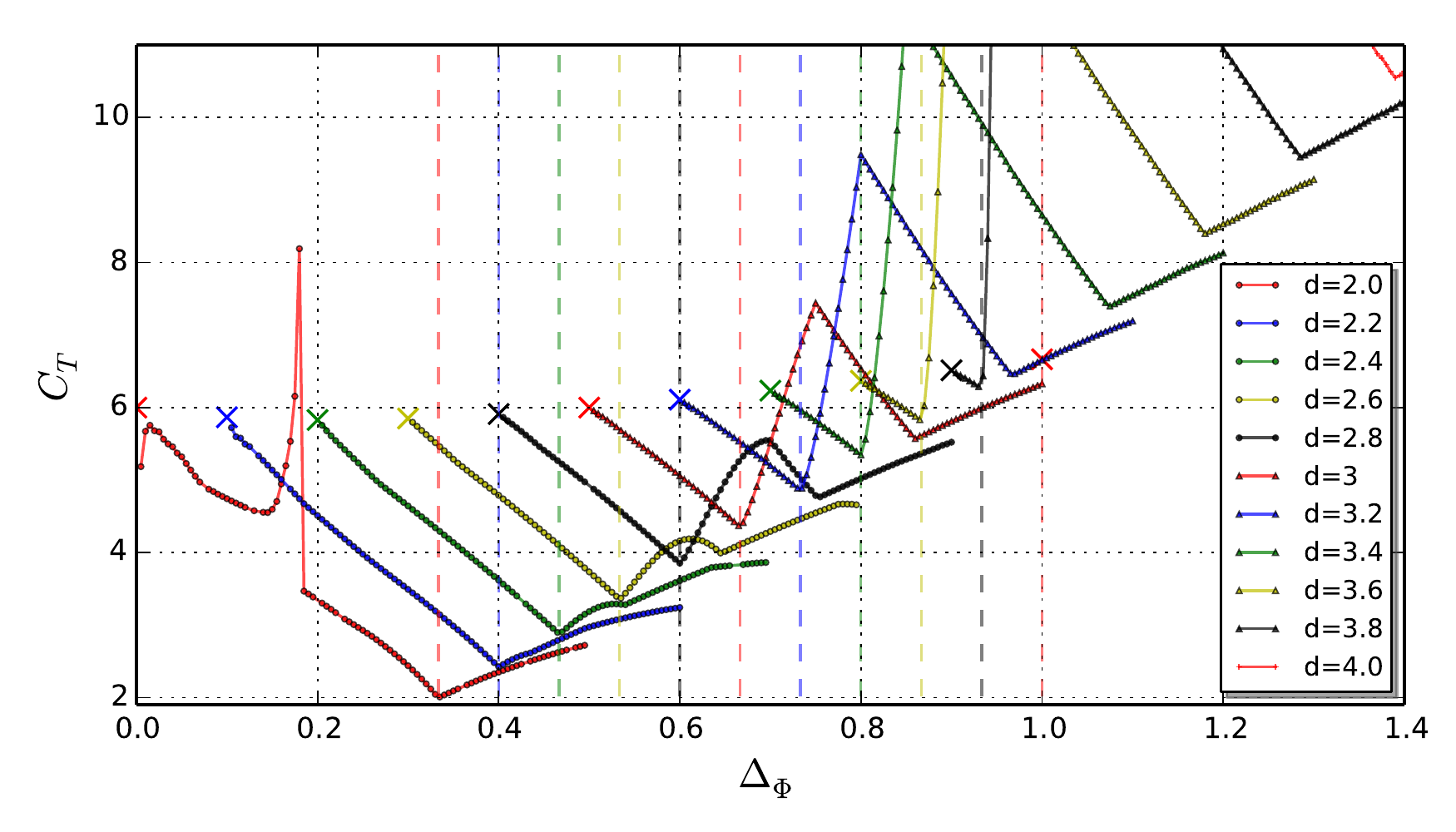}
\caption{A close-up of the curves in Figure \ref{fig:CT}. The minimum in every dimension exactly corresponds to $\Delta_\Phi=\frac{d-1}{3}$ (the locations of the vertical lines). Note that $C_T$ in $d=2$ lies precisely at 2, corresponding to the known value $c=\bar c = 1$ of the lowest $\mathcal{N}=2$ minimal model (see Section \ref{sec:2d_minmodels}).}
\label{fig:CT_zoom}
\end{figure} 
\end{center}

In the normalization given above, a free boson has $C^{(b)}_T = \frac{d}{d-1}$ while a free Dirac fermion has $C^{(f)}_T = d$, so for a free chiral multiplet we have
\begin{equation}
	C^\textrm{(free)}_T = 2 C_T^{(b)} + C_T^{(f)} = \frac{d (d+1)}{d-1}\,.
\end{equation}
The values of $C_T^{\textrm{(free)}}$ for $d=2,\dots,4$ are shown in Figures
\ref{fig:CT} and \ref{fig:CT_zoom} as large crosses which, as expected, sit at the
limiting value of $C_T$ as $\Delta_\Phi$ approaches the unitarity bound in dimension $d$.

The $C_T$ plots share a lot of the structure of the $\Delta_{[\Phi
\bar{\Phi}]}$ plots.  We find local minima (that are global minima within the
range of the plot) at $\Delta_\Phi=\frac{d-1}{3}$ corresponding to the exact
dimension of the chiral field in the $d$-dimensional cWZ model.  Moreover, a
sharp spike appears for $3 \leq d \leq 4$ at $\Delta_\Phi=\frac{d}{4}$. This
spike is a local maximum of the $C_T$ curve rather than a
minimum.  Once more, it is not clear if this last feature persists for $d < 3$.
There is also a third feature: another local minimum at the value of
$\Delta_\Phi$ corresponding to the third kink in the bounds plot.  This also
implies a local $C_T$ minimum in $d=4$ for the kink at $\Delta_\Phi\approx 1.4$, as
first observed in \cite{Poland:2011ey}.

It is important to emphasize that the curves depicted in Figures \ref{fig:CT} and \ref{fig:CT_zoom} are {\em not} the result of maximizing the stress-tensor OPE and hence are not, in any strict sense, lower bounds on $C_T$. However, a preliminary comparison of $\Delta_{[\Phi\bar{\Phi}]}$ maximization and $C_T$ minimization (analogous to the analysis in \cite{El-Showk2014}) suggests that these two are equivalent, at least in the region $\Delta_\Phi \lesssim \frac{d-1}{3}$.  A more thorough investigation of this question is left to future studies.

%%%%%%%%%%%%%%%%%%%%
\subsection{Two-dimensional $\mathcal{N}=2$ minimal models}
\label{sec:2d_minmodels}
%%%%%%%%%%%%%%%%%%%%

As there is a great deal known about two-dimensional superconformal minimal
models, we can use them as a benchmark to compare various exactly known
quantities with our numerical estimates. In Appendix \ref{App:minmod}, we summarize some of the salient features of these theories.
\begin{center}
\begin{figure}[h!]
\begin{center}
%\begin{subfigure}[t]{0.45\textwidth}
\includegraphics[scale=0.8]{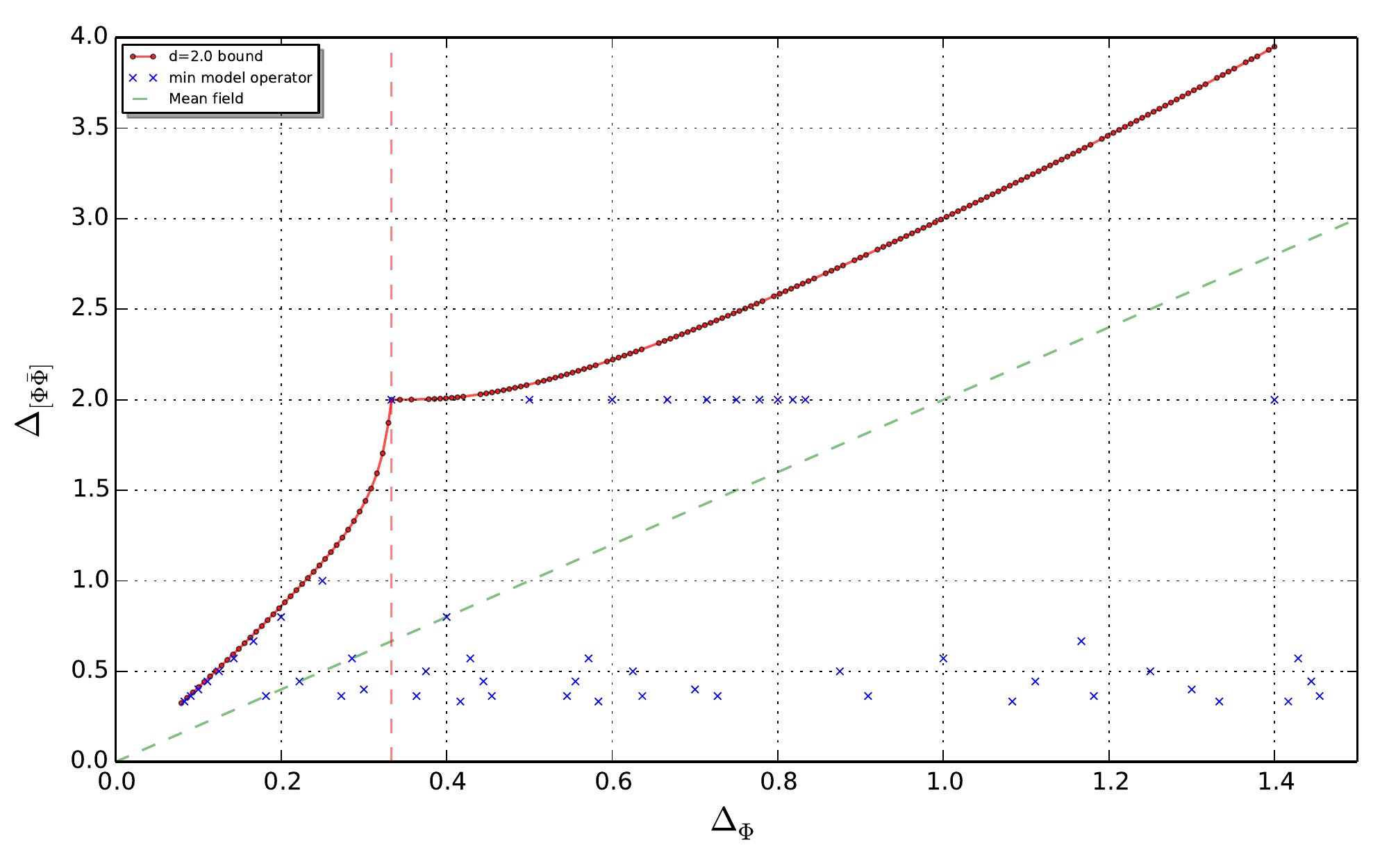}
\caption{
	An extended view of the upper bound on $\Delta_{[\Phi \bar{\Phi}]}$ in $d=2$ (with $n_{max}=9$). 
The blue crosses mark the exact dimensions of operators from various superconformal minimal models.  The cross at $(\frac{1}{3}, 2)$ corresponds to the super-Ising model (i.e. the $k=1$ super-Virasoro minimal model).  The dashed green like corresponds to $\Delta_{[\Phi\bar\Phi]}=2 \Delta_\Phi$, the expected value in mean field theory.  
}
\label{fig:2d_epsbound}
\end{center}
\end{figure} 
\end{center}
The ${\cal N}=2$ minimal models are labeled by a positive integer $k$, which determines their central charge via
\begin{equation}\label{eq:minmodel_c}
c= \frac{3\,k}{k+2}\,.
\end{equation}
Superconformal primaries in these models are labeled by two integers $n=0,\dots,k$ and $m=-n, -n+2,
\dots, n$ with (holomorphic) dimension $h$ and R-charge $\Omega$
\begin{equation}\label{eq:minmodel_hq}
	h_{n,m} = \frac{n\,(n+2) -m^2}{4\,(k+2)}\,, \qquad\qquad  \Omega=\frac{m}{\,(k+2)}\,.
\end{equation}
The chiral (antichiral) primaries have $m=\pm n$, respectively. In principle, one can apply the superconformal bootstrap to two-dimensional conformal theories with generic spectrum and only $(0,2)$ supersymmetry. However, in our analysis, we have restricted to theories with $(2,2)$ supersymmetry and a diagonal spectrum. Our conventions imply that $C_T = 2c$.

The model with $k=1$ has $C_T=2$ and two super-Virasoro primary operators of
dimension $\Delta_{1,\pm 1}=\frac{1}{3}$ and R-charge $q=\pm \frac{2}{3}$ (and of
course the identity $\Delta_{0,0}=0$).  The $\Phi_{1,1} \times \Phi_{1,-1}$ OPE
contains only the super-Virasoro family of the identity, so that the first primary of the global superconformal algebra appearing after the identity is $\Omega_{-1}\bar\Omega_{-1}|0\rangle$, which has $\Delta = 2$.
Indeed, this operator must appear in the OPE of any chiral primary and its conjugate in any local two-dimensional $\mathcal{N}=2$ SCFT. This immediately allows us to determine that all hypothetical CFTs saturating our bounds for $\df \gtrapprox 1/3$ cannot be local theories. It is
possible that adding more constraints (i.e. derivatives in the crossing
symmetry relations) will bring the bound down, but we know that at best, it can
asymptote to the line $\Delta_{[\Phi\bar{\Phi}]} = 2 \Delta_\Phi$, corresponding
to a supersymmetric version of mean field theory (also known as
generalized free field theory).  Note that the latter indeed does not have a
local stress tensor and hence does not benefit from the standard enhancement
to the infinite conformal symmetry in $d=2$.
\begin{center}
\begin{figure}[ht!]
\begin{center}
\includegraphics[scale=0.8]{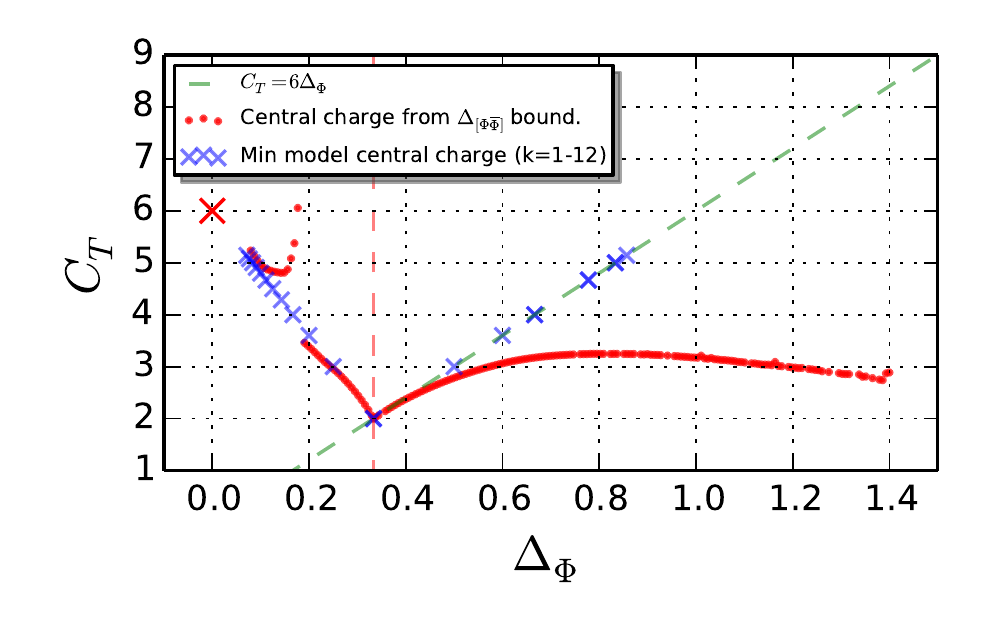}
\includegraphics[scale=0.8]{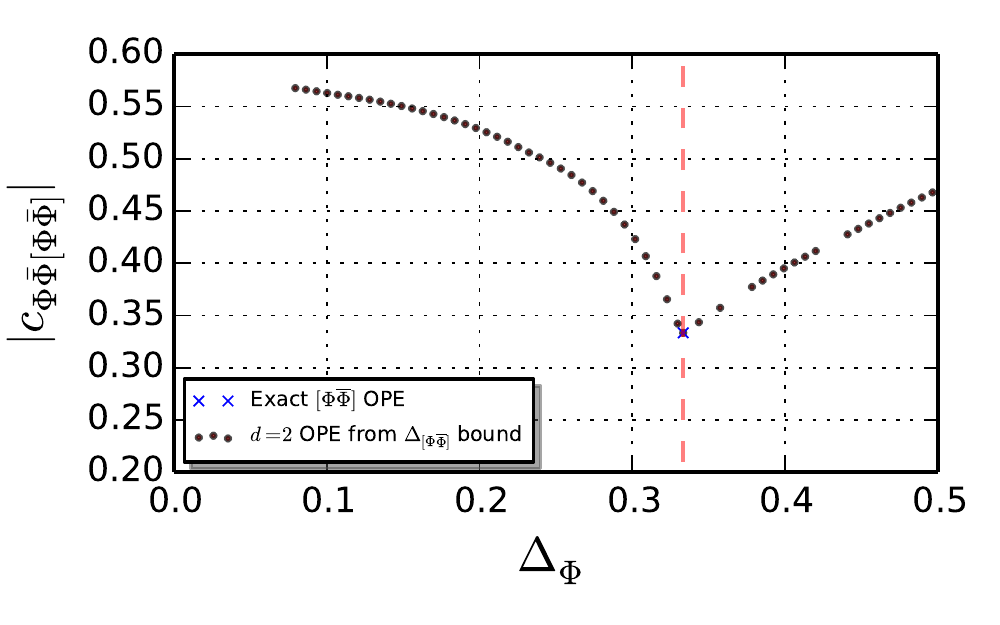}
\caption{Central charges {\em(left)}, and the OPE coefficient of $[\Phi\bar{\Phi}]$ {\em (right)}, for $d=2$, extracted from the boundary solution in Figure \ref{fig:2d_epsbound} . 
The blue crosses give the expected values of $C_T$ for the first few super minimal-models ($k=1, \ldots,11$). The dashed green line, $C_T = 6 \Delta_\Phi$, is the unitarity bound discussed in Appendix \ref{App:C1}. Both figures were made with $n_{max}=9$.} \label{fig:2d_CT_OPE}
\end{center}
\end{figure} 
\end{center}

In Figure \ref{fig:2d_epsbound}, we focus our attention on the $d=2$ bound and
superimpose the dimensions of known minimal model operators.  At $\Delta_{\Phi}
= \frac{1}{3}$, we find that the bound is very close to 2, suggesting that the $k=1$ minimal model saturates our bound.  This observation is further confirmed
in the left panel of Figure \ref{fig:2d_CT_OPE}, where we show that $C_T\approx2$ at this point as
expected. As a further check we plot, in the right panel of Figure \ref{fig:2d_CT_OPE},
the absolute value\footnote{Since OPE coefficients only appear squared in the crossing symmetry relations we consider, we only have access to their magnitude, not their sign.} of the OPE coefficient $|c_{\Phi \bar{\Phi} \,
[\Phi\bar{\Phi}]}|$.  There is clearly a cusp at $\Delta_\Phi=\frac{1}{3}$, $|c_{\Phi \bar{\Phi} \,[\Phi\bar{\Phi}]}|\approx \frac{1}{3}$, which is indeed the expected value for this OPE
coefficient in the $k=1$ model (see Appendix \ref{App:C1} for a derivation).  Let
us emphasize once more that the OPE coefficients appearing in our figures are not
computed by maximizing any OPE coefficient but rather are extracted
from the solutions saturating the $\Delta_{[\Phi\bar{\Phi}]}$ bound (see
\cite{ElShowk:2012hu}).

As mentioned above, $\Delta_\Phi=\frac{d-1}{3}$ is the expected dimension of the
protected operator $\Phi$ of the cWZ model, which can
be thought of as a super-symmetric generalization of the Ising model.  The
$k=1$ model fits naturally into this role, being the simplest super-Virasoro
minimal model. Moreover, it was shown in \cite{Vafa:1988uu,Lerche:1989uy} that precisely the minimal model with $k=1$ arises from an $\mathcal{N}=2$ Ginzburg-Landau theory with a cubic superpotential, i.e. the two-dimensional incarnation of the Wess-Zumino model.

To the left of $\df=1/3$, we see that the upper bound on $\Delta_{[\Phi\bar{\Phi}]}$ is very nearly saturated at points corresponding to $\Phi = \Phi_{1,1}$, $\bar\Phi = \Phi_{1,-1}$, $[\Phi\bar\Phi] = \Phi_{2,0}$ in the minimal models with $k\geq 2$, which lie at
\bea
\Delta_{\Phi}=\frac{1}{(k+2)}\,, \qquad\qquad \Delta_{[\Phi \bar \Phi]}=\frac{4}{(k+2)}\,,\qquad k\geq 2\,.
\eea
From the left panel in Figure \ref{fig:2d_CT_OPE} it seems, however, that for
$\Delta_\Phi < 1/3$, the central charges extracted from the boundary solutions
do not precisely match those of the $k > 1$ minimal models.  This suggests that
the latter do not exactly saturate our bound\footnote{The ``extremal functional
method'' advocated in \cite{ElShowk:2012hu} requires a very precise
determination of the maximal scalar gap in order to yield (generically) a
unique solution. Moreover, if this maximal value is sufficiently far from the
expected value of $\Delta_{[\Phi\bar\Phi]}$ in a particular theory, then the
resulting spectrum might be quite different.} at the given constraint level, a
phenomenon which has also been observed for the higher minimal models in the
non-supersymmetric case.  It may be that imposing further constraints, i.e. higher values of $n_{max}$, will improve the situation but, as this is not our focus here, we leave this question
for future explorations.

The blue crosses in Figure \ref{fig:2d_epsbound} to the right of the super-Ising point $(1/3,2)$ correspond to the fusion of $\Phi_{k,k}$, i.e. the chiral primary with the highest conformal dimension in the $k$-th minimal model, with its conjugate $\Phi_{k,-k}$. As noted above, in this case, $[\Phi\bar\Phi] = \Omega_{-1}\bar\Omega_{-1}|0\rangle$, and thus $\Delta_{[\Phi\bar\Phi]}=2$. Our numerical bound does show a short plateau with $\Delta_{[\Phi\bar\Phi]} = 2$ just to the right of the super-Ising kink. These boundary solutions are ruled out however in a full-fledged $\mathcal{N}=2$ SCFT with super-Virasoro symmetry. This can be seen by the virtue of the unitarity bound $C_T \geq 6\Delta_{\Phi_{\rm max}}$ (see Appendix \ref{App:C1}), which is shown in the left panel of Figure \ref{fig:2d_CT_OPE} as the green dashed line. However, it is reassuring that $C_T$ corresponding to the numerical solution of the crossing on the boundary asymptotes to $C_T = 6\Delta_{\Phi}$, and hence to the correct value in the minimal models. 

%%%%%%%%%%%%%%%%%%%%%%%%%%%%%%%%%%
\subsection{Bootstrapping the cWZ model in $2\leq d \leq 4$}
%%%%%%%%%%%%%%%%%%%%%%%%%%%%%%%%%%

In this section, we analyse in more detail the numerical bootstrap
results at $\Delta_\Phi=\frac{d-1}{3}$ for $2 \leq d \leq 4$.  As previously
noted, this value of $\Delta_\Phi$ is significant as it corresponds to the
protected dimension of a chiral primary operator in the $d$-dimensional cWZ model.  As the bounds for every $2 \leq d \leq 4$
in Figure \ref{fig:epsbound} have a kink precisely at this value of
$\Delta_\Phi$, we conjecture that the bounds are saturated by the operator
$[\Phi\bar{\Phi}]$ in the $d$-dimensional cWZ theory.

As argued in \cite{Hogervorst:2014rta}, it is likely that theories in
fractional dimension are non-unitary and may even suffer further pathologies.
Nonetheless, they provide a useful interpolation between theories in integer
dimension, allowing us to track critical exponents and other features as a function of the dimensions. This idea is similar in spirit to the $\epsilon$-expansion.  The literature
on the cWZ model is rather sparse and very few critical
exponents have been computed and only to leading order, see \cite{Thomas,
Lee:2006if, Grover:2013rc}.  This motivates a companion paper \cite{BEMP2}
where we conduct a more detailed numerical study of the phenomenologically
interesting case of $d=3$.
\begin{center}
\begin{figure}[h!]
\begin{center}
\includegraphics[scale=0.8]{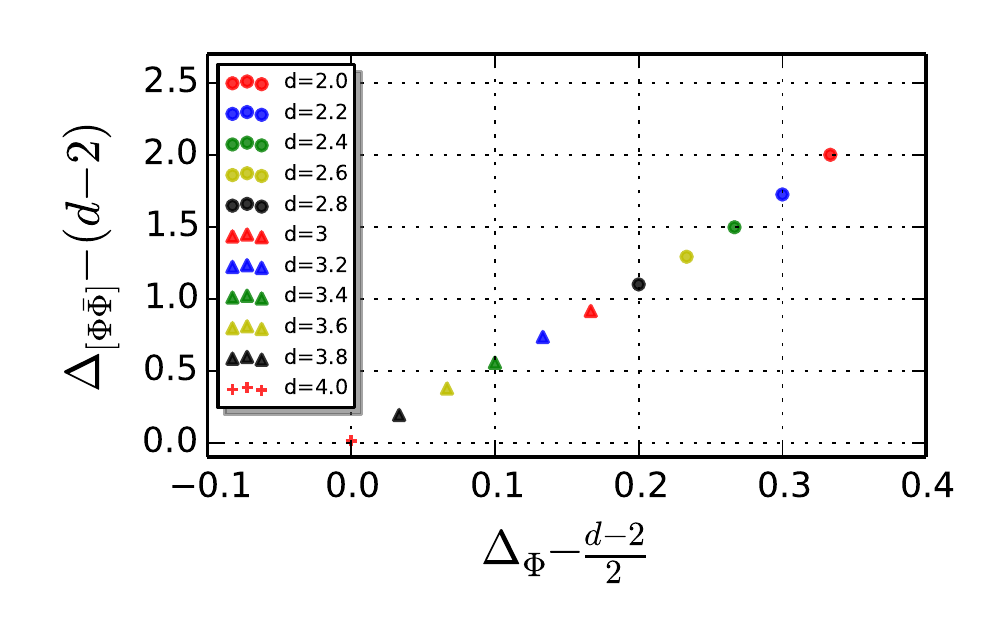}
\includegraphics[scale=0.8]{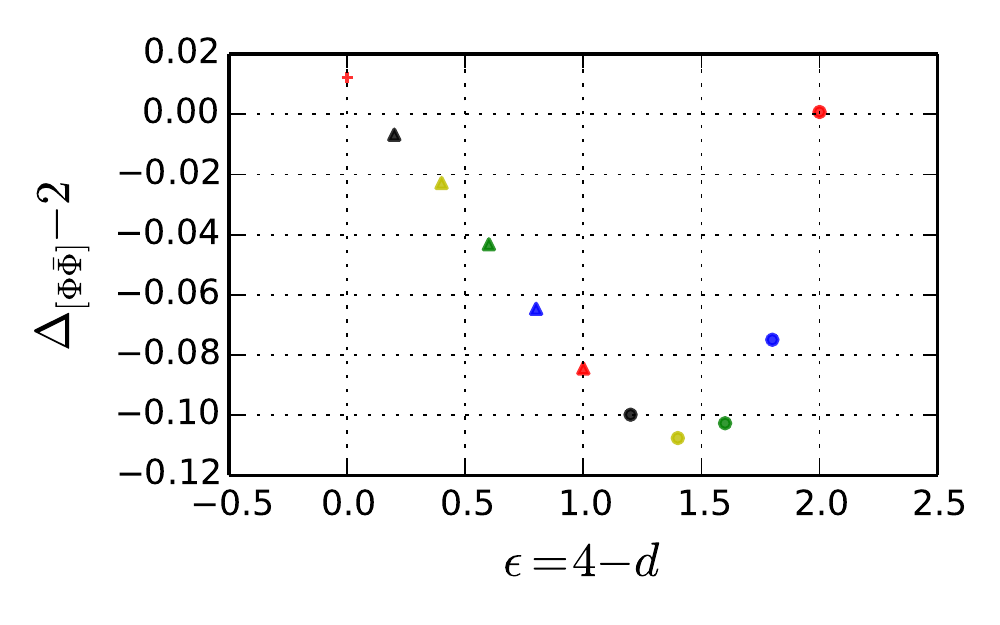}
\caption{Predictions for the anomalous dimension of $[\Phi \bar{\Phi}]$ in the $d$-dimensional cWZ model.   The $\epsilon$-expansion for this operator dimension is known to linear order and gives $\Delta_{[\Phi \bar{\Phi}]} = 2 + {\cal O}(\epsilon^2)$ so on the RHS we show $\Delta_{[\Phi \bar{\Phi}]} - 2$ as a function of $\epsilon$. }
\label{fig:IsingEps}
\end{center}
\end{figure} 
\end{center}

\begin{center}
\begin{figure}[h!]
\centering
\includegraphics[scale=0.8]{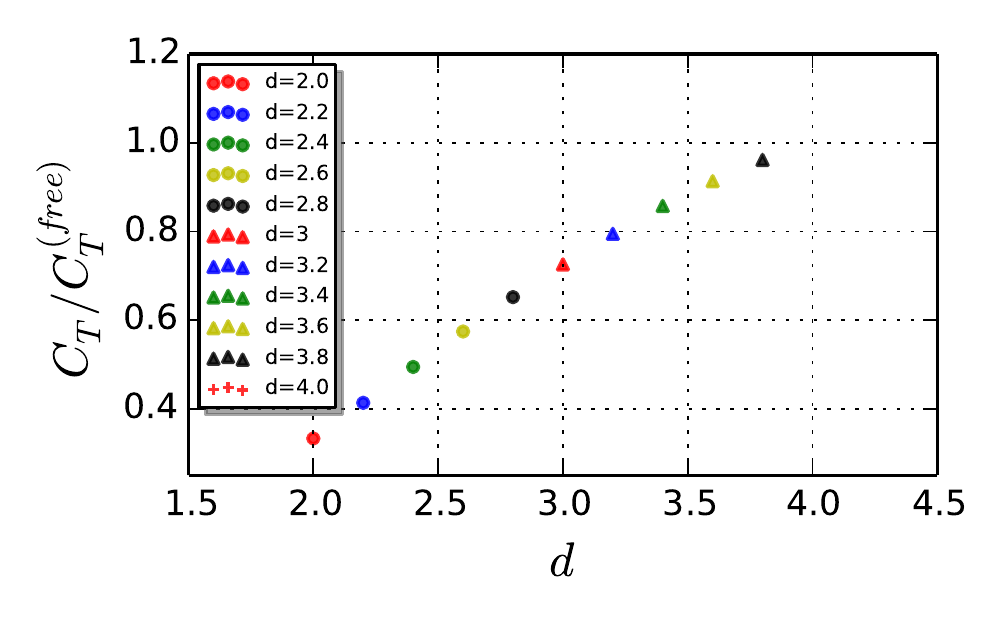}
\caption{Our prediction for the central charge, $C_T$, of the cWZ model in $d$ dimensions normalized by the value for a free chiral superfield, $C_T^{\text{free}}=\frac{d(d+1)}{d-1}$.  This data is extracted from the solution which saturates the bounds given in Figure \ref{fig:epsbound} at $\Delta_\Phi=\frac{d-1}{3}$.  The exact value in $d=3$ can be computed via localization to be $\simeq 0.7268$ while in this figure ($n_{max}=6$) we find $\sim 0.7260$ (see \cite{BEMP2} for a more precise determination).}
\label{fig:IsingCT}
\end{figure} 
\end{center}

As discussed in Section \ref{sec:WZ}, the dimension of $[\Phi \bar{\Phi}]$ in this theory has only been computed to the first order in the $\epsilon$-expansion
\begin{equation} \label{eq:dimeps}
	\Delta_{[\Phi\bar{\Phi}]} =  2 - \epsilon + \frac{1}{\nu} = 2+ {\cal O}(\epsilon^2),
\end{equation}
so we do not have precise estimates to compare with.  Our numerical results for the maximal value of $\Delta_{[\Phi\bar{\Phi}]}$ at
$\Delta_{\Phi}=(d-1)/3$ are presented in Figure \ref{fig:IsingEps}.   To give a
better sense for this quantity, we plot both the anomalous dimension
$\Delta_{[\Phi\bar{\Phi}]}-(d-2)$ against the anomalous dimension
$\Delta_{\Phi}-\frac{d-2}{2}$, and the difference
$\Delta_{[\Phi\bar{\Phi}]} -2$ as a function of $\epsilon$. The latter gives an estimate for the form
of the unknown ${\cal O}(\epsilon^2)$ corrections in (\ref{eq:dimeps}).
We also plot, in Figure \ref{fig:IsingCT}, the values of $C_T$ at
$\Delta_\Phi=(d-1)/3$, normalized with respect to $C_T$ for a free chiral
field.  Recall, from Figure \ref{fig:CT_zoom}, that these correspond to local
minima of $C_T$ which we conjecture to correspond to the $d$-dimensional cWZ
model.

The location of the kink and the fact that it corresponds to the exact result,
$\Delta_{[\Phi\bar{\Phi}]} = 2$, in $d=2$ supports our claim that we are indeed
studying the cWZ theory.  Moreover, equation (\ref{eq:dimeps}) is consistent with
what we observe in Figure \ref{fig:epsbound_zoom}; namely that
$\Delta_{[\Phi\bar{\Phi}]} \approx 2$ for $2 \leq d \leq 4$.

The strongest evidence for our conjecture comes, however, not from a critical
exponent, but from the computation of $C_T$.  As discussed in Section \ref{sec:WZ}, it is possible to
determine this quantity, in $d=3$, by taking derivatives of the squashed-sphere
partition function, a quantity that is exactly computable via localization.
This computation yields $C_T/C_T^{\textrm{(free)}} \simeq 0.7268$ while our
best numerical estimate (in \cite{BEMP2}) gives $0.72652(33)$, putting the
exact value just within our error bars. As noted in Section
\ref{subsec:centralcharge}, we have checked (in $d=3$) that for $\Delta_\Phi
\lesssim 2/3$, the value of $C_T$ extracted from the OPE coefficients of the
solution maximizing $\Delta_{[\Phi\bar{\Phi}]}$ does, indeed, correspond to
what one would get using $C_T$-minimization in the sense of
\cite{El-Showk2014a} (i.e. it is the {\em minimal} value of $C_T$, as a
function of $\Delta_\Phi$, consistent with unitarity and crossing symmetry
under the very mild additional assumption of not having additional scalars of
very low dimension).  Since the exact value of $C_T$ is close to saturating
this lower bound (which will only increase as we increase $n_{max}$) one could
conceivably turn this into a {\em proof} that the theory under consideration is
necessarily the cWZ model.

\begin{center}
\begin{figure}[h!]
\begin{center}
\includegraphics[scale=0.8]{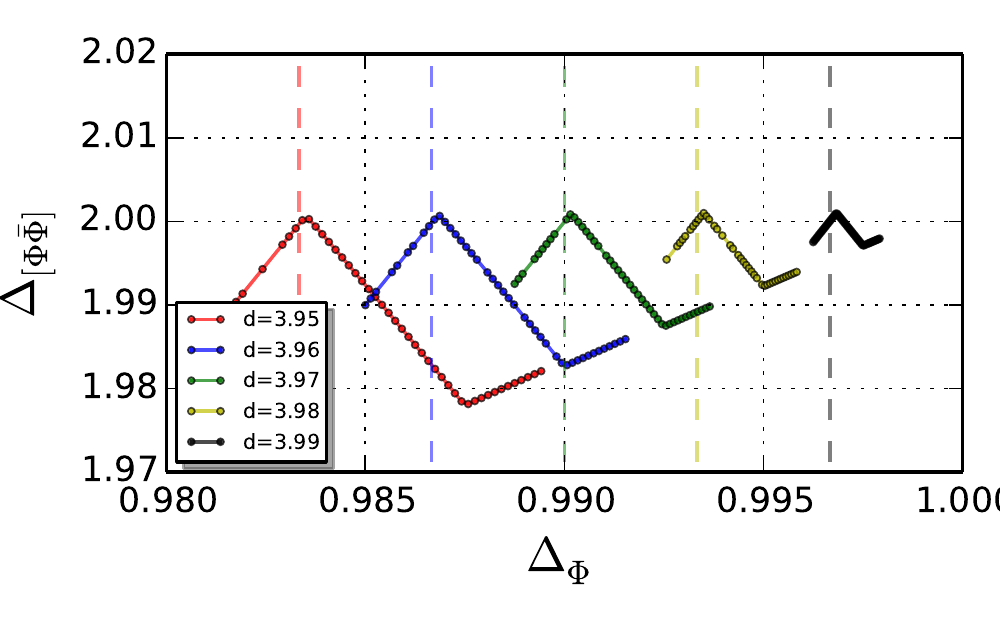}
\includegraphics[scale=0.8]{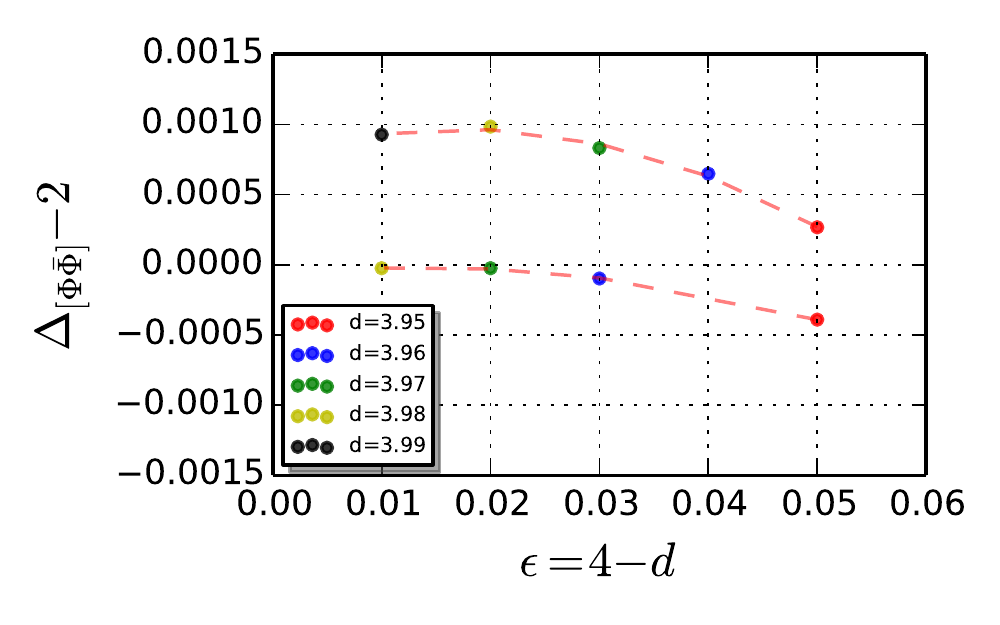}
\caption{Bound plots for $[\Phi \bar{\Phi}]$ near $d=4$ {\em (left)}.  On
closer inspection, the kink in the bound plot is slightly to the right of
$\Delta_\Phi = \frac{d-1}{3}$ (shown as dashed vertical lines).  As explained
in the text, we read off and plot  {\em(right)} $\Delta_{[\Phi \bar{\Phi}]}$ at
the bound both at the local maximum {\em (top curve)} and the value
$\Delta_\Phi=\frac{d-1}{3}$ {\em (bottom curve)}.  }
\label{fig:IsingEpsNear4}
\end{center}
\end{figure} 
\end{center}
Near $d=4$, we expect that the $\epsilon$-expansion should yield good numerical estimates so, as an additional test of our results, we would like to check the vanishing of the $\mathcal O(\epsilon)$ term in \eqref{eq:dimeps} by studying our bounds for small $\epsilon$.
In Figure \ref{fig:IsingEpsNear4}, we show the $\Delta_{[\Phi\bar{\Phi}]}$
bounds, now computed for $d=3.95-3.99$ in steps of $0.01$.  We expect that the
low-order $\epsilon$-expansion should yield reasonable results for these small
values of $\epsilon \sim 0.01-0.05$.  The first thing to note about the bounds
is that we see (at this resolution) that the kink does not exactly coincide
with $\Delta_\Phi=\frac{d-1}{3}$ but rather is very slightly to the right of
that value.  Although we know that the cWZ theory has an operator exactly at
$\Delta_\Phi=\frac{d-1}{3}$, we also know our bounds are not optimal (as we are
using a relatively small number of Taylor coefficients corresponding to
$n_{max}=6$), and the bound curve will move down as we increase the number of
constraints.  In fact, in \cite{BEMP2} we show that, for $d=3$, the minimum of the
$C_T$ curve does indeed correspond much more closely to
$\Delta_\Phi=\frac{d-1}{3}$, and that as we add more derivatives, the
kink in the $\Delta_{[\Phi\bar{\Phi}]}$ bound moves left towards
$\Delta_\Phi=\frac{d-1}{3}$ and towards the minimum of the $C_T$ plots.

We will nonetheless be conservative here and estimate the value of
$\Delta_{[\Phi\bar{\Phi}]}$ using two different procedures and show that our
results are relatively robust. In the first
approach, we simply extract the value of the bound at
$\Delta_\Phi=\frac{d-1}{3}$.  The second approach is to read off the value of
$\Delta_{[\Phi\bar{\Phi}]}$ at the local maximum in the left plot of Figure
\ref{fig:IsingEpsNear4}.  In both cases, we find a quadratic fit for
$\Delta_{[\Phi\bar{\Phi}]} - 2$ as a function of $\epsilon$ and read off the
subleading terms in equation (\ref{eq:dimeps}).  The two fits are shown in the
right plot in  Figure \ref{fig:IsingEpsNear4} with the lower curve corresponding to
the values at $\Delta_\Phi=\frac{d-1}{3}$.

The results for the two fits are:
\begin{align}
	\Delta_{[\Phi\bar{\Phi}]} -2 &= -0.283\, \epsilon^2 + 7.76 \times 10^{-3} \epsilon + 7.17 \times 10^{-5} \;, 
	&\qquad \Delta_{[\Phi\bar{\Phi}]} \textrm{ at }\Delta_\Phi=\frac{d-1}{3}\;, \label{eq:fitepsmax}\\
	\Delta_{[\Phi\bar{\Phi}]} -2 &= -0.648\,\epsilon^2 +  22.3 \times 10^{-3}\epsilon +   77.4\times 10^{-5}\;,
	&\qquad \Delta_{[\Phi\bar{\Phi}]} \textrm{ at local max}\;,\label{eq:fitepsIsing}
\end{align} 
It is clear that the quadratic coefficient depends on how we choose to extract
$\Delta_{[\Phi\bar{\Phi}]}$, meaning that our bounds have not converged sufficiently.
What does seem rather robust however, is that the constant and linear pieces
are orders of magnitude smaller than the quadratic piece, consistent with the
$\epsilon$-expansion prediction in equation (\ref{eq:dimeps}).

%%%%%%%%%%%%%%%%%%%%%%%%%%%
\subsection{Additional kinks}
\label{sec:kinks}
%%%%%%%%%%%%%%%%%%%%%%%%%%%

In every dimension in the range $2 \leq d \leq 4$, we clearly observe a kink at
$\Delta_\Phi=\frac{d-1}{3}$ which, as explained above, very likely corresponds
to the cWZ model.  For $3 \leq d \leq 4$, there is also a very clear kink at $\Delta_\phi=\frac{d}{4}$, but it stops being sharp below $d=3$. Moreover, for $2 \leq d \leq 4$, there is yet one more kink at some $\Delta_\Phi >
\frac{d}{4}$ that is an extension of the $d=4$ kink first observed in
\cite{Poland:2011ey}.  In this section, we initiate a very brief exploration of
these two structures.  We will
refer to them as the second and third kink even though the former may not exist
for $d < 3$, rendering the name ``third kink'' somewhat incorrect in those
dimensions.  Thus by ``third kink'', we will always mean the feature located at
$\Delta_\Phi > \frac{d}{4}$.
\begin{center}
\begin{figure}[ht!]
\includegraphics[scale=0.8]{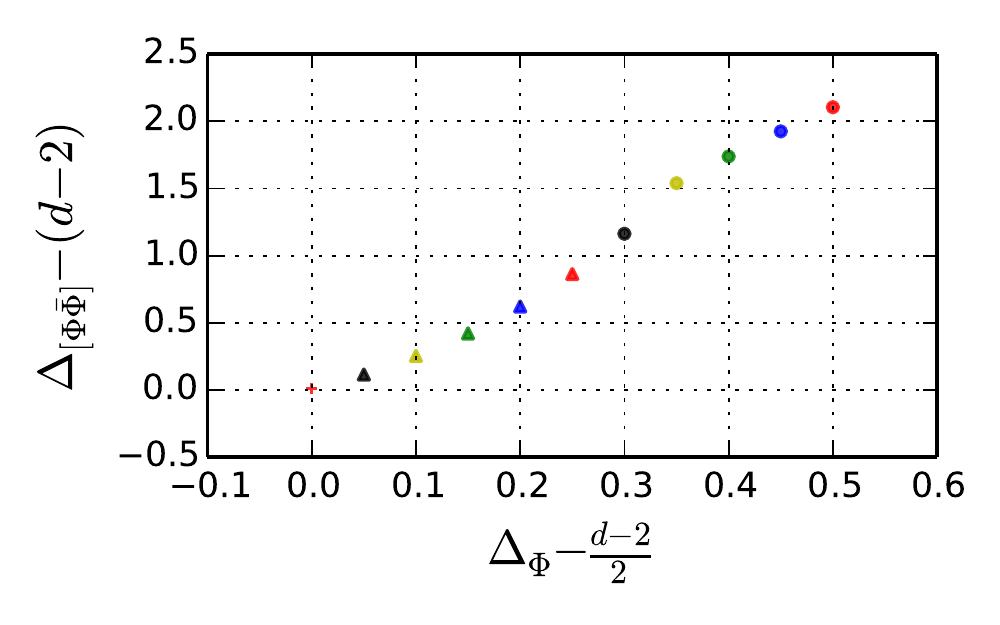}
\includegraphics[scale=0.8]{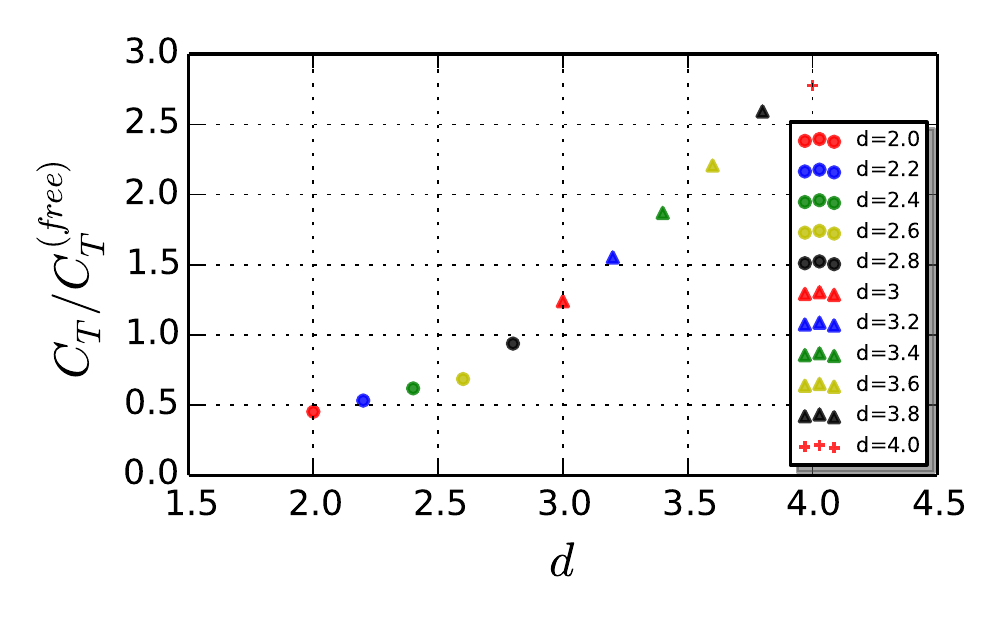}
\caption{Second kink: the anomalous dimension of $\Delta_{[\Phi \bar{\Phi}]}$ vs that of $\Delta_\Phi$ at $\Delta_\Phi=\frac{d}{4}$ for $2\leq d\leq 4$ {\em(left)}. The central charge, normalized by that of a free chiral superfield, at $\Delta_\Phi=\frac{d}{4}$ {\em(right)}.}
\label{fig:kink2}
\end{figure} 
\end{center}
In Figure \ref{fig:kink2}, we plot the dimension bound for
$\Delta_{[\Phi\bar{\Phi}]}$ and the central charge extracted from Figure
\ref{fig:epsbound} at $\Delta_\Phi=\frac{d}{4}$.   This kink is distinct from
the first and third kink, and from various other crossing symmetry kinks that
have appeared in the literature \cite{El-Showk2014, El-Showk2014a}, in two
important ways. First, as is clear from Figure \ref{fig:CT}, it corresponds to a local {\em maximum} of the central
charge rather than a minimum.  This statement is not
entirely accurate as Figure \ref{fig:CT} is not a central charge bound plot, in
the sense of \cite{El-Showk2014a}, but rather the central charge corresponding
to the saturating solution, which, {\em a priori}, may not minimize the central
charge.  Second, this kink occurs at a kinematically special point in terms of
the constraints imposed by supersymmetry. Precisely at $\Delta_\Phi=\frac{d}{4}$, the two additional scalar operators
allowed in the R-charged channel at dimensions  $\Delta=d - 2 \Delta_\Phi$ and
$\Delta=2\Delta_\Phi$ have equal dimensions, see \eqref{eq:opcontent}. For this reason,
one might suspect that the second kink is a kinematical feature of the
boundary solution that may not correspond to any physically interesting theory.
The fact that this structure does not continue below $d=3$, whereas the
coincidence of the two operator dimensions persists, might, however, suggest otherwise.
Motivated by this possibility, in Section \ref{sec:superhints}, we discuss some initial attempts to guess a physical theory corresponding to the second kink, and provide some guidance for others who would try their hand at this task.
\begin{center}
\begin{figure}[ht!]
\includegraphics[scale=0.8]{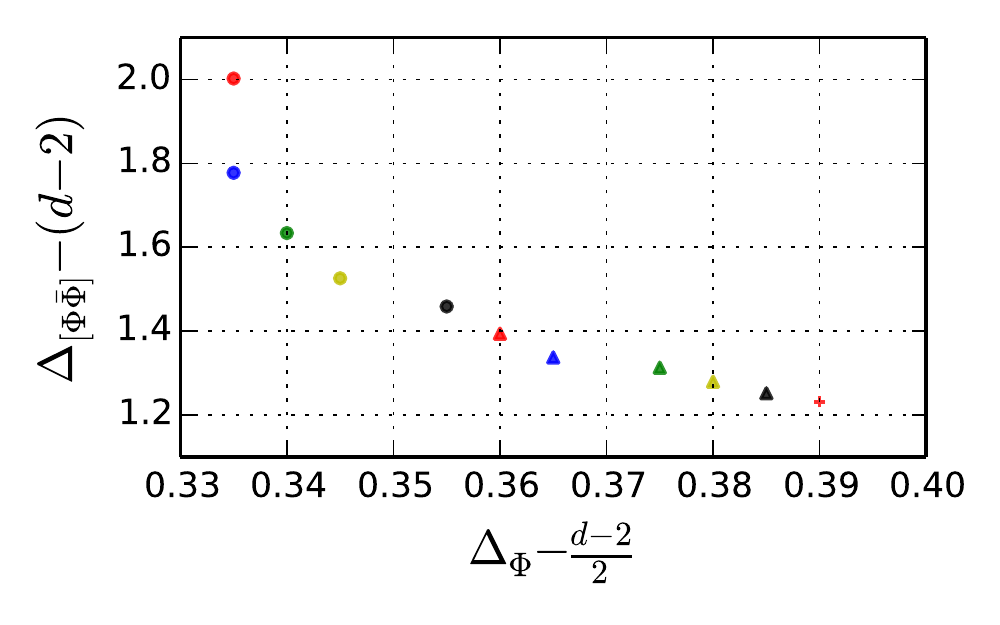}
\includegraphics[scale=0.8]{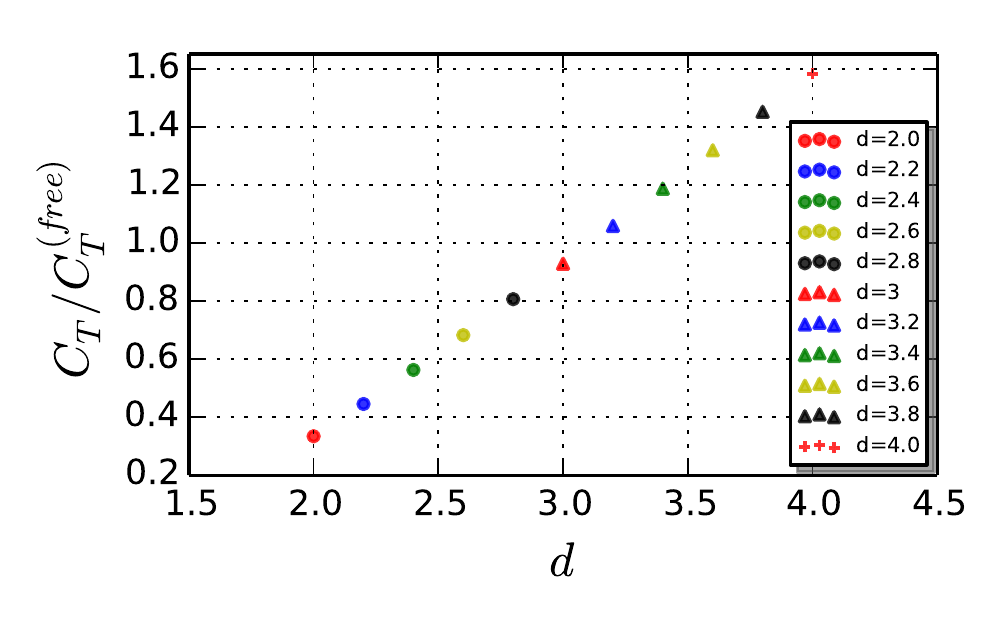}
\caption{Third kink: the anomalous dimension of $\Delta_{[\Phi \bar{\Phi}]}$ vs that of $\Delta_\Phi$ for the third kink {\em(left)}; the central charge, normalized by that of a free chiral superfield, for the third kink {\em(right)}.}
\label{fig:kink3_epssig_CT}
\end{figure} 
\end{center}
The third kink is much more ``traditional'', since it locally
minimizes $C_T$ and also appears at values of $\Delta_\Phi$ which do not enjoy any known significance.  As mentioned before, these kinks seem
to be a continuation of the one first observed at $d=4$ in
\cite{Poland:2011ey}. The third kink merges with the first in $d=2$, and thus becomes the $\mathcal{N}=2$ minimal model with $k=1$.
\begin{center}
\begin{figure}[ht!]
\begin{center}
\includegraphics[scale=0.8]{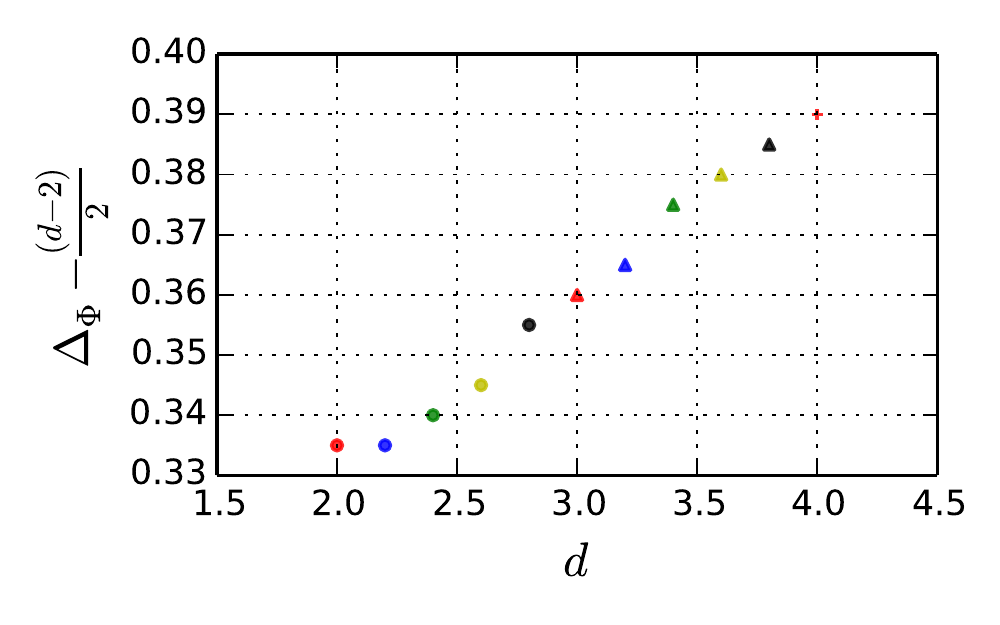}
\includegraphics[scale=0.8]{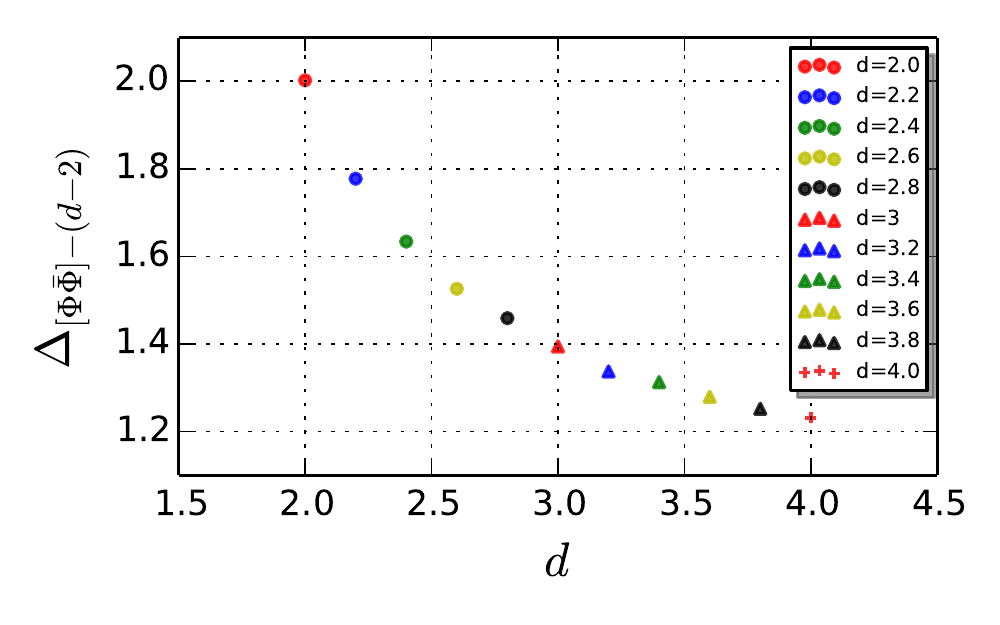}
\caption{Third kink: the anomalous dimension of $\Delta_\Phi$ as a function of $d$ for the third kink {\em(left)};
the anomalous dimension of $\Delta_{[\Phi \bar{\Phi}]}$ as a function of $d$ for the third kink {\em(right)}.}
\label{fig:kink3_sig_eps}
\end{center}
\end{figure} 
\end{center}
In Figure \ref{fig:kink3_epssig_CT}, we display the anomalous dimension of
$\Delta_{[\Phi\bar{\Phi}]}$ as a function of the anomalous dimension of $\Delta_\Phi$, as well as the ratio $C_T/C_T^{\text{free}}$ as a function of $d$, for the third
kink.  We determine the location of the kink by choosing the minimum of $C_T$
(or equivalently the location of the kink in the $\Delta_{[\Phi\bar{\Phi}]}$
bound) up to the resolution of Figure \ref{fig:CT}, which is\footnote{As we have
not conducted any systematic convergence estimate for our bounds, we do not make
any claim that this resolution bounds the error in any way.} $\sim 0.005$.  For
$d=2$, we do not see any distinct kink and since already at $d=2.2$, the location of the kinks seems to be merging, we assume that for $d=2$ the first and third kink coincide. To exhibit  the structure of the third kink in more detail we also provide, in Figure \ref{fig:kink3_sig_eps}, plots of the anomalous dimensions of $\Delta_\Phi$ and $\Delta_{[\Phi\bar{\Phi}]}$ at the kink as a function of $d$ .  

%%%%%%%%%%%%%%%
\subsubsection{Some speculations}
\label{sec:superhints}
%%%%%%%%%%%%%%%

In this section, we would like to offer some speculations about the nature of the second and third kinks.

\begin{center}
\begin{figure}[ht!]
\begin{center}
\includegraphics[scale=0.72]{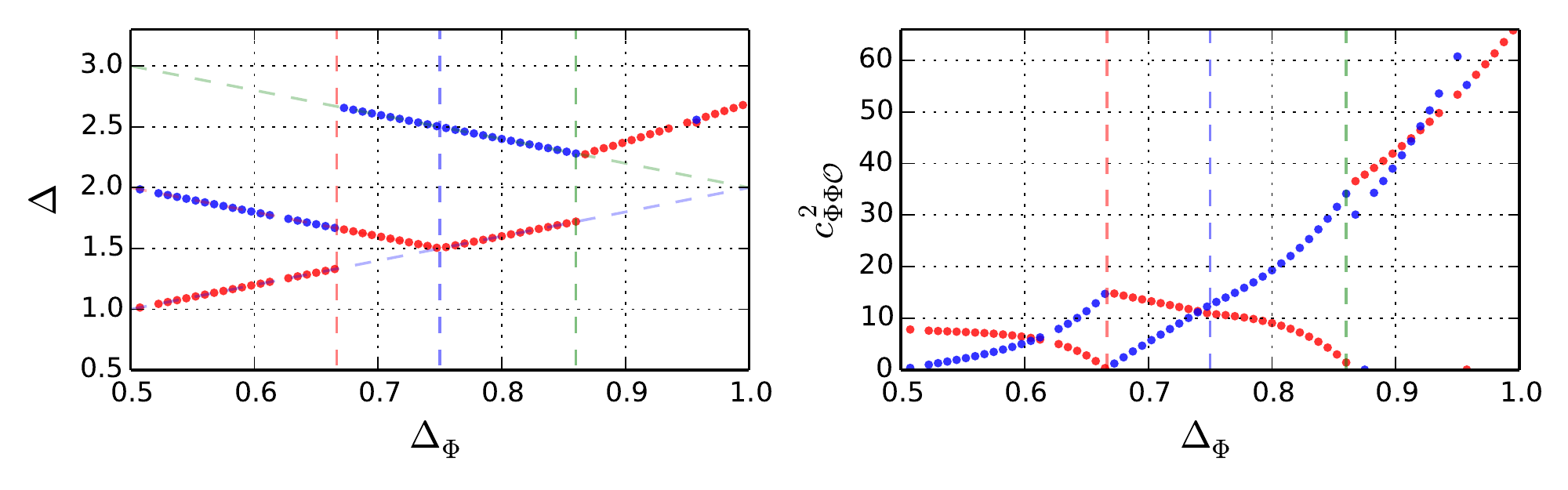}
\caption{Scaling dimensions {\em (left)} and OPE coefficients {\em (right)} for the first three scalar operators in the the $\Phi \times \Phi$ OPE, extracted from the saturating solution, for $d=3$ with $n_{max}=9$.  Operators appear in both plots with the same color (chosen according to ordering in the scaling dimension plot).  Note the decoupling of $\Phi^2$ at $\Delta_{\Phi} = 2/3$, corresponding to the cWZ model, as well as at the location of the third kink at $\Delta_\Phi\sim 0.86$.}
\label{fig:spin0OPE}
\end{center}
\end{figure} 
\end{center}

The second kink is kinematically special, since here the two candidate scalar conformal primaries in the $\Phi\times \Phi$ OPE ($\Phi^2$, with dimension $2\Delta_{\Phi}$, and $Q^2 \bar \Psi$, with dimension $d-2\Delta_{\Phi}$) have equal dimensions, see \eqref{eq:opcontent}. Between $\Delta_\Phi=(d-1)/3$ and this point, the bound on $[\Phi \bar \Phi]$ is linearly decreasing, and an analysis of the $\Phi\times \Phi$ OPE coefficients shows that all along this line, the $\Phi^2$ operator is not present, see Figure \ref{fig:spin0OPE}. At $\Delta_\Phi=d/4$, the chiral scalar field $\Psi$ becomes a free field, and so it should decouple from the spectrum. At this precise point, the $\Phi^2$ operator reappears, and it is this transition that marks the appearance of the second kink. It is interesting to note that the kink persists all the way to $d=4$, where it seems to lead to a very abrupt change in the central charge. Although our numerics present problems close to the free theory point in $d=4$, it seems then that the second kink describes free theory with more than one chiral superfields. A natural guess is three chiral superfields, since this gives\footnote{\label{free}
At precisely the free point, there are extra spin-1 and spin-2 currents which mix with the stress-tensor. Our numerics cannot disentangle these, hence the discontinuous jump to the single-field value of $C_T$ at the free point.} $C_T=20$, which seems to be very close to the asymptotic value in Figure \ref{fig:CT}.

%%%%%%%%%

We are then led to guess that the second kink describes a theory with three chiral superfields, $X,Y,Z$. Furthermore, we expect a superpotential term $X^2 Y$, which implies that if $Y$ becomes free at the fixed point, then one has $\Delta_X=d/4$ as required. In $d=3$, we can use F-maximization to find the scaling dimensions of the chiral fields \cite{Jafferis:2010un}. We have found two superpotentials which seem to have the right properties, namely $W=X^2 Y+X Z$, and $W=X^2 Y+Y^2 Z^2$. In the first case, the fixed-point conformal dimensions (which are equal to the R-charges in $d=3$), as fixed by $F$-maximization, come out to be $\Delta_X=3/4$, $\Delta_Y=1/2$, and $\Delta_Z=5/4$. In the second case, one finds that the dimension of $Z$ at the fixed point is naively below the unitarity bound. This signals the emergence of accidental flavor symmetries which mix with the $R$-symmetry, modifying the $F$-maximization procedure. This accidental flavor symmetry is accounted for by noting that the field $Z$ becomes free and thus $\Delta_Z=1/2$. This then leads to $\Delta_Y=1/2$ and $\Delta_X=3/4$.

To distinguish between the two guesses above, we need another observable. A convenient choice is the central charge $C_T$. As mentioned above, by computing the partition function on a squashed sphere, it is possible to determine $\tau_{RR}/\tau_{RR}^{\mbox{\tiny free}}=C_T/C_T^{\mbox{\tiny free}}$, with $\tau_{RR}$ the R-current two-point function coefficient. For the two superpotentials above we find 
\bea
\frac{\tau_{\mbox{\tiny RR}}(3/4)+\tau_{\mbox{\tiny RR}}(1/2)+\tau_{\mbox{\tiny RR}}(5/4)}{\tau_{\mbox{\tiny RR}}(1/2)}=1\;, &\qquad& W=X^2 Y+X Z\;,\label{eq:option1}\\
\frac{\tau_{\mbox{\tiny RR}}(3/4)+2\tau_{\mbox{\tiny RR}}(1/2)}{\tau_{\mbox{\tiny RR}}(1/2)}\simeq 2.5603\;, &\qquad& W=X^2 Y+Y^2 Z^2\label{eq:option2}\;.
\eea
This should be compared with the ratio $C_T/C_T^{\mbox{\tiny free}}\simeq 1.24$ that we obtain from Figure \ref{fig:kink2} (to avoid any confusion in the formulas above, we are normalizing by dividing by the values for a single free chiral field). Indeed, it appears we would need $\Delta_{Z}\simeq 1.14$, which seems hard to obtain from a polynomial superpotential with three chiral superfields.

To finish the discussion on the second kink, we should mention the intriguing possibility that the corresponding theory is in fact non-unitary. Violations of unitarity are not necessarily excluded by our bootstrap methods, as long as squares of OPE coefficients remain. This exotic suggestion is motivated by the observation that
\bea
\frac{4\tau_{\mbox{\tiny RR}}(3/4)-\tau_{\mbox{\tiny RR}}(1/2)}{\tau_{\mbox{\tiny RR}}(1/2)}\simeq 1.2413 \;.
\label{eq:nonunitary}
\eea
This suggest that the field theory actually contains five chiral superfields, but one of them has the wrong sign kinetic term, so that in terms of $C_T$, they effectively appear as three chiral superfields. Considering the superpotential $W=(X^2+Z^2+W^2+V^2)Y$, $F$-maximization leads to $\Delta_Y<1/2$, which is below the unitarity bound, and signals that the field $Y$ is actually free, i.e. $\Delta_{Y}=1/2$. After taking this into account, we find $\Delta_X=\Delta_Z=\Delta_W=\Delta_V=3/4$. Hence, it appears that this theory has all the right properties to match our second kink.

The attentive reader may have noticed a small sleight of hand here. When a chiral field becomes free, it decouples from the rest of the theory and hence stops contributing to the OPE coefficient of the conserved spin-2 current. Therefore, $C_T$ derived from numerical bootstrap measures the two-point function of the stress-tensor of the interacting part of the CFT only, and we should leave out the free contributions in \eqref{eq:option1}, \eqref{eq:option2}, \eqref{eq:nonunitary}. However, we expect that the extra field is free only precisely at the kink and not in its immediate neigbourhood, and thus we should include its contribution by continuity. We would then also expect that another spin-one superconformal primary approaches the unitarity bound as we approach the kink, providing the extra $U(1)$ symmetry of the free chiral. Unfortunately, preliminary numerical studies suggest that this is not so.

Let us focus now on the line of theories for $(d-1)/3< \Delta_\Phi<d/4$. The decoupling of $\Phi^2$ suggests a chiral ring relation $\Phi^2=0$. One particular such theory is the Wess-Zumino model with two chiral superfields $\Upsilon, \Lambda$ and a cubic superpotential of the form $W=\lambda \Upsilon^2 \Lambda$. Denoting the lowest components of the superfields by $\Phi$ and $\Psi$ respectively, this model yields the correct OPE $\Phi \times \Phi=Q^2 \bar \Psi$, i.e. the operator $\Phi^2$ is absent. In addition, we have the relation $\Delta_\Psi=(d-1)-2\Delta_\Phi$, which follows from chirality and R-charge conservation. The exact dimensions in this model can be determined in $d=3$ by $F$-maximization, as shown in \cite{Nishioka:2013gza}, giving $\Delta_\Phi\simeq 0.708$. Could it be that our bound is saturated by this theory? Unfortunately this is not so. In the same reference, the authors compute $\tau_{\mbox{\tiny RR}}\simeq 0.380 $, whence it follows that
\bea
\frac{C_T}{C_T^{\mbox{\tiny free}}}=\frac{\tau_{\mbox{\tiny RR}}}{\tau_{\mbox{\tiny RR}}(1/2)}\approx 1.52\,.
\label{CTUpsLam}
\eea
On the other hand, from Figure \ref{fig:CT_zoom}, we read off that at $\Delta_\Phi\simeq 0.708$, $C_T\simeq 6$, and hence $C_T/C_T^{\mbox{\tiny free}}\simeq 1$, very different from what we obtain above.

Consider now the third kink, which was first observed in $d=4$ \cite{Poland:2011ey}. Our analysis adds a few more pieces of information about a putative theory sitting there. First, the kink continues to exist all the way to $d=2$, where it apparently merges with the $C_T=2c=2$, $\mathcal N=2$ minimal model. Second, the chiral field $\Phi^2$ disappears from the spectrum also at this kink, as witnessed by Figure \ref{fig:spin0OPE}. In $d=2$, this corresponds to the non-existence of a dimension-2/3 Virasoro primary in the $c=1$ model. This is a strong hint that the chiral ring of the theory at the kink has a relation $\Phi^2=0$. Since the kink does not merge with the free theory in $d=4$, we do not expect it can be described by a Lagrangian for a collection of chiral superfields. It is concievable it arises as an IR fixed point of a non-abelian gauge theory in $d=4$, or even an abelian gauge theory in $d=3$. Note that the central charge $C_T$ in $d=4$ is rather low -- about 1.6 times that of the free chiral multiplet and only about a half of a single free vector multiplet.

%%%%%%%%%%%%%%%%%%%%%%%%%%%%%%%%%%%%%
\section{Discussion}\label{sec:discussion}
%%%%%%%%%%%%%%%%%%%%%%%%%%%%%%%%%%%%%

In this paper, we have investigated the constraints of the conformal bootstrap on superconformal field theories with four Poincar\'e supercharges in $d\leq 4$. The cases $d=2$ and $d=3$ have not been analyzed before and thus we provide new universal bounds on unitary SCFTs with $\mathcal{N} = 2$ supersymmetry in these dimensions. We have also shown that the bounds display three interesting features (kinks), one of which we have conjecturally identified as the infrared fixed point of the single-field Wess-Zumino model with cubic superpotential. This conjecture is supported by the matching of the protected dimension of the chiral field, comparison of the value of $C_T$ with an exact calculation by supersymmetric localization in $d=3$, the structure of the OPE in the chiral sector,  $\epsilon$-expansion computations, and the agreement with exact results in $d=2$. In \cite{BEMP2}, we take this conjecture at face value to provide a detailed study of the theory for $d=3$.

It is clearly of great interest to elucidate the remaining two kinks. We expect that at least the third kink corresponds to a physical theory, since it shares many features with the better-understood Ising-like kinks. Perhaps a good candidate theory can be found with the correct value of $C_T$, and a gauge-invariant chiral operator $\Phi$ with the right dimension and chiral ring relation $\Phi^2 = 0$. It could also be interesting to see if $C_T$ can be derived using localization in continuous $d$, in the spirit of \cite{Giombi:2014xxa}, and matched with our results for the cWZ model or used as a tool to probe the other kinks.

The crucial ingredient in this work was to formulate a dimension-independent approach to superconformal algebras with four Poincar\'e supercharges in $d\leq 4$. This allowed us, among other things, to write down the action of the superconformal Casimir on a four-point function as a differential equation, whose solutions in turn gave us the superconformal blocks relevant for the bootstrap analysis. This approach can be extended to superconformal theories with eight Poincar\'e supercharges in general dimension, the parent algebra being the $(1,0)$ superconformal algebra in six dimensions. Theories with this amount of supersymmetry are particularly suited for bootstrap analysis since, apart from the case $d=2,4$, they do not admit marginal deformations. Work on this is currently in progress.

It is intriguing that the supersymmetric conformal blocks can be recast as non-supersymmetric ones with shifted external dimensions. In this paper, we have extended this observation, previously noted in \cite{Fitzpatrick2014} for $d=2,4$, to any dimension and more general external operators. In the same reference, the authors showed that certain $\mathcal N=2$ superblocks in four dimensions are given by a similar expression, this time with a shift by two units. It would be interesting to see if there is any deep reason for this connection and if the latter result also extends to other spacetime dimensions.

We have only briefly touched upon the extension of our analysis to $d<2$. While the superconformal blocks we derived should be valid in any $d\leq 4$, it is not clear whether one can use the numerical bootstrap techniques to extract interesting information in $d=1$ \cite{Golden:2014oqa}. This certainly deserves further study since superconformal quantum-mechanical models are ubiquitous and should be dual to the $AdS_2$ near horizon regions of some extremal black holes.

Another interesting avenue for future exploration is to combine the constraints from superconformal symmetry studied here with the simplifications that occur in large $N$ CFTs, i.e. when correlation functions factorize. This was explored to some extent with $\mathcal{N}=4$ supersymmetry in $d=4$ in \cite{Alday2014} but much remains to be understood. The interest in this problem stems in part from the AdS/CFT correspondence and the fact that string theory leads to a vast landscape of holographic duals to SCFTs with four supercharges.
\bigskip
\bigskip

%%%%%%%%%%%%%%%%%%%%%%%%%%%%%%%%%%%%%
\noindent \textbf{ Acknowledgements }
%%%%%%%%%%%%%%%%%%%%%%%%%%%%%%%%%%%%%
\bigskip

\noindent We would like to thank Chris Beem, Davide Gaiotto, Simone Giombi, Igor Klebanov, Marco Meineri, Toine Van Proeyen, Silviu Pufu, Slava Rychkov, Balt van Rees, and Alessandro Vichi for useful discussions. The work of NB is supported in part by the starting grant BOF/STG/14/032 from KU Leuven, by the COST Action MP1210 The String Theory Universe, and by the European Science Foundation Holograv Network. The research of DM was supported by Perimeter Institute for Theoretical Physics. Research at Perimeter Institute is supported by the Government of Canada through Industry Canada and by the Province of Ontario through the Ministry of Research and Innovation. During this work MFP was supported by a Marie Curie Intra-European Fellowship of the European
Community's 7th Framework Programme under contract number PIEF-GA-2013-623606 and by US DOE-grant DE-SC0010010. Numerical calculations were performed on the CERN cluster and Brown University's CCV cluster. NB would like to thank the Perimeter Institute for Theoretical Physics and the Theory Division at CERN for hospitality during the course of this work. DM is grateful to the ITF at KU Leuven and the Theory Division at CERN for hospitality.

%%%%%%%%%%%%%%%%%%%%%%%%%%%%%%%%%%%%%
\begin{appendices}
%%%%%%%%%%%%%%%%%%%%%%%%%%%%%%%%%%%%%

%%%%%%%%%%%%%%%%%%%%%%%%%%%%%%%%%%%%%
\appendix

%%%%%%%%%%%%%%%%%%%%%%%%%%%%%%%%%%%%%
\section{OPE derivation of 3d $\mathcal{N}=2$ superconformal blocks}
\label{App:opeder}
%%%%%%%%%%%%%%%%%%%%%%%%%%%%%%%%%%%%%

In this Appendix, we provide further evidence for the formulae \eqref{eq:sbdecomposition}, \eqref{eq:sbcoefficients} by explicitly determining the coefficients $a_{i}$ from constraints imposed by $d=3$, $\mathcal{N}=2$ superconformal invariance on the OPE.\footnote{It is quite possible that these results can be derived also using techniques from superspace similar to the ones in \cite{Park:1999cw}.} As a first order of business let us present the explicit realization of the $d=3$, $\mathcal{N}=2$ algebra. The bosonic generators are just the conformal generators and the R-charge $\{R,M_{ij},D,P_i,K_i\}$. They satisfy the commutation relations already presented in \eqref{bosonicCA}. A realization of these commutation relations in terms of differential operators is given by
\begin{equation}
\begin{split}
M_{jk} &= -\im(x_j\partial_{k}-x_k\partial_j)\;, \\
K_j &= -\im (x_kx^k \partial_j -2 x_jx^k\partial_k)\;,\\
P_j &= -\im\partial_j\;, \qquad D = \im x^k\partial_{k}\;, \qquad R=r\;.
\end{split}
\label{CAdiff}
\end{equation}
Note that the action of the conformal generators on operators in the CFT picks up a minus sign relative to \eqref{CAdiff}. See the discussion around equations (2.28)-(2.31) in \cite{Minwalla:1997ka}.

The fermionic generators are $Q_{\alpha}^{\pm}$, $S^{\alpha \pm}$, where $\alpha=1,2$ is the Dirac index. The Dirac representation is self-dual, with the isomorphism with the dual representation provided by the antisymmetric tensor $\epsilon^{12}=-\epsilon^{21}=\epsilon_{21}=-\epsilon_{12}=1$. Thus in $d=3$ there is no real distinction between the $\alpha$ and $\dot{\alpha}$ index used in Section \ref{sec:superalgebras} and we will omit the dots in this Appendix. Hermitian conjugation acts as $(Q_{\alpha}^{\pm})^{\dagger}=S^{\alpha\mp}$. Let $(\sigma_i)^{\alpha}_{\phantom{\alpha}\beta}$ be the usual Pauli matrices
\begin{equation}
\sigma_1\equiv 
\begin{pmatrix}
 0 & 1\\ 1 & 0 
 \end{pmatrix}\,,\qquad
\sigma_2\equiv
\begin{pmatrix}
 0 & -\im\\ \im & 0 
 \end{pmatrix}\,,\qquad
\sigma_3\equiv
\begin{pmatrix}
 1 & 0\\ 0 & -1 
 \end{pmatrix}\,,
\end{equation}
and further define
\begin{equation}
\begin{aligned}
 (\sigma_i)_{\alpha\beta} = \epsilon_{\alpha\gamma}(\sigma_i)^{\gamma}_{\phantom{\alpha}\beta}\,,\quad
 (\sigma_i)^{\alpha\beta} = (\sigma_i)^{\alpha}_{\phantom{\alpha}\gamma}\epsilon^{\gamma\beta}\,,\quad
 (\sigma_i)_{\alpha}^{\phantom{\alpha}\beta} = \epsilon_{\alpha\gamma}(\sigma_i)^{\gamma}_{\phantom{\alpha}\delta}\epsilon^{\delta\beta}\,.
\end{aligned}
\end{equation}
The action of the bosonic generators on the fermionic ones is then
\begin{equation}
\begin{aligned}
 \lbrack R,Q_{\alpha}^{\pm}] &= \pm Q_{\alpha}^{\pm}\,, \qquad\qquad
 [R,S^{\alpha\pm}] = \pm S^{\alpha\pm}\,,\\
 [M_{ij},Q_{\alpha}^{\pm}] &= \frac{1}{2}\varepsilon_{ijk}(\sigma_k)^{\beta}_{\phantom{\alpha}\alpha}Q_{\beta}^{\pm}\,, \qquad
 [M_{ij},S^{\alpha\pm}] = \frac{1}{2}\varepsilon_{ijk}(\sigma_k)_{\beta}^{\phantom{\alpha}\alpha}S^{\beta\pm}\,,\\
 [D,Q_{\alpha}^{\pm}] &= -\frac{\im}{2} Q_{\alpha}^{\pm}\,, \qquad\qquad
 [D,S^{\alpha\pm}] = \frac{\im}{2} S^{\alpha\pm}\,,\\
 [P_i,S^{\alpha\pm}] &= - (\sigma_i)^{\beta\alpha}Q_{\beta\pm}\,, \qquad\qquad
 [K_i,Q_{\alpha}^{\pm}] = (\sigma_i)_{\beta\alpha}S^{\beta\pm}\,,
\end{aligned}
\end{equation}
with all other commutators vanishing. Note that $\varepsilon_{ijk}$ is the completely antisymmetric tensor in three dimensions. Finally, the anticommutation relations among the fermionic generators are
\begin{equation}
\begin{aligned}
 \{Q_{\alpha}^{+},Q_{\beta}^{-}\} &= P_i(\sigma_i)_{\alpha\beta}\,, \qquad\qquad
 \{S^{\alpha+},S^{\beta-}\} = K_i(\sigma_i)^{\alpha\beta}\,,\\
 \{S^{\alpha-},Q_{\beta}^{+}\} &= (\im D-R)\delta^{\alpha}_{\phantom{\alpha}\beta}+\frac{1}{2}\varepsilon_{ijk}M_{ij}(\sigma_k)^{\alpha}_{\phantom{\alpha}\beta}\,,\\
 \{S^{\alpha+},Q_{\beta}^{-}\} &= (\im D+R)\delta^{\alpha}_{\phantom{\alpha}\beta}+\frac{1}{2}\varepsilon_{ijk}M_{ij}(\sigma_k)^{\alpha}_{\phantom{\alpha}\beta}\,,
\end{aligned}
\end{equation}
with all other anticommutators vanishing. This algebra is of course in harmony with the general presentation in Section \ref{sec:superalgebras} of the superconformal algebras in $d\leq 4$.

%%%%%%%%%%%%%%%%%%%%%
\subsection*{Generalities}
%%%%%%%%%%%%%%%%%%%%%

Let us first consider a CFT without supersymmetry and review how a conformal multiplet, with primary $\mathcal{P}_{i_1\ldots i_s}$ of dimension $\Delta$ in the symmetric traceless representation of spin $s$, contributes to the four-point function, $\langle\phi_1\phi_2\phi_3\phi_4\rangle$, of scalar primaries $\phi_i$ of dimensions $\Delta_i$. Define the OPE coefficient, $c_{\phi_1\phi_2}^{\mathcal{P}}$, by writing the contribution of the conformal family of $\mathcal{P}$ to the $\phi_1 \times \phi_2$ OPE as\footnote{In this appendix we freely use the operator-state correspondence, which is valid in any CFT in Euclidean signature. The state corresponding to an operator $\phi(x)$ will be denoted by $|\phi\rangle$.}
\begin{equation}
 \phi_1(x)|\phi_2\rangle =\ldots +  c_{\phi_1\phi_2}^{\mathcal{P}}|x|^{-\Delta_1-\Delta_2+\Delta-s}x^{i_1}\ldots x^{i_s}\left[|\mathcal{P}_{i_1\ldots i_s}\rangle
 +\textrm{desc.}\right] +\ldots\,.
\end{equation}
The contribution of level-one descendants in the square bracket is
\begin{equation}
 \textrm{desc.} = \alpha(x\cdot P)|\mathcal{P}_{i_1\ldots i_s}\rangle+
 \beta x_{i_1}P^{j}|\mathcal{P}_{ji_2\ldots i_s}\rangle + \ldots\,,
 \label{eq:l1descendants}
\end{equation}
where
\begin{equation}
 \begin{aligned}
  \alpha &= -\frac{\im}{2}\frac{\Delta + \Delta_{12} + s}{\Delta + s}\,,\\
  \beta &= -\frac{\im}{2}\frac{s\Delta_{12}}{(\Delta + s)(\Delta -s -1)}\,.
 \end{aligned}
\end{equation}
The two-point function of $\mathcal{P}$ and its conjugate $\bar{\mathcal{P}}$ takes the form
\begin{equation}
 \left\langle\mathcal{P}_{i_1\ldots i_s}(x) \bar{\mathcal{P}}_{j_1\ldots j_s}(y)\right\rangle = \frac{f_{\mathcal{P}\bar{\mathcal{P}}}}{|x-y|^{2\Delta}}\left[\frac{1}{s!}\sum\limits_{\sigma\in S_s}\prod\limits_{n=1}^{s}I_{i_nj_{\sigma(n)}}(x-y)-\textrm{ traces }\right]\,,
 \label{eq:OPE}
\end{equation}
where $S_s$ is the permutation group on $s$ elements and
\begin{equation}
 I_{ij}(x)=\delta_{ij} - 2\frac{x_{i}x_{j}}{|x|^2}\,.
\end{equation}
It is useful to note that the coefficient $f_{\mathcal{P}\bar{\mathcal{P}}}$ also appears in the scalar product
\begin{equation}
 \left\langle \mathcal{P}_{i_1\ldots i_s}|\mathcal{P}_{j_1\ldots j_s}\right\rangle = f_{\mathcal{P}\bar{\mathcal{P}}}
 \left(\frac{1}{s!}\sum\limits_{\sigma\in S_s}\prod\limits_{n=1}^{s}\delta_{i_nj_{\sigma(n)}}-\textrm{ traces }\right)\,.
\end{equation}
With these normalizations, the contribution of the conformal family to the four-point function is
\begin{equation}
 \left\langle\phi_1(x_1)\phi_2(x_2)\phi_3(x_3)\phi_4(x_4)\right\rangle\!|_{\mathcal{P}} = \frac{
  f_{\mathcal{P}\bar{\mathcal{P}}}\,
 c_{\phi_1\phi_2}^{\mathcal{P}} c_{\phi_3\phi_4}^{\bar{\mathcal{P}}}
 }{|x_{12}|^{\Delta_1 + \Delta_2}|x_{34}|^{\Delta_3+\Delta_4}}\frac{|x_{24}|^{\Delta_{12}}|x_{14}|^{\Delta_{34}}}{|x_{14}|^{\Delta_{12}}|x_{13}|^{\Delta_{34}}}G^{\Delta_{12},\Delta_{34}}_{\Delta,s}(u,v)\,,
\end{equation}
where $G^{\Delta_{12},\Delta_{34}}_{\Delta,s}(u,v)$ is the conformal block whose normalization is determined as in \eqref{eq:cbasymptotics}. The following derivation of the superconformal blocks relies on the observation that when $\phi_i$ are superconformal primaries and $\phi_{1,3}$ chiral primaries, superconformal symmetry fixes the OPE coefficients and two-point functions of all contributing conformal primaries from the same superconformal multiplet in terms of those of the superconformal primary.

Consider now the correlator $\langle\phi_1\phi_2\phi_3\phi_4\rangle$ in a 3d, $\mathcal{N}=2$ SCFT, where $\phi_i$ are scalar superconformal primaries, with $\phi_{1,3}$ chiral, i.e. $[Q_{\alpha}^+,\phi_{1,3}] = 0$. We wish to determine which conformal primaries in the superconformal family of a superconformal primary $\mathcal{P}$ can appear in both the OPE of $\phi_1 \times \phi_2$ and the OPE of $\bar\phi_3 \times \bar\phi_4$. Only those conformal primaries can contribute to the above four-point function. Consider the OPE $\phi_1(x)|\phi_2\rangle$. It follows from the chirality of $\phi_1$ and the superconformal algebra that $[S^{\alpha+},\phi_1(x)] = 0$. Hence
\begin{equation}
 S^{\alpha+}\phi_1(x)|\phi_2\rangle = 0\,,
 \label{eq:scondition1}
\end{equation}
since $\phi_2$ is a superconformal primary. Similarly,
\begin{equation}
 S^{\alpha-}\bar\phi_3(x)|\bar\phi_4\rangle = 0\,.
 \label{eq:scondition2}
\end{equation}
Consequently, the conformal primary with the lowest dimension from a given superconformal family that contributes to both OPEs must be annihilated by $S^{\alpha\pm}$, and thus this operator is necessarily the superconformal primary. It follows that the superconformal primary has integer spin and its R-charge is given by $q=q_1 + q_2 = -q_3 - q_4$. Let us denote this operator with $\mathcal{P}^{(0)}_{i_1\ldots i_s}$, and its dimension and spin with $\Delta$ and $s$, respectively. All other contributing conformal primaries from the same supermultiplet must have integer spin and R-charge $q$. The conformal primaries in the multiplet have dimensions $\Delta + n/2$, with $n=0,\ldots,4$ labelling the number of $Q$ supercharges acting on $\mathcal{P}^{(0)}_{i_1\ldots i_s}$. These operators have integer spin only when $n$ is even. For $s>0$ and generic $\Delta - |q|$, there are four candidate conformal primaries with dimension $\Delta + 1$. $\mathcal{P}^{(1)}$ with spin $s+1$, $\mathcal{P}^{(2)}$ with spin $s-1$, and $\mathcal{P}^{+-}$, $\mathcal{P}^{-+}$, both with spin $s$. All four can be obtained by acting with linear combinations of the products $Q_{\alpha}^{\pm}Q_{\beta}^{\mp}$ on $\mathcal{P}^{(0)}$. Consider the action of spacetime parity $x^i\mapsto - x^i$. A proper tensor of spin $s$ transforms as $(-1)^s$, while a pseudotensor as $(-1)^{s+1}$. The supercharges $Q_{a}^{\pm}$ ``square to the momentum", and thus must transform such that any product transforms as $Q_{\alpha}^{\pm}Q_{\beta}^{\mp}\mapsto - Q_{\alpha}^{\pm}Q_{\beta}^{\mp}$. It follows that $\mathcal{P}^{(1)}$, $\mathcal{P}^{(2)}$ have the same parity as $\mathcal{P}^{(0)}$, while $\mathcal{P}^{+-}$, $\mathcal{P}^{-+}$ have the opposite. In theories invariant under parity, this gives an argument why only $\mathcal{P}^{(1)}$ and $\mathcal{P}^{(2)}$ can contribute. However, the Casimir approach from the main text does not require parity invariance and thus shows that our formula for superconformal blocks is valid in general. Finally, there is a conformal primary $\mathcal{P}^{(3)}$ with dimension $\Delta + 2$, spin $s$ and R-charge $q$, obtained from $\mathcal{P}^{(0)}$ by acting with a linear combination of products of four $Q$'s, which can also contribute to the four-point function. In the following subsections, we show how the constraints \eqref{eq:scondition1} and \eqref{eq:scondition2} fix the OPE coefficients of $\mathcal{P}^{(1)}$, $\mathcal{P}^{(2)}$, and $\mathcal{P}^{(3)}$ in terms of those of $\mathcal{P}^{(0)}$ and also compute the two-point functions $f_{\mathcal{P}^{(i)}\bar{\mathcal{P}}^{(i)}}$ for $i=1,2,3$ in terms of $f_{\mathcal{P}^{(0)}\bar{\mathcal{P}}^{(0)}}$. We find that
\begin{equation}
 a_i = \frac{f_{\mathcal{P}^{(i)}\bar{\mathcal{P}}^{(i)}}}{f_{\mathcal{P}^{(0)}\bar{\mathcal{P}}^{(0)}}} \frac{c_{\phi_1\phi_2}^{\mathcal{P}^{(i)}} c_{\phi_3\phi_4}^{\bar{\mathcal{P}}^{(i)}}} {c_{\phi_1\phi_2}^{\mathcal{P}^{(0)}} c_{\phi_3\phi_4}^{\bar{\mathcal{P}}^{(0)}}}\,,
\end{equation}
reproduce the results in \eqref{eq:sbcoefficients} for $d=3$.

%%%%%%%%%%%%%%%%%%%%%%%%%%%%
\subsection*{The contribution of $\mathcal{P}^{(1)}$}
%%%%%%%%%%%%%%%%%%%%%%%%%%%%

It is useful to define the operator
\begin{equation}
 T_{j}(x,y) \equiv (\sigma_{j})^{\alpha\beta}\left[(x-y)Q_{\alpha}^{+}Q_{\beta}^{-}-(x+y)Q_{\alpha}^{-}Q_{\beta}^{+}\right]\,.
\end{equation} 
An explicit expression for $\mathcal{P}^{(1)}$ is then given by
\begin{equation}
 |\mathcal{P}^{(1)}_{i_1\ldots i_{s+1}}\rangle =\frac{1}{s+1} \sum\limits_{n=1}^{s+1}T_{i_n}(\Delta+s,q)|\mathcal{P}^{(0)}_{i_1\ldots\hat{i}_{n}\ldots i_{s+1}}\rangle-\mathrm{traces}\,,
\end{equation}
where the notation $i_1\ldots\hat{i}_{n}\ldots i_{s+1}$ means that the index $i_{n}$ is omitted from the string of indices, and the traces are subtracted to make the resulting state traceless. Applying $S^+_{\alpha}$ to the OPE $\phi_1(x)|\phi_2\rangle$ mixes the contribution of level-one conformal descendants of $\mathcal{P}^{(0)}$, \eqref{eq:l1descendants}, with that of the conformal primary $\mathcal{P}^{(1)}$. Requiring that the result vanishes and looking at the coefficient of the highest power of $\bar z = x_1 - \im x_2$ leads to
\begin{equation}
 c_{\phi_1\phi_2}^{\mathcal{P}^{(1)}} =
 - \frac{\im(\Delta + \Delta_{12} + s)}{4(\Delta + s)(\Delta + s + 1)(\Delta + s + q)}c_{\phi_1\phi_2}^{\mathcal{P}^{(0)}}\,,
 \label{Cf1f2P1}
\end{equation}
and hence also
\begin{equation}
 c_{\phi_3\phi_4}^{\bar{\mathcal{P}}^{(1)}} =
 - \frac{\im(\Delta + \Delta_{34} + s)}{4(\Delta + s)(\Delta + s + 1)(\Delta + s - q)}c_{\phi_3\phi_4}^{\bar{\mathcal{P}}^{(0)}}\,.
 \label{Cf1f2Pb1}
\end{equation}
Note the opposite sign of the R-charge in the denominators of \eqref{Cf1f2P1}, \eqref{Cf1f2Pb1} resulting from the presence of the conjugate operator. The two-point function can be found by using the superconformal algebra
\begin{equation}
 f_{\mathcal{P}^{(1)}\bar{\mathcal{P}}^{(1)}} = 
 8(\Delta + s)(\Delta + s + 1)(\Delta  + s+q)(\Delta + s-q)f_{\mathcal{P}^{(0)}\bar{\mathcal{P}}^{(0)}}\,.
\end{equation}
Putting the pieces together, we find
\begin{equation}
 a_1 = \frac{f_{\mathcal{P}^{(1)}\bar{\mathcal{P}}^{(1)}}}{f_{\mathcal{P}^{(0)}\bar{\mathcal{P}}^{(0)}}} \frac{c_{\phi_1\phi_2}^{\mathcal{P}^{(1)}} c_{\phi_3\phi_4}^{\bar{\mathcal{P}}^{(1)}}} {c_{\phi_1\phi_2}^{\mathcal{P}^{(0)}} c_{\phi_3\phi_4}^{\bar{\mathcal{P}}^{(0)}}} =
 - \frac{(\Delta + \Delta_{12} + s)(\Delta + \Delta_{34} + s)}{2(\Delta + s)(\Delta + s + 1)}\,,
\end{equation}
in agreement with \eqref{eq:sbcoefficients}.

%%%%%%%%%%%%%%%%%%%%%%%%%%%%%%%%
\subsection*{The contribution of $\mathcal{P}^{(2)}$}
%%%%%%%%%%%%%%%%%%%%%%%%%%%%%%%%

The conformal primary $\mathcal{P}^{(2)}$ is given by the contraction
\begin{equation}
 |\mathcal{P}^{(2)}_{i_1\ldots i_{s-1}}\rangle =T_{j}(\Delta-s-1,q)|\mathcal{P}^{(0)}_{ji_1\ldots i_{s-1}}\rangle\,,
\end{equation}
so that the resulting state is automatically symmetric and traceless. Using again \eqref{eq:scondition1}, and this time looking at the next-to-leading power of $\bar{z} = x_1 - \im x_2$ fixes
\begin{equation}
 c_{\phi_1\phi_2}^{\mathcal{P}^{(2)}} =
 - \frac{\im s (\Delta + \Delta_{12} - s - 1)}
 {4(2s + 1)(\Delta - s)(\Delta - s - 1)(\Delta - s - 1 + q)}c_{\phi_1\phi_2}^{\mathcal{P}^{(0)}}\,,
\end{equation}
and similarly
\begin{equation}
 c_{\phi_3\phi_4}^{\bar{\mathcal{P}}^{(2)}} =
 - \frac{\im s (\Delta + \Delta_{34} - s - 1)}
 {4(2s + 1)(\Delta - s)(\Delta - s - 1)(\Delta - s - 1 - q)}c_{\phi_3\phi_4}^{\bar{\mathcal{P}}^{(0)}}\,.
\end{equation}
The norm of $|\mathcal{P}^{(2)}\rangle$ is
\begin{equation}
 f_{\mathcal{P}^{(2)}\bar{\mathcal{P}}^{(2)}} = 
 8\frac{(2s+1)}{(2s-1)}(\Delta - s)(\Delta - s - 1)(\Delta -s-1+ q)(\Delta - s - 1 - q)f_{\mathcal{P}^{(0)}\bar{\mathcal{P}}^{(0)}}\,,
\end{equation}
leading to
\begin{equation}
 a_2 = \frac{f_{\mathcal{P}^{(2)}\bar{\mathcal{P}}^{(2)}}}{f_{\mathcal{P}^{(0)}\bar{\mathcal{P}}^{(0)}}} \frac{c_{\phi_1\phi_2}^{\mathcal{P}^{(2)}} c_{\phi_3\phi_4}^{\bar{\mathcal{P}}^{(2)}}} {c_{\phi_1\phi_2}^{\mathcal{P}^{(0)}} c_{\phi_3\phi_4}^{\bar{\mathcal{P}}^{(0)}}} =
 - \frac{s^2(\Delta + \Delta_{12} - s - 1)(\Delta + \Delta_{34} - s -1)}{2(4s^2-1)(\Delta - s)(\Delta - s - 1)}\,,
\end{equation} 
in harmony with \eqref{eq:sbcoefficients} for $d=3$.

%%%%%%%%%%%%%%%%%%%%%%%%%%%%%%%%
\subsection*{The contribution of $\mathcal{P}^{(3)}$}
%%%%%%%%%%%%%%%%%%%%%%%%%%%%%%%%

In order to be able to write a relatively compact expression for $\mathcal{P}^{(3)}$ it is convenient to define the following operator, which takes symmetric traceless tensors of spin $s$ and dimension $\Delta$, into symmetric traceless tensors of spin $s$, dimension $\Delta+1$, and opposite parity
\begin{equation}
 U^{ab}_{\eta}:\quad|\psi_{i_1\ldots i_s}\rangle \mapsto \eta\, \epsilon^{\alpha\beta}Q^a_{\alpha}Q^b_{\beta}|\psi_{i_1\ldots i_s}\rangle + \im\sum\limits_{n=1}^{s}\varepsilon_{ji_{n}k}(\sigma_{j})^{\alpha\beta}Q_{\alpha}^{a}Q_{\beta}^{b}|\psi_{ki_1\ldots\hat{i}_n \ldots i_s}\rangle\,,
 \label{eq:u}
\end{equation}
where $a,b = \pm$ are R-charge indices, $\eta\in\mathbb{C}$, and $\varepsilon_{ijk}$ is the standard antisymmetric tensor with $\varepsilon_{123}=1$. This operator is useful also to write down the other four conformal primaries with dimension $\Delta + 1$. We find that
\begin{equation}
 \begin{aligned}
  |\mathcal{P}^{+-}\rangle &=U^{+-}_{\Delta - 1 + q}|\mathcal{P}^{(0)}\rangle\,,\\
  |\mathcal{P}^{-+}\rangle &=U^{-+}_{\Delta - 1 - q}|\mathcal{P}^{(0)}\rangle\,,\\
  |\mathcal{P}^{++}\rangle &=U^{++}_{\eta}|\mathcal{P}^{(0)}\rangle\,,\\
  |\mathcal{P}^{--}\rangle &=U^{--}_{\eta}|\mathcal{P}^{(0)}\rangle\,,
 \end{aligned}
\end{equation}
are conformal primaries, the first two of which have been discussed below \eqref{eq:scondition2}. The parameter $\eta$ is arbitrary for the last two cases since the second part of \eqref{eq:u} drops out by the symmetry of $(\sigma_{j})^{\alpha\beta}$, and the anticommutativity of supercharges of the same R-charge. We must remember that $U_{\eta}^{ab}$ acts not only on the Hilbert space, but also on the vector indices. The conformal primary of dimension $\Delta + 2$ and spin $s$ can then be written as
\begin{equation}
\begin{aligned}
 |\mathcal{P}^{(3)}\rangle =
 &(\Delta + s - q)(\Delta - s - 1 - q) \left(U^{++}_{\Delta}U^{--}_{\Delta}-U^{+-}_{\Delta+q}U^{+-}_{\Delta-1+q}-
 U^{-+}_{\Delta-q}U^{+-}_{\Delta-1+q}\right)
 |\mathcal{P}^{(0)}\rangle +\\
 +
 &(\Delta + s + q)(\Delta - s - 1 + q) \left(U^{--}_{\Delta}U^{++}_{\Delta}-
 U^{-+}_{\Delta-q}U^{-+}_{\Delta-1-q}-
 U^{+-}_{\Delta+q}U^{-+}_{\Delta-1-q}\right)
 |\mathcal{P}^{(0)}\rangle\,.
\end{aligned}
\end{equation}
Note that the second line is obtained from the first by flipping all R-charge indices and the sign of the R-charge. Using once again \eqref{eq:scondition1}, and looking at the leading power of $\bar z$ for the lowest scaling dimension where $|\mathcal{P}^{(3)}\rangle$ contributes, one can show that
\begin{equation}
 c_{\phi_1\phi_2}^{\mathcal{P}^{(3)}} =
 - \frac{(\Delta + \Delta_{12} + s)(\Delta + \Delta_{12} - s - 1)c_{\phi_1\phi_2}^{\mathcal{P}^{(0)}}}
 {16(4\Delta^2 - 1)(\Delta^2 - s^2)(\Delta^2 - (s + 1)^2)(\Delta + s + q)(\Delta - s - 1 + q)}\,,
\end{equation}
and so
\begin{equation}
 c_{\phi_3\phi_4}^{\bar{\mathcal{P}}^{(3)}} =
 - \frac{(\Delta + \Delta_{34} + s)(\Delta + \Delta_{34} - s - 1)c_{\phi_3\phi_4}^{\bar{\mathcal{P}}^{(0)}}}
 {16(4\Delta^2 - 1)(\Delta^2 - s^2)(\Delta^2 - (s + 1)^2)(\Delta + s - q)(\Delta - s - 1 - q)}\,.
\end{equation}
The norm is
\begin{equation}
 f_{\mathcal{P}^{(3)}\bar{\mathcal{P}}^{(3)}} = 
 64\Delta^2(4\Delta^2 - 1)(\Delta^2-s^2)(\Delta^2 - (s+1)^2)((\Delta+s)^2 - q^2)((\Delta-s-1)^2 - q^2)
 f_{\mathcal{P}^{(0)}\bar{\mathcal{P}}^{(0)}}\,,
\end{equation} 
so that
\begin{equation}
a_3 = \frac{f_{\mathcal{P}^{(3)}\bar{\mathcal{P}}^{(3)}}}{f_{\mathcal{P}^{(0)}\bar{\mathcal{P}}^{(0)}}} \frac{c_{\phi_1\phi_2}^{\mathcal{P}^{(3)}} c_{\phi_3\phi_4}^{\bar{\mathcal{P}}^{(3)}}} {c_{\phi_1\phi_2}^{\mathcal{P}^{(0)}} c_{\phi_3\phi_4}^{\bar{\mathcal{P}}^{(0)}}} =
  \frac{\Delta^2(\Delta + \Delta_{12} + s)(\Delta + \Delta_{12} - s - 1)(\Delta + \Delta_{34} + s)(\Delta + \Delta_{34} - s -1)}{4(4\Delta^2-1)(\Delta^2-s^2)(\Delta^2 - (s+1)^2)}\,,
\end{equation}
in complete agreement with \eqref{eq:sbcoefficients} for $d=3$.

%%%%%%%%%%%%%%%%%%%%%%%%%%%%%%%%%%%%%
\section{Decomposition of the generalized free chiral correlator}
\label{App:gff}
%%%%%%%%%%%%%%%%%%%%%%%%%%%%%%%%%%%%%

A natural solution to the crossing equation using the conformal blocks in \eqref{eq:superblock} in both channels corresponds to the supersymmetric analogue of the generalized free field. The elementary fields of this theory are a chiral scalar primary $\phi(x)$ of dimension $\Delta_{\phi}$, its supersymmetric descendants -- a fermion $\psi_a(x) = (Q^-_{a}\phi(x))$, and the auxiliary field $F(x) = \epsilon^{ab}Q^-_{a}Q^-_{b}\phi(x)$, as well as their conjugates. The correlators are computed using Wick's theorem, where each field only couples to its conjugate. The decomposition of the correlator $\langle\phi\bar\phi\phi\bar\phi\rangle$ into ordinary conformal blocks was given in \cite{Fitzpatrick:2011dm}. We can use this decomposition together with \eqref{eq:sbdecomposition} to find the decomposition into superconformal blocks
\begin{equation}
\begin{aligned}
  \left\langle\phi(x_1)\bar\phi(x_2)\phi(x_3)\bar\phi(x_4)\right\rangle &= \frac{1}{|x_{12}|^{2\Delta_\phi}|x_{34}|^{2\Delta_\phi}}\left[1+ \left(\frac{u}{v}\right)^{\Delta_{\phi}}\right] =\\
  &= \frac{1}{|x_{12}|^{2\Delta_\phi}|x_{34}|^{2\Delta_\phi}}\left[\mathcal{G}_{0,0}^{0,0}(u,v)+\sum\limits_{n,s\geq0}p_{n,s}\mathcal{G}^{0,0}_{2\Delta_{\phi} + 2n + s,s}(u,v)\right]\,,
\end{aligned}
\label{GFF4pt}
\end{equation}
where
\begin{equation}
 p_{n,s} = \frac{(-2)^s (\Delta_{\phi})_{n+s}^2\left(\Delta_{\phi}-\frac{d}{2}+1\right)_{n}^2}
 {n!s!\left(2\Delta_{\phi}+2n+s\right)_{s}\left(s+\frac{d}{2}\right)_{n}\left(2\Delta_{\phi}+n-d+2\right)_{n}\left(2\Delta_{\phi}+n+s-\frac{d}{2}+1\right)_{n}}\,,
 \label{GFFcoeff}
\end{equation}
and $(x)_n \equiv \Gamma(x+n)/\Gamma(x)$ is the Pochhammer symbol. This decomposition serves as a further test of the validity of the superconformal blocks in \eqref{eq:superblock}. The result in \eqref{GFF4pt}, \eqref{GFFcoeff} strongly resembles the decomposition of the four-point function of non-supersymmetric generalized free fields into ordinary conformal blocks given in equation (43) of \cite{Fitzpatrick:2011dm}. In fact the two results are identical to each other (up to an overall normalization) if one fixes $\Delta_1=\Delta_2=\Delta_{\phi}$ in equation (43) of \cite{Fitzpatrick:2011dm}, and performs the shift $2\Delta_{\phi} \mapsto 2\Delta_{\phi} -1$ in the denominator of \eqref{GFFcoeff}. This is reminiscent of the observations in Section \ref{ssec:susynosusy}.

%%%%%%%%%%%%%%%%%%%%%%%%%%%%%%%%%%%%%
\section{$\mathcal{N}=2$ minimal models}
\label{App:minmod}
%%%%%%%%%%%%%%%%%%%%%%%%%%%%%%%%%%%%%

%%%%%%%%%%%%%%%%%%%%%%%%%%%%%%%%%%%%%
\subsection*{Generalities}
\label{App:C1}
%%%%%%%%%%%%%%%%%%%%%%%%%%%%%%%%%%%%%

Here we collect some well-known facts about $\mathcal{N}=2$ minimal models in two dimensions, see for example \cite{Boucher:1986bh}. For a discussion on $\mathcal{N}=1$ minimal models see \cite{Friedan:1984rv}. The (holomorphic) infinite-dimensional $\mathcal{N}=2$ superconformal algebra in two dimensions is:
\begin{equation}\label{N2SCAinf}
\begin{aligned}
\, [L_m, L_n ] &= \frac{c}{12} \, (m^3 -m) \, \delta_{m+n,0} + (m-n) \, L_{m+n}\,, \\
[L_m, G_r^\pm] &= \Big( \frac m2 - r \Big) G^\pm_{m+r} \,,\\
[L_m, \Omega_n] &= - n \, \Omega_{m+n}\,, \\
\{ G_r^+, G_s^- \} &= \frac{c}3 \Big( r^2 - \frac14 \Big) \delta_{r+s,0} + 2 L_{r+s} + (r-s) \Omega_{r+s}\,,
\end{aligned}\qquad\begin{aligned}
\{ G_r^+, G_s^+ \} &= \{ G_r^-, G_s^- \} = 0\,, \\
[\Omega_n, G_r^\pm ] &= \pm G_{r+n}^\pm \,,\\
[\Omega_m, \Omega_n] &= \frac{c}3 \, m \, \delta_{m+n,0} \;.
\end{aligned}
\end{equation}
Here $m$ and $n$ are integers and in the NS sector $r$ and $s$ are half-integers. The modes of the energy momentum tensor are $L_{m}$, those of the superconformal R-symmetry are $\Omega_n$ and the two supercharges have modes $G^{\pm}_r$. The real number $c$ is the (left or right moving) central charge and it is related to the conformal anomaly of the CFT.

The finite, $sl(2|1)$, subalgebra of the superconformal algebra, given in \eqref{N=2algebra2d}, is obtained from \eqref{N2SCAinf} by restricting to the generators $\{L_{-1,0,1}, \Omega\equiv \Omega_0,G^{\pm}_{\pm 1/2}\}$.
Unitary representations of the infinite-dimensional $\mathcal{N}=2$ superconformal algebra exist for any real $c\geq 3$ and for the discrete values
\begin{equation}
c = \frac{3k}{k+2}\;, \qquad\qquad k=0,1,2,\ldots\;.
\end{equation}
This discrete series is usually referred to as the $\mathcal{N}=2$ minimal models. The dimensions and R-charges of the superconformal primary operators in the NS sector of the $k$-th minimal model are labeled by two integers $m$ and $n$
\begin{equation}
h = \frac{n(n+2)-m^2}{4(k+2)}\;, \quad \Omega = \frac{m}{k+2}\;, \quad 0\leq n \leq k\;, \quad -n\leq m \leq n\;, \quad m+n = \text{even}\;.
\end{equation}
The fusion rules for $\mathcal{N}=2$ minimal models are derived in \cite{Mussardo:1988av,Mussardo:1988ck} (see also \cite{Kiritsis:1987np}).\footnote{Notice that there is a factor of $1/2$ difference between our conventions for the R-charge and the ones in \cite{Mussardo:1988av,Mussardo:1988ck}.} The superconformal primaries in the $k$-th minimal model will be denoted by $\phi_{n,m}$. The fusion rules are then 
\begin{equation}
\phi_{n_1,m_1} \times \phi_{n_2,m_2} = \ds\sum_{n=\frac{m_2-n_2}{2}}^{\frac{m_2+n_2}{2}} \phi_{n_1-m_2+2n,m_1+m_2}\;.
\end{equation}
The $k$-th $\mathcal{N}=2$ minimal model has a $\mathbb{Z}_{k+2}$ symmetry generated by some of the primaries in the Ramond sector, see \cite{Mussardo:1988av}. Chiral, antichiral primaries are superconformal primaries also annihilated by $G_{-1/2}^{+}$, $G_{-1/2}^{-}$, respectively, which is equivalent to $\Omega = \pm 2 h$. In the minimal models, these are operators with $m = \pm n$, respectively.

One can derive a universal bound for the central charge of a two-dimensional $\mathcal{N}=2$ SCFT using the infinite-dimensional superconformal algebra, see for example \cite{Maldacena:1998bw}. Using unitarity and the algebra in \eqref{N2SCAinf} one finds
\begin{equation}
0 \leq \langle \phi| \{G^{+}_{-3/2},G^{-}_{3/2}\}|\phi \rangle =\langle \phi|\left(2L_{0}-3\Omega+\frac{2c}{3}\right)|\phi \rangle\;,
\end{equation}
for any superconformal primary $|\phi \rangle$. Thus we arrive at the following constraint for the dimension and R-charge of any superconformal primary
\begin{equation}
2h-3\Omega+\frac{2c}{3} \geq 0\;,
\end{equation}
which becomes $h \leq c/6$ for a chiral primary. The bound is saturated only if the state $G^{+}_{-3/2}|\phi \rangle$ is null. The highest-dimension chiral primary in every minimal model has $m=n=k$ and saturates the bound. If we have a unitary $(2,2)$ SCFT with a diagonal spectrum, i.e. $\bar{h}=h=\Delta/2$, we arrive at the following lower bound on the central charge
\begin{equation}
C_T \geq 6 \Delta_{\rm max}\;,
\label{stringexcl}
\end{equation}
where $\Delta_{\rm max}$ is the highest dimension of a superconformal primary in the theory. One can repeat the same analysis with the state $G^{+}_{-(2p+1)/2}|\phi \rangle$ with $p=1,2,3,\ldots$ and find the bound
\begin{equation}
C_T \geq \frac{12}{p+1} \Delta_{\rm max}\;,
\end{equation}
Clearly the strongest bound is obtained for $p=1$ as in \eqref{stringexcl}. 

In Section \ref{sec:2d_minmodels} we claimed that in every unitary CFT with $\cN=(2,2)$ supersymmetry there is always an operator (or state) of dimension $\Delta=2$ which is a superdescendant of the identity. This state is given by
\begin{equation}
\Omega_{-1}\bar{\Omega}_{-1}|0\rangle\;,
\end{equation}
where $|0\rangle$ is the NS vacuum and $\Omega_{-1}$ is defined in \eqref{N2SCAinf}. This state clearly has dimension $\Delta = h+\bar{h} = 2$ and R-charge $q=\Omega+\bar\Omega = 0$. Moreover its norm is given by
\begin{equation}
\langle0|\Omega_{1}\bar{\Omega}_{1}\Omega_{-1}\bar{\Omega}_{-1}|0\rangle = \frac{c^2}{9}\;.
\end{equation}
Therefore in a unitary theory the state is never null since $c>0$.

%%%%%%%%%%%%%%%%%%%%%%%%%%%%%%%%%%%%%
\subsection*{Super-Ising in $d=2$}
\label{App:C2}
%%%%%%%%%%%%%%%%%%%%%%%%%%%%%%%%%%%%%

The theory with $c=1$ can be realized in terms of a single compact boson, $\varphi$, at a specific radius $R=\sqrt{3}$ \cite{Waterson:1986ru,Kiritsis:1987np}. There are three superconformal primary operators
\begin{equation}
\phi_{1,\pm1} = :e^{\pm\frac{\im}{\sqrt{3}}\varphi}:\;, \qquad\qquad \phi_{0,0} =1\;.
\end{equation}
The operator $\phi_{1,1}$ is chiral with $\Delta=q/2=\frac{1}{3}$, and is identified with the (holomorphic part of the) operator $\Phi$ in Section \ref{sec:2d_minmodels}. Similarly the operator $\phi_{1,-1}$ is antichiral with $\Delta=-q/2=\frac{1}{3}$, and is identified with the operator $\bar{\Phi}$. One can now use the formula
\begin{equation}
:e^{\im a\varphi(z_1)}::e^{\im b\varphi(z_2)}: = (z_1-z_2)^{ab} :e^{\im a\varphi(z_1)+\im b\varphi(z_2)}:\;, 
\end{equation}
where $a$ and $b$ are some constants, to find the OPE
\begin{equation}
\phi_{1,1}(z_1)\phi_{1,-1}(z_2) \sim \frac{1}{(z_1-z_2)^{1/3}}+\frac{\im}{\sqrt{3}}\partial_{z_2}\varphi(z_2)(z_1-z_2)^{2/3}+\ldots \,.
\end{equation}
We normalize all two point functions in the theory to have coefficients $1$ and define the operator $\mathcal{O}_{\epsilon}(z) \equiv \im\partial_{z}\varphi(z)$, which has dimension $h=1$. The operator of dimension $\Delta=2$, which should be identified with $[\Phi\bar{\Phi}]$ from Section \ref{sec:2d_minmodels}, is obtained by taking $\mathcal{O}_{\epsilon}(z)\bar{\mathcal{O}_{\epsilon}}(\bar{z}) = \partial_{z}\varphi(z)\partial_{\bar{z}}\bar{\varphi}(\bar{z})$. Another useful OPE is given by
\begin{equation}
\im\partial_{z_1}\varphi(z_1) :e^{\im a\varphi(z_2)}: \sim a\frac{:e^{\im a\varphi(z_2)}:}{z_1-z_2} + \ldots\;.
\end{equation}
With these OPEs at hand one finds the following three point function
\begin{equation}
\langle \phi_{1,1}(z_1)\phi_{1,-1}(z_2)\mathcal{O}_{\epsilon}(z_3)\rangle = \frac{1}{\sqrt{3}} \frac{1}{(z_1-z_2)^{-2/3}(z_2-z_3)(z_1-z_3)}\;.
\end{equation}
Combining the left and right-moving sectors one finds the three-point function
\begin{equation}
\langle \Phi(z_1,\bar{z}_1)\bar{\Phi}(z_2,\bar{z}_2)[\Phi\bar{\Phi}](z_3,\bar{z}_3)\rangle = \frac{1}{3} \frac{1}{|z_1-z_2|^{-4/3}|z_2-z_3|^2|z_1-z_3|^2}\;.
\end{equation}
Thus we find that the OPE coefficient denoted by $c_{ \Phi\bar{\Phi}[\Phi\bar{\Phi}]}$ is given by
\begin{equation}
c_{ \Phi\bar{\Phi}[\Phi\bar{\Phi}]} = \frac{1}{3}\;.
\end{equation}
This matches nicely with the numerical value at the kink in the right panel of Figure \ref{fig:2d_CT_OPE}.

\subsection*{A comment on two supercharges}

Finally, let us mention a tangential observation about bootstrap of $(1,1)$ SCFTs in $d=2$, complementing the results of \cite{Bashkirov:2013vya} with $\mathcal{N}=1$ supersymetry in $d=3$. Analogously to that study, also with $(1,1)$ supersymmetry in $d=2$, the superconformal blocks are equal to the conformal blocks, so that there are no additional constraints from crossing symmetry besides the numerical bounds obtained in \cite{Rattazzi:2008pe,ElShowk:2012ht}. However, it may happen that the leading scalar appearing in the $\sigma \times \sigma$ OPE is the superdescendant of $\sigma$ itself. In this case, we have the extra constraint $\Delta_{[\sigma \sigma]}=\Delta_\sigma+1$. The result is that the two lines intersect at $\Delta_{\sigma} \approx 1/5$ and $\Delta_{\epsilon}\approx6/5$ for $d=2$. These dimensions correspond to the Virasoro minimal model with central charge $c=7/10$, i.e. the tricritical Ising model. This is in fact the first $\mathcal{N}=1$ minimal model \cite{Friedan:1984rv}, in harmony with the analogous results of \cite{Bashkirov:2013vya} in $d=3$.

%%%%%%%%%%%%%%%%%%%%%%%%%%%%%%%%%%%%%
\end{appendices}
%%%%%%%%%%%%%%%%%%%%%%%%%%%%%%%%%%%%%

\bibliography{Bib}
\bibliographystyle{JHEP}

\end{document}